\documentclass[12pt
]{article}

\usepackage{graphicx}
\usepackage{amsfonts}
\usepackage{amsmath}
\usepackage[usenames]{color}
\usepackage{amssymb}
\usepackage[mathscr]{eucal} 

\usepackage[all]{xy}

\def\Z{\mathbb{Z}}

\def\R{\mathbb{R}}
\def\C{\mathbb{C}}
\def\P{\mathbb{P}}

\begin{document}
\baselineskip 0.6cm
\newcommand{\gsim}{ \mathop{}_{\textstyle \sim}^{\textstyle >} }
\newcommand{\lsim}{ \mathop{}_{\textstyle \sim}^{\textstyle <} }
\newcommand{\vev}[1]{ \left\langle {#1} \right\rangle }
\newcommand{\bra}[1]{ \langle {#1} | }
\newcommand{\ket}[1]{ | {#1} \rangle }
\newcommand{\Dsl}{\mbox{\ooalign{\hfil/\hfil\crcr$D$}}}
\newcommand{\nequiv}{\mbox{\ooalign{\hfil/\hfil\crcr$\equiv$}}}
\newcommand{\nsupset}{\mbox{\ooalign{\hfil/\hfil\crcr$\supset$}}}
\newcommand{\nni}{\mbox{\ooalign{\hfil/\hfil\crcr$\ni$}}}
\newcommand{\EV}{ {\rm eV} }
\newcommand{\KEV}{ {\rm keV} }
\newcommand{\MEV}{ {\rm MeV} }
\newcommand{\GEV}{ {\rm GeV} }
\newcommand{\TEV}{ {\rm TeV} }

\def\diag{\mathop{\rm diag}\nolimits}
\def\tr{\mathop{\rm tr}}

\def\Spin{\mathop{\rm Spin}}
\def\SO{\mathop{\rm SO}}
\def\O{\mathop{\rm O}}
\def\SU{\mathop{\rm SU}}
\def\U{\mathop{\rm U}}
\def\Sp{\mathop{\rm Sp}}
\def\SL{\mathop{\rm SL}}

\def\change#1#2{{\color{blue}#1}{\color{red} [#2]}\color{black}\hbox{}}


\begin{titlepage}
 
\begin{flushright}
 LTH 829 \\
 UT-09-10 \\
 IPMU09-0047
\end{flushright}
 
\vskip 1cm
\begin{center}
 
 {\large \bf Right-handed Neutrinos in F-theory Compactifications }
 
 \vskip 1.2cm
 
 Radu Tatar$^1$, Yoichi Tsuchiya$^2$ and Taizan Watari$^{3}$

 \vskip 0.4cm
 
 {\it 
  $^1$Division of Theoretical Physics, Department of Mathematical
  Sciences, The University of Liverpool, Liverpool, L69 3BX, England,
  U.K. \\[2mm]

 $^2$Department of Physics, University of Tokyo, Tokyo 113-0033, Japan  
 \\[2mm]
 
 $^3$Institute for the Physics and Mathematics of the Universe, University of Tokyo, Kashiwano-ha 5-1-5, 277-8592, Japan
 }
 \vskip 1.5cm
 
\abstract{F-theory is one of the frameworks where up-type Yukawa
couplings of SU(5) unified theories are naturally generated. As charged matter
fields have localized zero modes in F-theory, a study of flavor
 structure could be easier in F-theory than in
Heterotic string theory. In a study of flavor structure in the 
lepton sector, however, an important role is played by right-handed 
neutrinos, which are not charged under the SU(5) unified gauge group. 
It is therefore solicited to find out what right-handed neutrinos are 
in F-theory compactifications and how their Majorana mass terms are
generated together with developing a theoretical framework where 
effective Yukawa couplings involving both SU(5)-neutral and charged 
fields can be calculated.
We find that the complex structure moduli chiral multiplets of F-theory
compactifications are good candidates to be right-handed neutrinos, 
and that their Majorana masses are automatically generated in flux 
compactifications. The mass scale is predicted to be somewhat below 
the GUT scale, which is in nice agreement with the $\Delta m^2$ of 
the atmospheric neutrino oscillation through the see-saw mechanism. 
We also discuss various scenarios of solving the dimension-4 proton 
decay problem in supersymmetric F-theory compactifications, along 
with considering the consequences of those scenarios in the nature 
of right-handed neutrinos. 
} 
 
 \end{center}
 \end{titlepage}



\section{Introduction}

The Standard Model of particle physics, including neutrino
masses, have 20--22 parameters associated with flavor, depending 
on whether the neutrino masses are Dirac or Majorana. The flavor 
parameters constitute the dominant fraction of the parameters 
appearing in the effective action of the Standard Model and this  calls 
for a better theoretical understanding. Super Yang--Mills theories 
in higher-dimensional spacetime realized in geometric 
compactifications of superstring theory can   yield charged massless 
matter fields as well as their Yukawa interactions by further
compactification down to 3+1 dimensions. The possible forms 
of trilinear couplings of the matter fields are determined by 
the Lie algebra of the microscopic super Yang--Mills theory, and 
a detailed pattern of the Yukawa couplings follows from geometry. 
Compactifications of superstring theory, therefore, have a chance 
to be a framework for understanding of the flavor physics.

If one takes $\SU(5)_{\rm GUT}$ unification seriously,\footnote{
quark doublets and lepton doublets belong to different 
irreducible representations} then one has to wonder what higher dimensional
super Yang--Mills theory would 
give rise to the up-type Yukawa couplings
\begin{equation}
 \Delta {\cal L} = \lambda^{(u)} {\bf 10}^{ab} {\bf 10}^{cd} 
 H({\bf 5})^e \epsilon_{abcde}.
\label{eq:up-Yukawa}
\end{equation}
This contraction by $\epsilon_{abcde}$ arises from the structure 
constant of $E_6$ Lie algebra and a $E_{7,8}$ that contains $E_6$. 
Heterotic $E_8 \times E_8$ string theory, M-theory and F-theory 
are the candidates for such a theoretical framework \cite{TW-1}.

In M-theory compactifications on manifolds with $G_2$ holonomy and 
in F-theory compactifications on elliptic fibered Calabi--Yau 4-folds,the 
charged matter fields are localized in the internal spaces. This
localized picture of matter fields enables us to discuss the flavor 
pattern of Yukawa matrices in rather intuitive ways. Explanations 
for hierarchical eigenvalues \cite{AS} as well as generation structures 
in the quark sector \cite{HSW} have been achieved in phenomenological 
models by using higher-dimensional gauge theory with localized matter
wavefunctions. An intuitive picture of localized matter fields 
and the mechanism of generating Yukawa couplings in M-theory
compactifications has also been exploited in \cite{TW-1}, to see 
that the up-type Yukawa matrix in (\ref{eq:up-Yukawa}) tends to 
have suppressed diagonal entries in M-theory compactifications 
on manifolds with $G_2$ holonomy. That is not in very good 
agreement with the real world.  It is also worth studying 
flavor pattern of generic F-theory compactifications, exploiting 
the fact that charged matter fields have localized wavefunctions. 

It is by now known how many massless charged matter fields are 
in the low-energy spectrum, and how to determine their zero-mode 
wavefunctions, once a geometry for supersymmetric F-theory 
compactification is given \cite{Curio, DI, DW-1, BHV-1, Hayashi-1, Hayashi-2}. 
A prescription for how to calculate Yukawa couplings for three charged 
matter fields in F-theory compactifications has also been written 
down \cite{BHV-1, DW-1, BHV-2, Hayashi-2}. A field theory formulation 
capturing a certain sector of dynamical degrees of freedom 
in F-theory \cite{KV, DW-1, BHV-1} turns out to be very useful 
for this purpose \cite{Hayashi-2}. 

Nevertheless, the Yukawa couplings involving only 
$\SU(5)_{\rm GUT}$-charged matter fields
do not account for neutrino Yukawa couplings. In order to study 
flavor physics in the lepton sector, it is crucial that we have 
a theoretical formulation dealing with Yukawa couplings that involve 
neutral matter fields. Therefore, in this article, we develop the crucial issue of
identifying the ($\SU(5)_{\rm GUT}$-neutral) right-handed 
neutrinos in the F-theory compactifications. We develop a general prescription of 
calculating the Yukawa couplings involving neutral fields, with a direct 
application for neutrino Yukawa couplings. 

In section \ref{sec:Majorana}, we will discuss a possibility 
that complex structure moduli of F-theory compactifications are 
identified with chiral multiplets of right-handed neutrinos. 
Those moduli fields acquire large masses in flux compactifications.
A main result of our paper is that  the typical mass scale in flux
compactifications, being below the GUT scale, is just about right 
for the mass scale of the right-handed neutrinos predicted by
experimental data.

The neutral fields of unified theories such as right-handed neutrinos are 
better captured in generic F-theory compactifications as moduli fields 
of Calabi--Yau 4-fold compactifications. On the other hand, field theory 
models for local geometry of Calabi--Yau 4-folds are better tools in 
calculating Yukawa couplings such as neutrino Yukawa couplings, where
both charged fields and these moduli fields are involved. We therefore 
explain in section \ref{sec:nuYukawa-general} how to combine the two
descriptions of F-theory compactifications to calculate Yukawa couplings 
involving neutral fields. As a digression, we discuss the cubic term of 
a neutral field in the next-to-minimal supersymmetric standard model 
in section \ref{sssec:3-singlets}.

The dimension-4 proton decay is a serious phenomenological problem in 
compactifications with low-energy supersymmetry. The absence of 
rapid proton decay implies that the complex structure moduli 
for our real world is somewhat special. Thus, right-handed 
neutrinos should be regarded as fluctuations of complex structure 
moduli from a special choice of complex structure moduli in our vacuum.
In section \ref{sec:Dim4}, we present some scenarios solving the 
dimension-4 proton decay problem, and discuss the consequences 
in the physics of right-handed neutrinos. 

We noticed that recent articles \cite{DW-3, BHSV, Randall, HV-CP,
Caltech, Bourjaily} cover similar subjects. 

\section{Singlet Masses from Moduli Stabilization}
\label{sec:Majorana}

The right-handed neutrinos $\bar{N}$ are not charged under $\SU(5)_{\rm GUT}$ 
unification group, and they are supposed to have trilinear 
couplings 
\begin{equation}
 \Delta {\cal L} = \lambda^{(\nu)}_{ij} \bar{N}_i l_j h_u + {\rm h.c.}. 
\label{eq:nu-Yukawa}
\end{equation}
Here, $l_j$ are lepton doublets of the Standard Model, and 
$h_u$ the Higgs doublet. This is the only certainty from phenomenology. 
Therefore, in string phenomenology, any 
light degrees of freedom that are neutral under $\SU(5)_{\rm GUT}$ 
are qualified to be considered right-handed neutrinos, as long as they have 
the coupling (\ref{eq:nu-Yukawa}).

Neutrino masses indicated by neutrino oscillation experiments are 
much smaller than the mass eigenvalues of the quarks and charged 
leptons of the Standard Model, this being a natural prediction 
of the see-saw mechanism. If right-handed neutrinos have mass terms, 
\begin{equation}
 \Delta {\cal L} = M_{ii'} \bar{N}_i \bar{N}_{i'} + {\rm h.c.},
\end{equation}
with the eigenvalues of $M_{ii'}$ much larger than the electroweak 
scale, then the Majorana masses of the left-handed neutrinos 
are generated, with their mass eigenvalues much smaller than 
the electroweak scale. From the measured value  
$\Delta m^2 \simeq 2\mbox{--}3 \times 10^{-3} \; \EV^2$
of the atmospheric neutrino oscillation \cite{PDG}, one can conclude that 
the lightest right-handed neutrino is not heavier than about 
\begin{equation}
 \frac{(v \lambda^{(\nu)})^2}{\sqrt{\Delta m^2}} = 
 (\lambda^{(\nu)})^2 \times (5.5\mbox{--}6.7) \times 10^{14} \; \GEV.
\label{eq:MR-bound}
\end{equation}
Here, $v \simeq 174 \; \GEV$ is the Higgs vacuum expectation value (vev). 
The neutrino Yukawa couplings $\lambda^{(\nu)}$ should not be 
much larger than unity, otherwise the perturbative field theory 
description would immediately become invalid because of the 
renormalization group flow of the couplings $\lambda^{(\nu)}$.  
Thus, the mass scale of right-handed neutrinos is below the energy 
scale of gauge coupling unification $M_{\rm GUT} \sim 10^{16} \; \GEV$ 
(simply called the GUT scale) with a safe margin. 

We will now address the issues:

$\bullet$ What are the right-handed neutrinos in string phenomenology, and 
how many of them are there?  

$\bullet$ Where does the energy scale of 
the Majorana masses $M_{ii'}$ come from 

$\bullet$ Where do the 
neutrino Yukawa couplings (\ref{eq:nu-Yukawa}) come from?

\subsection{Complex Structure Moduli as Right-Handed Neutrinos}

The compactification of F-theory to ${\cal N} = 1$ supersymmetry in 
4-dimensions is described by a set of data, $(X,G^{(4)})$. $X$ 
is a Calabi--Yau 4-fold that is an elliptic fibration over 
a base 3-fold $B_3$:
\begin{equation}
 \pi_X: X \rightarrow B_3.
\end{equation}
$G^{(4)} = dC^{(3)}$ is a 4-form flux on $X$, and $C^{(3)}$ is the 3-form 
potential in the language of eleven-dimensional supergravity.
The 4-form flux $G^{(4)}$ has to take its value only in the (2,2) 
component, and be primitive in order to leave an 
unbroken ${\cal N} = 1$ supersymmetry \cite{Becker}.
This class of compactifications has $h^{3,1}$ complex structure moduli, 
$h^{1,2}$ moduli associated with the configuration of $C^{(3)}$ that is 
not reflected in the field strength $G^{(4)} = dC^{(3)}$,  and 
K\"{a}hler moduli chiral multiplets. 

The complex structure moduli have interactions in the superpotential 
\cite{GVW} 
\begin{equation}
 \Delta W = W_{\rm GVW} = \int_X \Omega \wedge G^{(4)}.
\label{eq:GVW}
\end{equation}
In the absence of a flux $G^{(4)}$ on $X$, complex structure moduli 
parametrizing $\Omega$ would have remained massless (in the absence 
of supersymmetry breaking). 
Once a generic flux $G^{(4)}$ is introduced, however, potential 
is generated for the moduli, and the complex structure of $X$
dynamically sets itself to a minimum of the potential, so that 
$G^{(4)}$ has vanishing (1,3) components. All the complex structure 
moduli generically have mass terms around such a minimum \cite{flux-review}.

When the Calabi--Yau 4-fold $X$ is given by an equation 
$f=0$ on a space with a set of local coordinates $(x_1,x_2,\cdots,x_5)$, 
then $\Omega$ has an expression 
\begin{equation}
 \Omega = {\rm Res}_{f=0} \frac{dx_1 \wedge \cdots \wedge dx_5}{f}
  = \frac{dx_1 \wedge \cdots \wedge dx_4}{\partial f/\partial x_5} =
  \cdots. 
\end{equation}
Coefficients in the defining equation $f=0$---collectively denoted 
by $a$---set the complex structure of $X$.
The vacuum choice $\Omega(a=a_0)$ is a pure (4,0)-form when evaluated on 
a holomorphic coordinates corresponding to the vacuum value $a=a_0$. 
When $\Omega(a)$ is expanded in fluctuations from vacuum 
$\delta a \equiv a - a_0$, 
$\Omega(a)$ at order ${\cal O}(\delta a)$ stays within 
$H^{(4,0)}(X; \C) \oplus H^{(3,1)}(X; \C)$ with respect to the 
vacuum holomorphic coordinates at $a=a_0$ \cite{Candelas}. 
This is why non-vanishing flux $G^{(4)}$ in the (1,3)-component 
when evaluated at vacuum $a=a_0$ would have meant non-vanishing F-term 
vev $\vev{\delta W/ \delta a}|_{a=a_0}$ (and supersymmetry breaking).
The 4-form $\Omega(a)$ at order ${\cal O}((\delta a)^2)$ remains within 
$H^{(4,0)}(X; \C) \oplus H^{(3,1)}(X; \C) \oplus H^{(2,2)}(X; \C)$
evaluated the vacuum complex structure, and is expressed 
as a sum of the form 
\begin{equation}
 \Omega(a) = \Omega(a_0) + \left(k_a \Omega(a_0) + \chi_a \right)
  (\delta a)_a + 
 \left(k_{ab} \Omega(a_0) + l^c_{ab} \chi_c + \psi_{ab} \right) 
  (\delta a)_a (\delta a)_b.
\label{eq:Omega-2nd}
\end{equation}
Here, $\chi_a$'s are basis of $H^{(3,1)}(X_{a=a_0}; \C)$, and 
$\psi_{ab}$ elements of $H^{(2,2)}(X_{a=a_0}; \C)$.
The Gukov--Vafa--Witten superpotential with generic 
flux $G^{(4)} \in H^{(2,2)}(X_{a=a_0})$ gives rise to the non-vanishing 
mass terms (quadratic term)
\begin{equation}
 W_{\rm GVW} = \int_{X} \vev{\Omega} \wedge \vev{G^{(4)}} + 
   \left(\int_{X|_{a=a_0}} \psi_{ab} \wedge \vev{G^{(4)}} \right) 
   (\delta a)_a (\delta a)_b
 + \cdots
\label{eq:M-mass-via-GVW}
\end{equation}
for the fluctuations of the complex structure moduli $(\delta a)$.

This is a standard story of flux compactification and 
stabilization of complex structure moduli in Type IIB string theory / 
F-theory. Now, as discussed at the beginning of this section, any light degrees of freedom that are neutral under 
$\SU(5)_{\rm GUT}$ have a chance to be identified with right-handed
neutrinos. In this article the complex structure moduli are identified with 
the right-handed neutrinos so the moduli masses
from the flux compactification immediately become the Majorana masses
of right-handed neutrinos. 

Note that the need for right-handed neutrinos in the see-saw scenario 
does not motivate phenomenologically an unbroken $\SO(10)$ symmetry 
in the effective theory in 3+1 dimensions, or its realization on a 
stack of coincident branes. What used to be right-handed neutrinos 
in the spinor representation of SO(10) GUT just becomes neutral 
vector-bundle/brane-configuration moduli 
of SU(5) GUT, as one can see immediately by following the Higgs cascade 
studied in \cite{KachVafa, 6authors}. The idea of identifying right-handed
neutrinos with vector bundle moduli in Heterotic string compactification 
with SU(5) GUT dates back (at least to our knowledge) to
\cite{WittenSU(3)}. Under the 
duality between the Heterotic string and F-theory, all of vector bundle 
moduli and complex structure moduli in Heterotic string theory 
correspond to the complex structure moduli $H^{3,1}(X; \C)$ (and
$H^{1,2}(X; \C)$, which we mention later) in F-theory
compactification \cite{MV1, MV2, 6authors, FMW, Het-F-4D}. 
Since the vector bundle moduli in Heterotic string 
theory have the desired Yukawa couplings for neutrinos 
\begin{equation}
 \Delta W = \lambda \; \bar{\bf 5} \; {\bf 1} \; {\bf 5},  
\label{eq:Yukawa-515}
\end{equation}
we expect that the complex structure moduli in F-theory also have 
(at least qualitatively) the same interactions. 
Therefore, it is quite natural to identify the complex structure moduli 
with right-handed neutrinos in F-theory compactification. 
Their Majorana masses are generated by the Gukov--Vafa--Witten
superpotential, as discussed before.

In F-theory compactifications, the number of complex structure moduli, 
$h^{3,1}(X)$, is usually much larger than three, the number of
``generations'' of $\SU(5)_{\rm GUT}$-charged quarks and leptons. 
This is not a contradiction 
from phenomenological perspectives. All we know for sure 
is that at least two right-handed neutrinos are necessary in order to 
account for all the data of neutrino oscillation experiments 
in the see-saw mechanism \cite{FGY}. There is no upper bound from 
phenomenology on the number of right-handed neutrinos. In fact, 
there is even an indication \cite{HSW-2} that the number of right-handed 
neutrinos may be much larger than just three. Thus, a higher number of 
right-handed neutrinos possibly obtained from complex structure moduli 
is not problematic at all, and may even be a blessing in disguise.

\subsection{Estimation of the Majorana Mass Scale}

Let us now estimate the energy scale of the Majorana masses 
of right-handed neutrinos, assuming that the right-handed neutrinos 
are complex structure moduli, and that the Majorana masses derive 
from the superpotential (\ref{eq:GVW}, \ref{eq:M-mass-via-GVW}) in 
F-theory compactifications. We begin with 
a review of a similar problem in Type IIB orientifold
compactifications. 

In Type IIB string compactification on Calabi--Yau orientifolds 
on $B_3 = \widetilde{B}/\Z_2$,
the complex structure moduli acquire masses $m_{cs}$ with an order of magnitude
estimate given by \cite{KST}
\begin{equation}
 m_{cs}^2 \sim m_{\rm KK}^6 l_s^4 = 
 \left[ m_{\rm KK} \times \left(\frac{l_s}{R_6} \right)^2 \right]^2.
\label{eq:cpx-str-mass-IIB}
\end{equation}
Here, $m_{\rm KK} = 1/R_6$ is the ``Kaluza--Klein scale'', assuming 
that the Calabi--Yau 3-fold $\widetilde{B}$ for an orientifold 
compactification of 
Type IIB string theory is almost isotropic and its radius characterized 
by a single parameter $R_6$. We define the string length $l_s$ as 
$l_s \equiv (2\pi) \sqrt{\alpha'}$. 

This expression is understood as follows. 
The quantum fluctuations of the complex structure moduli fields correspond 
to $(0,2)$-type fluctuations of the metric on the Calabi--Yau 3-fold. 
Simple dimensional reduction leads to kinetic terms
\begin{equation}
 \Delta {\cal L}_{\rm kin} \sim \frac{1}{l_s^8 g_s^2} 
   \int_{\widetilde{B}} d^6 y \; R \sim 
  M_{\rm Pl}^2 \; |\partial \phi |^2
  \sim \frac{R_6^6}{l_s^8 g_s^2} \; |\partial \phi|^2.
\label{eq:cpxstr-kin}
\end{equation}
$\phi$ are complex scalar fields in the effective field theory on 
3+1 dimensions and correspond to the complex structure moduli fields.
Mass dimension of the $\phi$ fields is set to zero here.
On the other hand, such fluctuations of the metric change the complex
structure of the K\"{a}hler manifold, and the imaginary-self-dual 
3-form flux configuration in a vacuum is no longer imaginary-self-dual, 
costing potential energy. 
To obtain an estimate of the field strength of the 3-form flux 
$G^{(3)} \equiv F^{(3)} - \tau H^{(3)}$, note that
\begin{equation}
 \frac{1}{l_s^4} \int_{\widetilde{B}} H^{(3)} \wedge F^{(3)} 
 = \frac{g_s}{2 l_s^4} \int_{\widetilde{B}} d^6 y \; |G^{(3)}|^2
\label{eq:rewrite-HF}
\end{equation}
for imaginary-self-dual flux $* G^{(3)} = i G^{(3)}$.
Since the left-hand side of (\ref{eq:rewrite-HF}) is quantized, 
and hence so is the right-hand side, the typical value of the field 
strength will be of order\footnote{The $1/\sqrt{g_s}$ dependence is
missing in some literature papers. Interestingly, 
the estimate (\ref{eq:G3-estimate}) corresponds to a naive geometric
mean of $\vev{F^{(3)}} \sim l_s^2/R_6^3$ and 
$\vev{g_s^{-1} H^{(3)}} \sim l_s^2/(R_6^3 g_s)$. }
\begin{equation}
\vev{G^{(3)}} \sim \frac{l_s^2}{R_6^3 \sqrt{g_s}}.
\label{eq:G3-estimate}
\end{equation}
Thus, one finds that the potential energy of 
the complex structure moduli field $\phi$ is of order 
\begin{equation}
 V_{cs} (\phi) \sim \frac{1}{l_s^8} \; 
   \int_{\widetilde{B}} d^6 y \; |G^{(3)}|^2 \sim 
  \frac{1}{l_s^4 g_s} \times {\rm fcn}(\phi).
\end{equation}
This is why the mass-square of canonically normalized $\phi$ at a vacuum 
are typically of order\footnote{In (\ref{eq:cpxstr-kin}) and in all the
rest of this article, all the lengths, volumes, Kaluza--Klein scales
etc. are measured with a metric in the string frame. 
It is conventional in Type IIB orientifolds that the
string frame metric $g^{(S)}$ and the Einstein frame metric $g^{(E)}$ 
in 10 dimensions are related by Weyl rescaling 
$g^{(S)} = e^{\phi/2} g^{(E)}$, where the dilaton field $\phi$ includes 
both its vacuum value $\vev{\phi} = g_s$ and fluctuation from the vev, 
$(\phi - \vev{\phi})$.
All physical observables $M_{(d)}$ with mass dimension $d$ (except
$l_s$) in string frame and Einstein frame are related by 
$M_{(d)}^{(S)} = g_s^{-d/4} M_{(d)}^{(E)}$. 
Thus, this dependence on the frame or convention of Weyl
rescaling cancels as long as we talk of
dimensionless ratios. $(m_{cs}/m_{\rm KK})^2 \sim (l_s/R_6^{(E)})^4$ in
(\ref{eq:cpx-str-mass-IIB}) is expressed in terms of $R_6$ in the Einstein
frame, and this is equivalent to $(g_s l_s^4)/(R_6^{(S)})^4$, which is
the same as (\ref{eq:cpx-str-mass-IIB-2}).} 
\begin{equation}
m_{cs}^2 \sim 
 \frac{1}{l_s^4 g_s} \times \frac{l_s^8 g_s^2}{R_6^6} = 
 \frac{g_s l_s^4}{R_6^6}. 
\label{eq:cpx-str-mass-IIB-2}
\end{equation}

When an F-theory compactification on a Calabi--Yau 4-fold $X$ allows 
an interpretation as a Calabi--Yau orientifold $B_3 = \widetilde{B}/\Z_2$ 
compactification of the Type IIB string theory, then the complex structure 
moduli of $X$ consist of complex structure moduli of the Calabi--Yau
3-fold $\widetilde{B}$ and moduli describing the locus of D7-branes 
in $\widetilde{B}$, in addition to the axio-dilaton chiral multiplet $\tau$. 
In a generic F-theory compactifications, which do not necessarily 
correspond to simple Calabi--Yau orientifold compactifications of 
Type IIB string theory, there is no distinction between the complex 
structure of $\widetilde{B}$ and the moduli of D7-brane configuration 
in $\widetilde{B}$; we just have complex structure moduli of $X$ 
as a whole, and generically, all of these moduli are stabilized. 
The estimate (\ref{eq:cpx-str-mass-IIB}) was obtained for Calabi--Yau 
orientifold compactifications of Type IIB string theory with D7-branes
(but without D5-branes, O5-planes), which is a special subclass of 
F-theory compactifications. It is not immediately clear whether the 
result for the complex structure moduli of $\widetilde{B}$ in a special 
subclass of F-theory compactifications is readily applied for all the complex 
structure moduli of $X$ of generic F-theory compactification. 
We therefore perform an analysis separately in the following 
for generic F-theory compactifications.

An F-theory compactification on an elliptically fibered 
Calabi--Yau 4-fold $X$ is regarded as an M-theory compactification 
on the same $X$ in the limit the volume of the elliptic fiber is zero.
We use the supergravity language of M-theory compactifications, to
estimate the mass scale of the complex structure moduli fields. 
Let us first remind ourselves of dictionary between parameters 
in the duality of M-theory compactification on $T^2$ and 
Type IIB string theory on 10 dimensions. The $T^2$ compactification 
of 11-dimensional supergravity has two compactification parameters 
$\rho_\alpha$ and $\rho_\beta$, the size of the two edges of $T^2$, 
in addition to the unique theoretical parameter 
$l_{11}^9 \equiv (4\pi) \kappa_{11}^2$ appearing in the action of the 
11-dimensional supergravity. The $T^2$ compactification of the M-theory 
is dual to the Type IIB string theory compactified on $S^1$. 
The Type IIB string theory has two theoretical parameters 
$l_s$ and $g_s$, and one compactification parameter, the circumference 
$R_3$ of the compact $S^1$.
The dictionary between the two descriptions is 
\begin{equation}
 i \frac{1}{g_s}  =  i \frac{\rho_\beta}{\rho_\alpha}, \qquad
 M_*^4 \equiv \frac{1}{g_s l_s^4} = \frac{\rho^2}{l_{11}^6}, \qquad 
 R_3 = \frac{l_s^2}{\rho_\beta} = \frac{l_{11}^3}{\rho^2}.
\label{eq:dict-M-IIB}
\end{equation}
where $\rho \equiv \sqrt{\rho_\alpha \rho_\beta}$. The F-theory limit 
corresponds to $\rho \rightarrow 0$ and $l_{11} \rightarrow 0$
(relatively to the Kaluza--Klein radius or to the horizon size 
of the universe), while keeping $M_*^2 = \rho/l_{11}^3$ finite. 
In $T^2$-fibered compactification of 11-dimensional supergravity, 
the value of $\rho_\beta/\rho_\alpha$ may vary over the base manifold. 
In F-theory language, $i/g_s$ varies over the base space $B_3$. 
The two combinations, $M_*$ and $R_3$, however, depend only on 
$l_{11}$ and $\rho$, but not on $\rho_\beta/\rho_\alpha$, 
and remain constant over the base.

We now consider an M-theory compactification on a Calabi--Yau 4-fold $X$
that is an elliptic fibration $\pi: X \rightarrow B_3$ on a 3-fold
$B_3$. Let the typical size of the base $B_3$ be $R_6$, and that of 
the elliptic fiber be $\rho$. Then the effective action of the complex 
structure moduli of $X$ in 3+1 dimensions is of the form 
\begin{equation}
 R_3 \Delta {\cal L}_{3+1} = 
 \Delta {\cal L}_{2+1}; \qquad 
 \Delta {\cal L}_{2+1} \sim \frac{R_6^6 \rho^2}{l_{11}^9} \times 
  \left[ | \partial \phi |^2 - \vev{G^{(4)}}^2 \times {\rm fcn}(\phi) \right].
\end{equation}
The overall factor $R_6^6 \rho^2/l_{11}^9$ comes from integration 
over the real 8-dimensional manifold $X$, and 
$R_3$ is factored out from $\Delta {\cal L}_{2+1}$ to obtain 
the effective action in the 3+1-dimensions.\footnote{The overall factor 
in the effective action in 3+1 dimensions becomes 
$R_6^6 M_*^8 \sim M_{\rm Pl}^2$, and remains finite. } The overall
factor, however, is irrelevant to the masses of the complex structure 
moduli fields $\phi$. 
The value of the field strength is directly relevant to the mass scale 
of the moduli; $\vev{G^{(4)}}$ is the typical value of the 4-form field 
strength $G_{KLMN}^{(4)}$, which is of mass dimension 1 because we treat 
the 3-form potential field $C_{LMN}^{(3)}$ as dimensionless. 
As noted in \cite{DOPT}, the field strength of the 4-form is typically 
of order 
\begin{equation}
 \vev{G^{(4)}} \sim \frac{l_{11}^3}{R_8^4}
\label{eq:DOPT}
\end{equation}
in Calabi--Yau 4-fold compactification of M-theory; $R_8$ is the 
typical size of the real 8-dimensional manifold $X$. In F-theory 
compactifications that leave SO(3,1) unbroken Lorentz symmetry, 
only a limited class of components of the 4-form field strength 
$G^{(4)}$ is allowed. The vacuum value of $G^{(4)}$ can be introduced only 
in components with one leg in the elliptic fiber, and the remaining 
three legs in the base $B_3$ \cite{DRS}. Therefore, we conclude 
that the moduli masses are typically of order 
\begin{equation}
 m_{cs} \sim \vev{G^{(4)}} \sim \frac{l_{11}^3}{R_6^3 \rho} \sim 
  \frac{1}{R_6^3 M_*^2},
\label{eq:cs-mass-F}
\end{equation}
where we used the dictionary (\ref{eq:dict-M-IIB}).\footnote{
\label{fn:gs1}
In the estimation (\ref{eq:cs-mass-F}), there is no strong argument 
for taking $R_8^4 \sim R_6^3 \rho_\alpha$, $\sim R_6^3 \rho_\beta$ or 
$\sim R_6^3 \rho$. Thus, the estimate (\ref{eq:cs-mass-F}) contains 
an uncertainty at least of order $g_s^{+1/2}\mbox{--}g_s^{-1/2}$.
In generic F-theory compactifications, however, the value of $g_s$ 
varies over $B_3$, and cannot stay much larger or less than unity 
over the entire $B_3$, because of the non-trivial ${\rm SL}(2, \Z)$ 
monodromy. It will be smaller than unity in some region of $B_3$, and 
will be larger in other. Thus, despite the uncertainty and varying value
of $g_s$, it will be safe to infer conclude that some of complex 
structure moduli will have masses below the mass scale (\ref{eq:cs-mass-F}).
} This estimate turns out to be the same as the 
result (\ref{eq:cpx-str-mass-IIB-2}) of the Type IIB Calabi--Yau 
orientifolds.\footnote{There is no extra $g_s$ dependence in the
Type IIB Calabi--Yau orientifolds.} 

Therefore, the Majorana masses of complex structure moduli (and 
hence of right-handed neutrinos) are in the energy scale just below 
Kaluza--Klein scale $1/R_6$, as we consider in a regime where 
$R_6$ is larger than the string length $l_s$. The SU(5) GUT symmetry 
can be broken by introducing U(1)$_Y$ flux on the locus of 
$A_4 \simeq \SU(5)$ singularity \cite{BHV-2, TW-2, DW-2},\footnote{
See \cite{Buican, WY-IIB} for Type IIB versions.} in which case, 
the GUT scale is identified with the Kaluza--Klein scale. 
Thus, the Majorana masses of the right-handed neutrinos are 
typically somewhat below the GUT scale. This is a perfect agreement 
with the phenomenological requirement that the Majorana masses 
should be below the GUT scale. 

Let us refine the estimate of the mass scale a bit more, by 
making a distinction between the Kaluza--Klein scale and the GUT scale. 
In the language of Type IIB orientifold 
compactifications, three observables---the unified gauge coupling 
constant $\alpha_{\rm GUT} \equiv g^2_{\rm GUT}/(4\pi)$, energy scale 
of gauge coupling unification $M_{\rm GUT}$ and the reduced Planck
scale\footnote{
The reduced Planck scale is defined as the coefficient in the 
Einstein--Hilbert term ${\cal L} = (M_{\rm Pl}^2/2) \sqrt{-g} R +
\cdots$ in the effective action in the 3+1 dimensions. The reduced 
Planck scale is different from 
$1/\sqrt{G_N} \simeq 1.2 \times 10^{19}\; \GEV$.} 
$M_{\rm Pl} = 1/\sqrt{8\pi G_N}$, are given by compactification 
parameters as follows: 
\begin{eqnarray}
 \frac{1}{\alpha_{\rm GUT}} & = & \frac{R_{\rm GUT}^4}{g_{s} \; l_s^4} 
 = (R_{\rm GUT} M_*)^4 \simeq 24,  \label{eq:val-alpha_GUT}\\
 M_{\rm GUT} & = & \frac{c}{R_{\rm GUT}} \sim 10^{16} \; \GEV, 
  \label{eq:val-M_GUT}\\
 M_{\rm Pl}^2 & = & \frac{(4\pi) R_6^6}{g_s^2 l_s^8} = 
  (4\pi) R_6^6 M_*^8 \simeq 
  \left(2.4 \times  10^{18}  \; \GEV \right)^2.
  \label{eq:val-M-Pl}
\end{eqnarray}
Since $g_{s}$ and $l_s$ come in in the three
equations above only as a combination $(g_s l_s^4) \equiv 1/M_*^4$, 
we consider that all the three equations above perfectly makes sense 
in generic F-theory compactifications without an interpretation as 
Type IIB Calabi--Yau orientifold.\footnote{Here, we do not pay attention 
to a possible factor 2 associated with orientifold projection.} 
The volume of the locus $S$ of $A_4$ singularity is $R_{\rm GUT}^4$, 
and the volume of $B_3$ is $R_6^6$.
Thus, the energy scale of gauge coupling unification $M_{\rm GUT}$ 
should roughly be the same as $1/R_{\rm GUT}$, but they can be
different, for example by a factor $c=2\pi$ in case $S = T^4$.
The numerical coefficient $c$ can be regarded as a factor of order 
unity in general, ranging in between $c\sim (1\mbox{--}2\pi)$.  
We assumed that there are no extra $\SU(5)_{\rm GUT}$-charged particles 
much below the unification scale other than the particles in the minimal 
supersymmetric standard model; otherwise, the value of 
$\alpha_{\rm GUT}$ should be a little larger. 
One should also note that the value of $M_{\rm GUT}$ is accompanied 
by uncertainty of the half order of magnitude or so, coming from 
unknown tree-level and 1-loop threshold
corrections.\footnote{A beautiful study of GUT scale threshold
corrections is found in \cite{DW-2}. 
See also \cite{HN, TW-2}, \cite{Blhg-threshold} and \cite{FW-G2}.} 
The three observables $\alpha_{\rm GUT}$, $M_{\rm GUT}$ and $M_{\rm Pl}$ 
are expressed in terms of the same number of fundamental parameters 
$M_*$, $R_{\rm GUT}$ and $R_6$ in
(\ref{eq:val-alpha_GUT}--\ref{eq:val-M-Pl}).
Thus, the values of these parameters can be determined from the values 
of the observables in our world, which are also given in 
(\ref{eq:val-alpha_GUT}--\ref{eq:val-M-Pl}).

Following \cite{BHV-2}, we introduce a dimensionless parameter\footnote{
The two energy scales in the effective theory in 3+1 dimensions, 
$M_{\rm GUT}$ and the Planck scale $1/\sqrt{G_N}$ have a hierarchy 
of order 
$M_{\rm GUT}\sqrt{G_N} \simeq 0.8 \times 10^{-3} \times 
(M_{\rm GUT}/10^{16} \; \GEV)$, and some people have taken it seriously. 
This apparent hierarchy in the effective theory, however, is expressed 
in terms of more fundamental parameters as 
\begin{equation}
 M_{\rm GUT} \sqrt{G_N} \sim \frac{\alpha_{\rm GUT}}{\sqrt{32\pi^2}}
  \times \left[c (\epsilon^\gamma)^{\frac{1}{\gamma}}\right] 
 = (2.3 \times 10^{-3}) \times \left[0.35 \times (M_{\rm GUT}/10^{16} \;
				\GEV) \right].
\label{eq:deg-decouple}
\end{equation} 
The apparent hierarchy of three orders of magnitude in the effective 
theory is mostly due to the first factor, a combination of small 
value of $\alpha_{\rm GUT}$ and some powers of $\pi$ in the
denominator. This hierarchy has little to do with a moderately small 
parameter $\epsilon = (R_{\rm GUT}/R_6)^3$, which directly parametrizes 
the degree of ``decoupling''. It is therefore misleading to take 
the small value of $M_{\rm GUT}\sqrt{G_N} \simeq 10^{-3}$ as an  
indication of ``decoupling''. Unlike in the original context of 
decoupling \cite{decoupling} where a large hierarchy between 
the electroweak scale 
and the Planck scale was discussed, the hierarchy under consideration 
here is not that large, and most of this hierarchy is accounted for by the
first factor of (\ref{eq:deg-decouple}). 
}
\begin{equation}
 \epsilon \equiv \left(\frac{R_{\rm GUT}}{R_6}\right)^3 
 = \frac{\sqrt{4\pi} M_{\rm GUT}}{\alpha_{\rm GUT} c M_{\rm Pl}} \sim
 0.35 \times \left(\frac{M_{\rm GUT}}{c \; 10^{16} \; \GEV} \right).
\end{equation}
%
\begin{figure}[tb]
\begin{center}
 \begin{tabular}{ccc}
  \includegraphics[width=.25\linewidth]{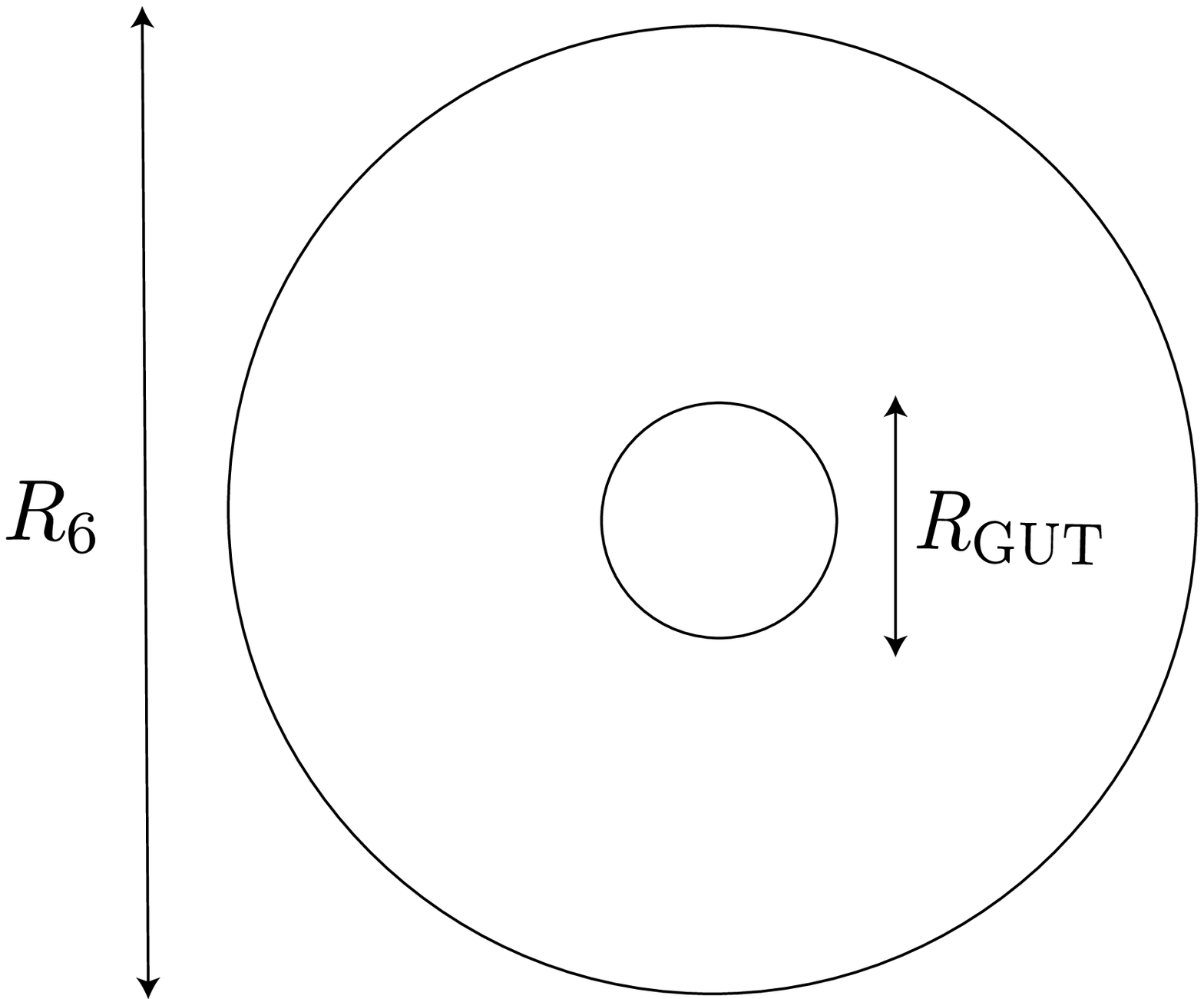}  
  & $\qquad \qquad $&
  \includegraphics[width=.4\linewidth]{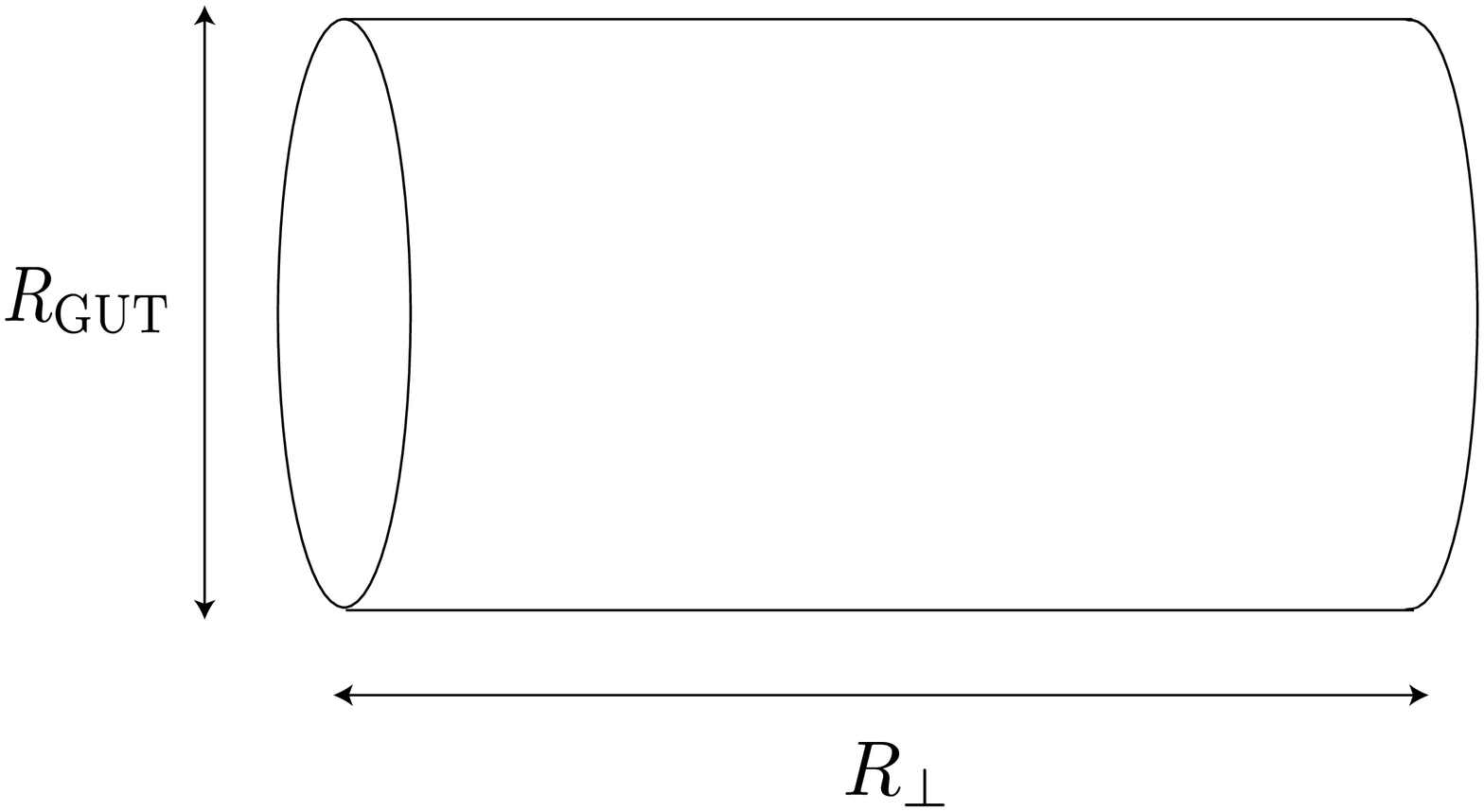}  \\
 (a) & & (b)
 \end{tabular}
\caption{\label{fig:(an)-isotropic} Cartoon picture of $B_3$ in 
the homogeneous model (a) and in the tubular model (b). This is a
 reproduction of essentially the same figure in \cite{BHV-2} just 
for readers' convenience.}
\end{center}
\end{figure}
In a homogeneous model, where $B_3$ looks like 
Figure~\ref{fig:(an)-isotropic}~(a), the 4-form field strength 
$\vev{G^{(4)}}$ in (\ref{eq:DOPT}) becomes 
$l_{11}^3/(\rho R_6^n R_{\rm GUT}^{3-n})$ for some $n = 0,1,2,3$.
The unspecified parameter $n$ remains because some of the 4-form fluxes 
are associated with 4-cycles around $S$, others are not.
Using $\epsilon^{\gamma = 1/3} = (R_{\rm GUT}/R_6)$, the estimate 
of the masses of complex structure moduli is rewritten as 
\begin{equation}
 m_{cs} \sim M_{\rm GUT} \times \frac{\sqrt{\alpha_{\rm GUT}}}{c} \times 
  \left(\epsilon^{\gamma = 1/3}\right)^{n},
\end{equation}
where we have used (\ref{eq:val-alpha_GUT}, \ref{eq:val-M_GUT}). 
All the factors 
$\sqrt{\alpha_{\rm GUT}}$, $(1/c)$ and $(\epsilon^\gamma)^n$
make the moduli mass smaller than the unification scale. 
In a tubular model where $B_3$ is like a product of $A_4$ singularity locus $S$
and a real 2-dimensional space, like in Figure~\ref{fig:(an)-isotropic}
(b), $R_6^3 = R_{\rm GUT}^2 R_{\perp}$, and one finds
that $R_{\rm GUT}/R_{\perp} = \epsilon^{\gamma = 1}$ \cite{BHV-2}. 
Depending on the topological cycles $G^{(4)}$ are associated with, 
the field strength of (\ref{eq:DOPT}) may vary as 
$l_{11}^3/(\rho R_{\perp}^n R_{\rm GUT}^{3-n})$ for $n = 0,1,2$ 
(and maybe $3$).
Thus the estimate of the moduli masses become 
\begin{equation}
 m_{cs} \sim M_{\rm GUT} \times \frac{\sqrt{\alpha_{\rm GUT}}}{c} \times 
\left(\epsilon^{\gamma = 1}\right)^{n}.
\end{equation}
In either one of the two models in Figure~\ref{fig:(an)-isotropic}, 
we can safely conclude that there are complex structure moduli fields 
(right-handed neutrinos) whose Majorana masses are smaller than 
the GUT scale at least by one order of magnitude. 

To recapitulate, we have four fundamental parameters,  
$l_{11}$, $\rho$, $R_6$, $R_{\rm GUT}$ for this compactifications. 
As we consider the F-theory limit with $\rho \rightarrow 0$ and 
$l_{11} \rightarrow 0$ (while keeping $l_{11}^3/\rho$ finite),  
there are only three relevant parameters. From these three parameters, 
six observable parameters\footnote{If the fundamental parameters 
$\rho$ and $l_{11}$ are kept small but non-zero, then there is another 
observable parameter, the $S^1$ radius $R_3/(2\pi)$ of the
compactification of this 3+1 dimensional ``Minkowski'' space. 
Thus, the parameter counting argument for predictability in the
following does not change.} can 
be derived: $M_{\rm Pl}$, $\alpha_{\rm GUT}$, the energy scale of 
gauge coupling unification $M_{\rm GUT}$, right-handed neutrino 
masses $M_R = m_{cs}$ from flux compactification, 
Kaluza--Klein scale of bulk modes $M_{\rm KK}$ ($1/R_6$ or
$1/R_{\perp}$), and the energy scale $M_{str} \sim 1/l_s$ where stringy 
excitations appear.
The last two of them, though, may not be practically observable, except 
that those mass scales may set some limitations on inflation 
models in F-theory compactifications. We have determined all the three
relevant fundamental parameters $R_{6}$, $R_{\rm GUT}$ and $M_*$ 
in (\ref{eq:val-alpha_GUT}--\ref{eq:val-M-Pl})
by using the known value of the three observable parameters 
$M_{\rm GUT}$, $\alpha_{\rm GUT}$ and $M_{\rm Pl}$, and then 
the value of the remaining one observable 
$M_R = m_{cs}$ is predicted by (\ref{eq:cs-mass-F}). 
Although its upper bound has been
indicated phenomenologically by the observed value of $\Delta m^2$ 
in neutrino oscillation, its value is theoretically independent of 
the GUT scale or Kaluza--Klein scale in effective field theory in
general (except in individual models). In generic F-theory 
compactifications with fluxes, however, an upper bound of the energy 
scale of right-handed neutrinos is predicted, and what is more, the 
prediction implies that the phenomenological upper bound is fortunately satisfied. 

\section{Yukawa Couplings Involving Singlets}
\label{sec:nuYukawa-general}

Now that we have seen the complex structure moduli of 
F-theory compactifications are good candidates for the right-handed 
neutrinos, let us study how we should compute their Yukawa couplings. 

Yukawa couplings of quarks and charged leptons are trilinear 
interactions where all the three fields are charged under the 
GUT gauge group $\SU(5)_{\rm GUT}$. Such charged matter fields 
are associated with rank-1 enhancement of singularity \cite{6authors,
KV}, and the Yukawa couplings with rank-2 enhancement \cite{BHV-1, DW-1,
Hayashi-1, Hayashi-2}. In the absence of microscopic formulation 
of F-theory, it would not be practical to study such Yukawa couplings by using the
singular geometry of $X$. Instead, 
the field theory formulation of \cite{KV, DW-1,
BHV-1} provides an approximate description of local geometry of $X$, 
whether $X$ is singular or not. Such field theory local models\footnote{
In this article, as in \cite{Hayashi-2}, we mean by a ``field
theory local model'' a field theory that models a local geometry of $X$;
to be more precise, that is a field theory on a local patch $U$ of 
($A_4$-)singularity locus $S$ that provides an approximate description
of physics associated with local geometry of $X$ containing $U \subset S
\subset X$. Thus, the ``local model'' does not imply local geometry of
$X$ along the entire $S$ that models the physics of visible elementary 
particles or engineering of the Standard Model.} were 
used to study the Yukawa couplings of {\it charged} matter fields 
\cite{BHV-2, HV-Nov08, Hayashi-2}.

Neutrino Yukawa couplings (\ref{eq:nu-Yukawa}, \ref{eq:Yukawa-515}), 
on the other hand, involve two $\SU(5)_{\rm GUT}$-charged fields and 
one $\SU(5)_{\rm GUT}$-{\it neutral} field. Such trilinear couplings also 
appear in the next-to-minimal supersymmetric standard model (NMSSM), 
$\Delta W = \lambda \; S H_u H_d$. $\SU(5)_{\rm GUT}$-neutral fields 
are, in general, identified with moduli of Calabi--Yau 4-fold $X$, 
say, $H^{3,1}(X; \C)$ or $H^{1,2}(X; \C)$. As the 
field theory formulation has been very useful in studying Yukawa 
couplings, we will study in the following the way to include 
the moduli fields of Calabi--Yau 4-fold in the field theory
language.\footnote{This study has been initiated  
by \cite{DW-1}, but we will extend the results in \cite{DW-1}.} 

\subsection{Capturing the Moduli Fields in the Field Theory Formulation}

Suppose that a local geometry of $X$ is given by an Weierstrass 
equation 
\begin{equation}
 y^2 = x^3 + x f(z,u,v) + g(z,u,v),
\label{eq:Weierstrass}
\end{equation}
where $(x,y)$ are the coordinates of the elliptic fiber, $(u,v)$ 
are local coordinates of a divisor $S$ in $B_3$, and $z$ a normal coordinate 
of $S$ in $B_3$. Let us further assume that the surface with $(x,y,z)$
coordinates have $A_4$ singularity at $z = 0$. $S$ is identified with 
the locus of $A_4$ singularity. Four vanishing 2-cycles $C_A$ 
($A=1,2,3,4$) are in the $(x,y,z)$ plane, and have intersection form 
that is negative of the Cartan matrix of $A_4$. That is the local
condition for the SU(5)$_{\rm GUT}$ theories. 

Let us focus on a local region of $S$. Through a projection 
$(x,y,z,u,v) \mapsto (u,v)$, $X$ is now regarded locally as a fibration 
over $S$. By focusing further on a region of $X$ near the $A_4$
singularity locus, the geometry of $X$ in this local region may be 
regarded approximately as 
an ALE fibration on $S$, with non-vanishing 2-cycles $C_P$ and the
vanishing 2-cycles $C_A$. When the intersection form of the space 
spanned by $C_A$ and $C_P$ is that of negative of Cartan matrix of 
Lie algebra $\mathfrak{g}$, then a field theory local model can be 
constructed for this local region of $X$, with the gauge group $G$ 
\cite{DW-1, BHV-1}. 
The structure group of a Higgs bundle on $S$ has a structure 
group $G'$, which is the commutant of the unbroken symmetry group 
$G'' = \SU(5)_{\rm GUT}$ in $G$. See \cite{Hayashi-2, DW-3} 
for the use of Higgs bundle in F-theory compactifications; references 
on mathematics of Higgs bundle are also provided there.

\subsubsection{Determination of the Field Theory Background}

The compactification data $(X, G^{(4)})$ are encoded in 
the 4-form\footnote{
The coefficient of $du \wedge dv$ transforms like a section 
of ${\cal O}(K_S)$, because $dx/y$ transforms as sections of 
${\cal O}(K_{B_3}) = {\cal O}(K_S) \otimes N_{S|B_3}^{-1}$, and 
$dz$ as those of $N_{S|B_3}$. } $\Omega$ 
\begin{equation}
 \Omega(a) = \frac{dx \wedge dz \wedge du \wedge dv}{y} 
  = \frac{dx \wedge dz \wedge du \wedge dv}{\sqrt{x^3 + x f + g}},
\end{equation}
and the 3-form potential $C^{(3)}$ on $X$.
Here, $a$ collective denotes the coefficients in the defining 
equation (\ref{eq:Weierstrass}). Using the (Poincare dual of the)
2-cycles $C_A$ and $C_P$ in the fiber direction, $\Omega(a)$ and 
$C^{(3)}$ can be expanded as 
\begin{eqnarray}
 \Omega(a) & = & \sum_{P} C_P \otimes \varphi^P + \cdots , 
    \label{eq:Omega-varphi}\\
 C^{(3)} & = & \sum_P C_P \otimes A^P + \cdots,  
   \label{eq:C3-A}
\end{eqnarray}
where ellipses stand for a linear combination running over non-compact 
2-cycles. Coefficients for the 2-cycles $C_A$ should be dropped for 
a background configuration $(\Omega(a_0), \vev{C^{(3)}})$, because 
we want to preserve the $G'' = \SU(5)_{\rm GUT}$ symmetry.\footnote{
$\vev{C^{(3)}}$ can be introduced in the $C_A$ components as well, 
in order to break the $\SU(5)_{\rm GUT}$ symmetry to the Standard Model 
gauge symmetry.} A background configuration 
$(\Omega(a_0), \vev{C^{(3)}})$ therefore determines
$(\mathfrak{h}'\otimes \C)$-valued (2,0)-form $\varphi(a_0)$ and 
$\mathfrak{h}'$-valued 1-form $\vev{A}$; here, $\mathfrak{h}'$ is the 
Cartan subalgebra of the structure group $\mathfrak{g}'$.

\begin{figure}[t]
 \begin{center}
    \includegraphics[width=.5\linewidth]{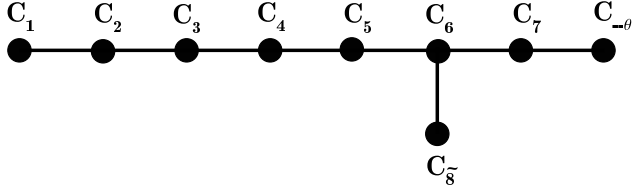}  
 \caption{\label{fig:E8-dynkin} An extended Dynkin diagram of $E_8$. 
We follow the conventions of \cite{Hayashi-1}.}
 \end{center}
\end{figure}

It will be useful to have some specific examples in mind. 
In F-theory compactifications with Heterotic dual, 
$X \rightarrow S$ can be regarded as a K3-fibration globally 
over $S$. The K3-fiber has 22 topological 2-cycles, 8 of which 
have intersection form that is the negative of the Cartan matrix of $E_8$.
We can choose a basis $C_A$ ($A=1,2,3,4$) and $C_P$ 
($P=\tilde{8},6,7,-\theta$), so that their intersection form is 
described by the Dynkin diagram in Figure~\ref{fig:E8-dynkin}. 
$G'' = \SU(5)_{\rm GUT}$ is generated by $C_{A}$'s, and 
the structure group $G' = \SU(5)_{\rm str}$ by $C_P$'s. 
When the complex structure of $X$ is in the stable degeneration 
limit, the size of all the $C_P$'s is small relatively to all other
2-cycles in the K3 fiber, and a field theory model with 
$G = E_8$ gauge group with an $G' = \SU(5)_{\rm str}$ Higgs bundle 
given by $(\varphi^P(a_0), \vev{A}^P)$ provides a good approximation 
to some parts of physics in the F-theory compactification by 
$(X, G^{(4)})$. In this case, the $E_8$ field theory model with an 
$G' = \SU(5)_{\rm str}$ Higgs bundle is defined globally on $S$, 
not just in a local patch of $S$. The (2,0)-form field background 
$\varphi^P$ in $\mathfrak{su}(5)_{\rm str} \otimes \C$ becomes 
\begin{equation}
 \vev{\varphi} = \varphi(a_0) = \diag ( - \varphi^{\tilde{8}}, 
 \varphi^{\tilde{8}} - \varphi^6  , 
 \varphi^6 - \varphi^7, \varphi^7 - \varphi^{-\theta}, \varphi^{-\theta})(a_0)
\label{eq:varphi-vac} 
\end{equation}
in the fundamental ($\SU(5)_{\rm str}$-{\bf 5}) representation. 
This is the background configuration of the (2,0)-form field 
in the field theory on $S \times \R^{3,1}$ \cite{DW-1}.

Another example is a case where local geometry of $X$ is approximately 
given by 
\begin{equation}
 y^2 \simeq x^2 + z^5 (z^2 + 2 u z + v ),
\end{equation}
after dropping higher order terms in local coordinates $(u,v,z,x,y)$.
The degree-7 polynomial in $z$ can be locally expressed as 
$\prod_{i = 1,\cdots,5,6,7} (z-z_i)$, with $z_i = 0$ ($i = 1, \cdots
5$), $z_6 = z_-$, $z_7 = z_+$, and 
\begin{equation}
 z_\pm(u,v) = - u \pm \sqrt{u^2 - v}. 
\label{eq:z-pm}
\end{equation}
We take the 2-cycles $C_I$ ($I = 1,\cdots, 6$) as $S^1$-fibrations over 
intervals from $z = z_I$ to $z = z_{I+1}$. 
The intersection form of $C_I$'s is negative of Cartan matrix of
$A_6$, and hence the gauge group of the field theory model for this
local geometry is $G = \SU(7)$.  
$C_A$ ($A=1,2,3,4$) have vanishing size, whereas the others $C_{5,6}$ have 
non-vanishing sizes, and hence $G'' = \SU(5)_{\rm GUT}$ generated by 
$C_A$'s is the unbroken symmetry, and the structure group is 
$G' = \SU(2) \times \U(1)$.
The period integral over these 2-cycles 
determine the (2,0)-form field backgrounds $\varphi^P(a_0)$. 
The period integral over the 2-cycles $C_I$ can be carried out 
explicitly, and 
\begin{equation}
 \int_{C_I} \Omega = - 2\pi i (z_{I+1} - z_I) \; du \wedge dv.
\end{equation}
The background (2,0)-form configuration \cite{Hayashi-2}
\begin{equation}
 \varphi(a_0) = 2\pi i \diag (\overbrace{0,\cdots,0}^5, z_-(u,v),
  z_+(u,v)) \; du \wedge dv
\label{eq:varphi-vac-2}
\end{equation} 
reproduces the period integral above for positive roots $C_I$'s. 

We have provided an explicit procedure of obtaining 
$(\varphi, A)$ background for a given set of data $(X,G^{(4)})$.
Although the dictionary between the deformation of ADE singularity 
and Cartan vev of $\varphi$ has been known in principle, some ambiguity
remained . The deformation parameters of 
ADE singularity are certainly identified with 
$(\mathfrak{h} \otimes \C)/W$, where $\mathfrak{h}$ is the Cartan
subalgebra of the corresponding ADE Lie algebra, and $W$ its Weyl group,
but the dictionary still has an overall scaling ambiguity \cite{KM}. 
The overall scaling ambiguity comes in at each point $(u,v)$ in the
base. There is a discrepancy between the choice of $\varphi$ 
in \cite{Hayashi-2} and one in \cite{HV-Nov08-rev}, and it comes exactly 
from the point-wise overall scaling ambiguity (see the appendix for
more). The explicit procedure above, which 
uses the period integral, fixes this ambiguity. 

Enhancing the rank of the  gauge group by 1 is the minimal choice of the field theory 
local models for regions along matter curves,but the gauge group 
should be chosen at least by rank-2 higher for regions containing 
codimension-3 singularities where singularity types are enhanced by 
rank-2. These choices are minimal, and one can choose larger gauge groups 
of the field theory model larger, by maintaining higher order 
terms in the local defining equation of the geometry. 
It is a matter of the level of approximation how many higher order 
terms are taken into account.

In the two examples above, we set the background $\varphi$ field 
configuration in the diagonal entries, fixing all the gauge 
transformation non-commuting with $\vev{\varphi}$.  
The gauge symmetry of the field theory local models remains  
(when $\varphi$ is set diagonal) only in the $\SU(5)_{\rm GUT}$
subgroup (and some U(1) factors for now). Since all other gauge 
degrees of freedom were redundant from the beginning, and were 
even absent in the Calabi--Yau 4-fold description of F-theory 
compactification, we never need to introduce such gauge degrees 
of freedom in any descriptions. 

\subsubsection{Moduli (Massless) Fluctuations}

Although we have so far only talked about the vacuum configuration 
$\varphi^P(a_0) = \vev{\varphi^P}$ in (\ref{eq:varphi-vac}, 
\ref{eq:varphi-vac-2}), the same process defines a set of corresponding 
$\varphi^P(a)$ for any deformed complex structure moduli 
$a = \delta a + a_0$ (where $\delta a$ need not be infinitesimal). 
Because of the procedure above, the resulting $\varphi^P(a)$'s 
determines a field configuration $\varphi(a)$---a (2,0)-form 
on $S$---taking its value in the Cartan part of the structure group $G'$. 
Thus, the deformation of the complex structure of $X$ of an F-theory 
compactification, $a$, corresponds to those of (2,0)-form $\varphi$ on $S$ 
in the Cartan part of $G' \subset G$, not the entire $G'$-{\bf adj.} 
representation. Similarly, (\ref{eq:C3-A}) defines 
a Cartan-valued 1-form field $A$.

This may seem odd at first sight, because bosonic fields $(A, \varphi)$
and fermionic fields $(\eta, \psi, \chi)$ are introduced for all 
the roots of the gauge group $G$, except for the first component 
in the decomposition 
\begin{equation}
 {\rm Res}^G_{\vev{G'} \times G''} G\mbox{-}{\bf adj.} \rightarrow 
 ({\bf adj.}, {\bf 1}) + ({\bf 1}, {\bf adj.}) + \oplus_i (U_i, R_i).
\label{eq:decomp}
\end{equation}
In the $({\bf adj.}, {\bf 1})$ component, only the Cartan part 
was restored from metric and 3-form potential by the reduction 
procedure (\ref{eq:Omega-varphi}, \ref{eq:C3-A}). 
The entire dynamical fields 
other than those in the Cartan part are missing. 
This is not surprising, however, because only the Cartan part 
is obtained by dimensional reduction of metric and $C^{(3)}$ 
in geometric engineering of an ADE gauge group, and 
all other dynamical degrees of freedom for the roots of the ADE 
gauge group originate from M2-branes wrapping on vanishing topological 
2-cycles. ``W-bosons'' corresponding to blown-up/deformed 2-cycles 
become massive, and can be integrated out of an effective theory. 
Since all the $\vev{\varphi}$ eigenvalues (e.g. in (\ref{eq:varphi-vac},
\ref{eq:varphi-vac-2})) in $\mathfrak{g}'$ are different generically, 
all the 2-cycles for the roots in $G'$-${\bf adj.}$ are blown up, 
and the corresponding fields become massive. In the field theory
language, this is understood as Higgs mechanism by the Higgs vev 
$\vev{{\rm adj.}(\varphi)}$. Thus, apart from a neighborhood of 
branch loci, where two eigenvalues of $\vev{\varphi}$ become 
degenerate, we do not need to keep the dynamical fields for the 
roots of $\mathfrak{g}'$ in an effective theory on $S$.

The BPS condition for the field configuration is \cite{BHV-1}\footnote{
See \cite{Hayashi-2} for details of the conventions. 
$\alpha \in \C$ is a coefficient associated with the ambiguity in 
the normalization of $\varphi$ field.} 
\begin{equation}
 \omega \wedge F - \frac{|\alpha|^2}{2}[ \varphi, \overline{\varphi}] 
 = 0, \qquad 
 F^{(0,2)} = 0, \qquad \bar{\partial}_A \varphi = 0, 
\label{eq:BPS}
\end{equation}
and zero-mode equations are obtained by imposing the same 
set of conditions for the background field configuration 
$(\vev{A}_{\bar{m}},\vev{\varphi})$ plus infinitesimal deformation 
$(\psi_{\bar{m}}, \chi)$. This is applied to any irreducible components 
in (\ref{eq:decomp}), and the $({\bf adj.}, {\bf 1})$ component 
is not an exception. 
The zero mode equations for the Cartan part of $({\bf adj.}, {\bf 1})$ are 
\begin{equation}
 \omega \wedge d \psi^{\rm tot} = 0, \qquad \bar{\partial} \psi = 0, \qquad 
 \bar{\partial} \chi = 0. 
\label{eq:Cartan-0}
\end{equation}
Here, $d = \partial + \bar{\partial}$ and 
$\psi^{\rm tot}$ is a 1-form $\psi + \bar{\psi}$.
We understand that $\vev{A}^{(0,1)/(1,0)}$, 
$\vev{\varphi}$ and $\vev{\overline{\varphi}}$ can be made diagonal at 
generic points on $S$ (other than at branch loci) which lead to the simplification of  
the zero-mode equation.
These equations tell us that the fluctuations in $\chi = \delta \varphi$ 
and $\psi = \delta A^{(0,1)}$ can be chosen independently.
Furthermore, remembering that there is a monodromy around branch 
loci, and that fields are twisted by the Weyl group of $G'$ \cite{Hayashi-2}, 
the Cartan-valued fields (which means that there are rank-$\mathfrak{g}'$ 
components) $\psi$ and $\chi$ on $S$ are better understood as 
single-valued fields $\psi$ and $\chi$ on a spectral surface $C_V$ 
of a certain representation.\footnote{
When we are referring to matter curves, Higgs/vector bundles, spectral 
surfaces and other things associated with an irreducible component 
$(U_i, R_i)$, it is sometimes convenient to use $R_i$ as a label, but 
$U_i$ is more useful in other situations. We therefore introduce the following 
convention in this article: when a representation of the unbroken 
symmetry is used as a subscript, we put it as ${}|_{(R_i)}$, but when 
we use a representation of the structure group, the subscript becomes 
just ${}|_{U_i}$.}  
Thus, the zero-mode equations (\ref{eq:Cartan-0}) implies that 
the zero modes from the $({\bf adj.}, {\bf 1})$ sector are characterized as 
\begin{equation}
 \chi = \delta \varphi \in H^0(C_V; K_{C_V}), \quad {\rm and} \quad 
 \psi = \delta A^{(0,1)} \in H^{0,1}(C_V; \C) \simeq H^1(C_V; {\cal O}_{C_V}), 
\label{eq:moduli-cohomology-specsurf}
\end{equation}
because $\chi$ is a holomorphic (2,0)-form, and 
$\psi^{\rm tot}$ a harmonic 1-form. Note the argument so far does not provide 
justification for the holomorphicity of $\chi$ or harmonic nature 
of $\psi$ at the ramification locus\footnote{To our knowledge, 
configuration of $\varphi$, $\overline{\varphi}$ and gauge fields 
$A^{(0,1)/(1,0)}$ around the ramification locus has not been discussed 
in sufficient details in the literature so far.} 
of the spectral surface $C_V$ over $S$, as the eigenvalues 
of $\vev{\varphi}$ have been assumed to be non-degenerate so far. 
Thus, 
the characterization of the moduli (\ref{eq:moduli-cohomology-specsurf}) 
should not be regarded as something derived already in the field theory
formulation of F-theory. But (\ref{eq:moduli-cohomology-specsurf}) 
is the same as the characterization of vector bundle moduli 
in Heterotic string compactification: moduli of the spectral surface 
and Wilson lines on the spectral surface.\footnote{A slight difference 
is, though, that only a subspace of the moduli of the spectral surface 
is actually the moduli of the vector bundle in Heterotic string theory; 
the subspace is characterized as the kernel of 
$H^0(S; R^1 \pi_* {\rm ad}V) \rightarrow H^2(S; R^0 \pi_* {\rm ad}V)$. 
We have not figured out where this discrepancy comes from.
We will come back to this issue shortly.}
Thus, we consider that the argument above as quite reasonable. 
In short, moduli characterized as $H^{3,1}(X; \C)$ globally in $X$ 
are captured as $H^0(C_V; K_{C_V})$ in local models, and 
the $H^{1,2}(X; \C)$ moduli in global picture are treated as 
$H^1(C_V; \C)$ in field theory local models. 

Let us digress for a moment to provide a little mathematical 
characterization of the moduli fields 
(\ref{eq:moduli-cohomology-specsurf}) as a whole. 
The following observation is theoretically interesting 
on its own, but it also turns out to be useful later in this article 
in understanding 
the moduli zero modes in a limit of complex structure where 
the spectral surface $C_V$ is singular.
Let us first note this relation:
\begin{equation}
 0 \rightarrow H^1 (C_V; {\cal O}_{C_V}) \rightarrow 
  {\rm Ext}^1(i_{C_V *}{\cal O}, i_{C_V *} {\cal O}) \rightarrow 
  H^0(C_V; K_{C_V}) \rightarrow H^2(C_V; {\cal O}_{C_V}),
\end{equation} 
where $i_{C_V} : C_V \hookrightarrow \mathbb{K}_S$, and 
$\mathbb{K}_S$ is the total space of the canonical bundle $K_S$ of $S$.
See \cite{KS, DKS} for an almost similar discussion in Type IIB
language. The extension group picks up only the kernel 
of the last map from the moduli of the spectral surface 
$H^0(C_V; K_{C_V})$, just as in the Heterotic
dual. Thus, the extension group above can be regarded as the 
characterization of the moduli fields. Since it does not matter 
if we replace both of ${\cal O}_{C_V}$'s in 
${\rm Ext}^1(i_{C_V *}{\cal O}_{C_V}, i_{C_V *}{\cal O}_{C_V})$ by 
a line bundle on $C_V$ simultaneously, we take 
\begin{equation}
 ({\bf adj.}, {\bf 1}): \quad 
 {\rm Ext}^1 (i_{C_V *} {\cal N}_V, i_{C_V *} {\cal N}_V)
\label{eq:moduli-ext}
\end{equation}
as the characterization of the moduli fields. 
${\cal V} \equiv i_{C_V *} {\cal N}_V$ is the Higgs sheaf\footnote{See 
\cite{Hayashi-2, DW-3} for the use of Higgs sheaf in F-theory 
compactifications. The notion of Higgs sheaf was already introduced to 
physics in the context of Type IIB compactifications \cite{DKS, Pantev}.} of 
the F-theory compactification. 

The characterization of the massless modes from the 
$({\bf adj.}, {\bf 1})$ component (\ref{eq:moduli-ext}) goes completely 
in parallel with those of charged matter fields \cite{DW-1, BHV-1, 
Hayashi-1, Hayashi-2}
\begin{eqnarray}
  ({\bf 1}, {\bf adj.}): & &  
      {\rm Ext}^1 (i_{\sigma *} {\cal L}, i_{\sigma *} {\cal L}), \\
  (U_i, R_i): & & 
 {\rm Ext}^1 (i_{\sigma *} \rho_{R_i}^\times ({\cal L}), 
              i_{C_{U_i}*} {\cal N}_{U_i}).
\end{eqnarray}
All of these matter fields are described by extension groups 
in $\mathbb{K}_S$, and now so is the moduli fields in the 
$({\bf adj.}, {\bf 1})$ component. All of these zero-modes are given 
interpretations as if they were matters arising from intersection 
of a pair of D7-branes, although completely generic F-theory 
compactifications are being discussed here. 

It is true at the moment that (\ref{eq:moduli-ext}) is nothing more
than (\ref{eq:moduli-cohomology-specsurf}). One should note, however, 
that we have arrived at (\ref{eq:moduli-cohomology-specsurf}) by 
assuming implicitly that the spectral surface is smooth. When 
the spectral surface becomes singular, it is not that the zero modes 
split into fluctuations of $\varphi$ and those of $A_{\bar{m}}$.
The expression (\ref{eq:moduli-ext}), however, still remains an 
appropriate way to characterize the moduli zero modes, as we will 
see some examples in section \ref{sec:Dim4}. The expression 
(\ref{eq:moduli-ext}) also makes it possible to understand the 
Yukawa couplings of $H^{1,2}(X; \C)$ zero modes and $H^{3,1}(X; \C)$ 
zero modes in an ``unified'' way, as we see shortly.

\subsection{Yukawa Couplings involving Singlets}

\subsubsection{Neutral-Charged-Charged Yukawa Couplings}

Now that we have understood how to deal with the moduli fields 
of Calabi--Yau 4-fold compactification within the field theory local 
models of \cite{KV, DW-1, BHV-1, Hayashi-2}, we can use the field 
theory local models to study the Yukawa couplings involving the 
moduli fields. The mode expansion 
\begin{equation}
 \Phi = \sum_I \chi^{(2,0)}_I \phi_I(x,\theta), \qquad 
 A = \sum_I \psi^{(0,1)}_I \phi_I(x,\theta)
\end{equation}
with chiral multiplets $\phi_I(x,\theta)$ of ${\cal N} = 1$
supersymmetry in 3+1 dimensions and their mode functions 
$(\psi_I, \chi_I)$ is inserted in the superpotential \cite{DW-1, BHV-1}\footnote{
This superpotential can be regarded as a non-Abelian extension 
of the Gukov--Vafa--Witten superpotential
(\ref{eq:GVW}).}\footnote{We did not pay attention to the overall
normalization of the superpotential (\ref{eq:GVW}, \ref{eq:F-super},
\ref{eq:Het-super}), or in the reduction 
(\ref{eq:Omega-varphi}, \ref{eq:C3-A}). Reference \cite{Conlon} 
may be useful when determining the superpotential. } 
\begin{equation}
 \Delta W_{\rm DW-BHV} = \int_S \tr \left(\Phi \wedge F \right),
\label{eq:F-super}
\end{equation}
and the overlap integration of the mode functions over $S$ yields 
the coupling constants of the chiral multiplets.

Let us first study how the Yukawa couplings of the form 
(\ref{eq:Yukawa-515}) are generated. Neutrino Yukawa couplings 
and the $S H_u H_d$ interaction of the next-to-minimal theory 
are in this form.

The $\SU(5)_{\rm GUT}$ singlet field in the Yukawa couplings 
(\ref{eq:Yukawa-515}) may be fluctuations from the vacuum in 
$H^{1,2}(X; \C)$, or in $H^{3,1}(X; \C)$. 
In the former case, the Yukawa coupling $\lambda$ 
is calculated by an overlap integration 
\begin{equation}
 \lambda \sim 2 \int_S \tr \left(\chi_{U} \wedge 
   \psi_{{\bf adj.}(U)} \wedge \psi_{\overline{U}} \right), 
\label{eq:Yukawa-12-55bar-A}
\end{equation}
where $(\psi_{U},\chi_{U})$ is the zero mode wavefunction for 
a massless chiral multiplet from the $(U, \bar{\bf 5})$
irreducible component of $G' \times \SU(5)_{\rm GUT}$, 
and $(\psi_{\overline{U}},\chi_{\overline{U}})$ is for 
the $(\overline{U},{\bf 5})$ irreducible component. 
$\psi_{{\bf adj.}(U)}$ is the wavefunction for the $({\bf adj.}, {\bf 1})$ 
component. In local models with gauge group $G = \SU(6)$, for example, 
$U$ is a line bundle whose structure group is the commutant of 
$\SU(5)_{\rm GUT}$ within this $G = \SU(6)$ (possibly along with
$\U(1)_Y$). When a local model with $G = E_8$ is chosen, then 
$U = \wedge^2 {\bf 5}$.
The zero mode wavefunctions satisfy a relation 
$(\psi_{\overline{U}},\chi_{\overline{U}}) = 
 (- \psi_{U}, \chi_{U})$  (e.g., \cite{Hayashi-2}) when we make 
an approximation that 
the 4-form flux is ignored,\footnote{
This is a good approximation in the large Kaluza--Klein radius limit.} 
and hence the factor $2$ follows. 
Unless the singlet wavefunction $\psi_{{\bf adj.}(U)}$ is always pointing 
towards the normal direction of the matter curve of 
$\SU(5)_{\rm GUT}$-${\bf 5}+\bar{\bf 5}$ representations, this trilinear 
coupling does not vanish. This overlap integration has contributions 
from everywhere along the matter curve $\bar{c}_{(\bar{\bf 5})}$. 

The overlap integration (\ref{eq:Yukawa-12-55bar-A}) is carried out 
over the surface $S$, but there might be an alternative expression 
where the overlap integration is done over the matter curve. 
Note that the matter fields in the $\SU(5)_{\rm GUT}$-${\bf 5}$ and 
$\bar{\bf 5}$ representations are characterized 
by \cite{DW-1, BHV-1, Hayashi-1}
\begin{equation}
 f_{({\bf 5})} \in H^0(\tilde{\bar{c}}_{(\bar{\bf 5})}; 
    {\cal L}_{(\bar{\bf 5})}^{-1} \otimes K_{\tilde{\bar{c}}_{(\bar{\bf
    5})}}^{1/2}),   \qquad 
 f_{(\bar{\bf 5})} \in H^0(\tilde{\bar{c}}_{(\bar{\bf 5})}; 
    {\cal L}_{(\bar{\bf 5})} \otimes K_{\tilde{\bar{c}}_{(\bar{\bf
    5})}}^{1/2}),   
\end{equation}
and hence $f_{({\bf 5})} \cdot f_{(\bar{\bf 5})}$ is in 
$H^0(\tilde{\bar{c}}_{(\bar{\bf 5})}; K_{\tilde{\bar{c}}_{(\bar{\bf 5})}})$; 
here, ${\cal L}_{(\bar{\bf 5})}$ is a line bundle for the fields 
in $\bar{\bf 5}$ representation, coming from the $C^{(3)}$ potential.
A Wilson line $\psi_{{\bf adj.}(U)} \in H^1(\widetilde{C}_U; \C)$ on 
$\widetilde{C}_U$ also 
defines a Wilson line $\psi^{(0,1)}$ on the covering matter curve   
$\tilde{\bar{c}}_U = \tilde{\bar{c}}_{(\bar{\bf 5})}$. An overlap integration 
\begin{equation}
  \int_{\tilde{\bar{c}}_{(\bar{\bf 5})}} f_{(\bar{\bf 5})}^{(1/2,0)} \cdot 
     \psi^{(0,1)} \cdot f_{({\bf 5})}^{(1/2,0)}
\label{eq:Yukawa-12-55bar-B}
\end{equation}
is well-defined, because the integrand can be regarded as a (1,1) form 
on $\tilde{\bar{c}}_{(\bar{\bf 5})}$; all of 
$f_{({\bf 5})}$, $f_{(\bar{\bf 5})}$ and the 
Wilson line $\psi_{{\bf adj.}(U)}$ are objects on the same spectral 
surface $\widetilde{C}_U$ \cite{Hayashi-2}, and taking a product 
as above is a natural operation. Although we have not shown that 
(\ref{eq:Yukawa-12-55bar-A}) is the same as the new expression 
(\ref{eq:Yukawa-12-55bar-B}), it they are the same, 
the latter expression will allow us 
to short-cut the process of calculation, and to save time in computing 
the wavefunctions $(\psi_U,\chi_U)$ from 
$f_{(\bar{\bf 5})}$ and carrying out overlap integration 
on $S$ numerically. 

Changing the vev of $C^{(3)}$ on $X$ infinitesimally by 
$H^{1,2}(X; \C)$ corresponds to turning on an infinitesimally small 
vev in the $\SU(5)_{\rm GUT}$ singlet chiral multiplets ${\bf 1}$'s 
in the Yukawa couplings (\ref{eq:Yukawa-515}). Thus, 
if a pair of vector-like ${\bf 5}$+$\bar{\bf 5}$ states are in the 
low-energy spectrum, and if they have a Yukawa coupling with a 
$H^{1,2}(X; \C)$ moduli, then the vev of $C^{(3)}$ is 
chosen in our vacuum is such that the ${\bf 5}$+$\bar{\bf 5}$ 
pair happens to appear in the low-energy spectrum.

Yukawa couplings of the form (\ref{eq:Yukawa-515}) with a singlet
from $H^{3,1}(X; \C)$, on the other hand, are calculated by 
an overlap integral\footnote{\label{fn:kin-mix}
Since both $H^{1,2}(X; \C)$ and $H^{3,1}(X; \C)$ moduli fields can 
have Yukawa couplings of the form (\ref{eq:Yukawa-515}), kinetic 
mixing between the two types of the moduli fields is generated at 1-loop
level. At quantitative level, however, the mixing terms may still 
be loop-suppressed.} 
\begin{equation}
 \lambda = - 2 \int_S \tr \left(\psi_{U} \wedge 
   \chi_{{\bf adj.}(U)} \wedge \psi_{\overline{U}} \right).
\label{eq:Yukawa-31-55bar}
\end{equation}
The (0,1)-form valued wavefunctions $\psi_{U}$ and
$\psi_{\overline{U}}$ differ only by a phase $(-1)$ 
in the approximation of ignoring the 4-form fluxes. 
Thus, the overlap integration vanishes at the level of this 
approximation. The Yukawa couplings of this type may, therefore, 
be suppressed at least by some positive powers of a ratio 
$(l_s/R_{\rm GUT}) < 1$. The $H^{3,1}(X; \C)$ moduli fields 
under consideration here correspond to the deformation 
of spectral surface in Heterotic dual, and it was shown 
in \cite{Penn5} that this type of singlets has a Yukawa coupling
with ${\bf 5}$--$\bar{\bf 5}$ pairs in Heterotic compactifications, 
so that the deformation of the spectral surface change the number 
of massless states. 
It is hard to believe that the same couplings vanish in 
F-theory compactifications. This overlap integration in F-theory 
description is not necessarily localized at codimension-3 singularity 
points either.

The neutral-charged-charged Yukawa couplings
(\ref{eq:Yukawa-12-55bar-A}) and (\ref{eq:Yukawa-31-55bar}) 
are given an alternative expression.
A product 
\begin{equation}
 {\rm Ext}^1(i_{\sigma *} {\cal O}_S, i_{C_U *}{\cal N}_U) \times 
 {\rm Ext}^1(i_{C_U *} {\cal N}_U, i_{C_U *} {\cal N}_U) \times 
 {\rm Ext}^1(i_{C_U *}{\cal N}_U, i_{\sigma *} {\cal O}_S) \rightarrow
 {\rm Ext}^3(i_{\sigma *} {\cal O}_S,i_{\sigma *} {\cal O}_S)
\end{equation}
is well-defined, and here, both $H^0(C_U; K_{C_U})$ ($H^{3,1}(X; \C)$) 
moduli and $H^1(C_U; \C)$ ($H^{1,2}(X; \C)$) are treated at once. 
Because 
\begin{equation}
 {\rm Ext}^3 (i_{\sigma *} {\cal O}_S, i_{\sigma *} {\cal O}_S) 
\simeq H^2(S; K_S) = H^{2,2}(S; \C) \simeq \C, 
\end{equation}
the product above returns a complex number. 

\subsubsection{Trilinear Yukawa Couplings among Neutral Fields}
\label{sssec:3-singlets}

It is also of phenomenological interest whether 
the singlet chiral multiplet $S$ of the next-to-minimal supersymmetric 
Standard Model (NMSSM) has trilinear coupling $\Delta W = \kappa \; S^3$ 
or not. If this coupling is of order unity, then that is an ordinary 
NMSSM. If it is small, then an accidental U(1) global Peccei--Quinn 
symmetry exists, and there exists a light pseudo Goldstone boson in the 
spectrum \cite{HW}. 
The Higgs mass bound from the LEP experiment is relaxed in the presence 
of such a pseudo Goldstone boson, and the Higgs detection strategy at 
the LHC should be may also have to be different \cite{Higgs-LEP}.  
Such a light boson may also be relevant to some cosmic ray signals \cite{NT}. 

Following the arguments above, we consider the $H^{1,2}(X; \C)$ moduli and 
$H^{3,1}(X; \C)$ moduli as the candidates for such a singlet field $S$. 
If $S$ is a $H^{3,1}(X; \C)$ moduli, then such trilinear couplings 
may appear from the cubic term of $(\delta a)$ in the expansion in 
(\ref{eq:M-mass-via-GVW}); the 4-form $\Omega(a = \delta a + a_0)$ 
up to the order of $(\delta a)^3$ remains within 
$H^{4,0}(X; \C) \oplus 
H^{3,1}(X; \C) \oplus H^{2,2}(X; \C) \oplus H^{1,3}(X; \C)$ 
in the vacuum complex structure, and can be expressed as 
\begin{equation}
 \Omega (a) = \Omega(a_0) + \cdots + 
  (k_{abc} \Omega(a_0) + l_{abc}^d \chi_d + \tilde{\psi}_{abc} + 
   \tilde{l}_{abc;d}\tilde{\chi}^d) (\delta a)_a (\delta a)_b (\delta
   a)_c
  + {\cal O}\left( (\delta a)^4 \right), 
\end{equation}
where $\tilde{\psi}_{abc} \in H^{2,2}(X_{a = a_0}; \C)$, 
$\tilde{\chi}^d$ form a basis of $H^{1,3}(X|_{a=a_0}; \C)$, and 
ellipses stand for the terms in (\ref{eq:Omega-2nd}) that are 
linear or quadratic in the fluctuations in the complex structure moduli 
$(\delta a)$. The trilinear coupling for the chiral multiplets $S_a$
corresponding to $(\delta a)_a$ is given by 
\begin{equation}
 \Delta W  = \lambda_{abc} \; S_a S_b S_c;  \qquad 
 \lambda_{abc} = \int_X \tilde{\psi}_{abc} \wedge \vev{G^{(4)}}.
\label{eq:singl-trilinear}
\end{equation}
Generically, this coupling does not vanish. It should be reminded, though, 
that the moduli fields from $H^{3,1}(X; \C)$ generically obtain 
masses in the presence of 4-form fluxes, and the energy scale of 
their masses in section \ref{sec:Majorana} is quite large. Without 
a special reason, such moduli fields are not expected to appear 
at low energy scale such as the electroweak scale. 

If the singlet field $S$ is from $H^{1,2}(X; \C)$ moduli, on the other
hand, it has a good reason to not have a mass term. This class of moduli 
originates from the 3-form potential field $C^{(3)}$, and the gauge 
symmetry 
shifting $C^{(3)}$ by an exact 3-form prevents the moduli of this class 
from having arbitrary form of superpotential, other than (\ref{eq:GVW}) 
and its non-Abelian extension (\ref{eq:F-super}).
The zero modes of $C^{(3)}$ have the Yukawa couplings 
(\ref{eq:Yukawa-12-55bar-A}, \ref{eq:Yukawa-12-55bar-B}), because 
the charged matter fields originate from M2-branes, which carry electric 
charges of $C^{(3)}$.

As we have already noted in footnote \ref{fn:kin-mix}, the moduli chiral 
multiplets $\phi_{(3,1)}$ from $H^{3,1}(X; \C)$ and the
chiral multiplets $\phi_{(1,2)}$ from $H^{(1,2)}(X; \C)$ tend to have 
kinetic mixing of the form 
\begin{equation}
 \Delta K_{4D, {\rm eff.}} 
= \left(\phi_{(3,1)}^\dagger, \; \phi_{(1,2)}^\dagger
		       \right)
 \left(\begin{array}{cc}
  1 & \epsilon^* \\ \epsilon & 1
       \end{array}\right)
  \left( \begin{array}{c}
   \phi_{(3,1)} \\ \phi_{(1,2)}
	 \end{array}\right); 
\end{equation}
where the mixing coefficients $\epsilon$ may be loop-suppressed. 
This kinetic term and the mass term 
\begin{equation}
 \Delta W_{\rm eff.} = \frac{1}{2}
 \left(\phi_{(3,1)}, \; \phi_{(1,2)} \right)
 \left(\begin{array}{cc}
  M_R & 0 \\ 0 & 0
       \end{array}\right)
  \left( \begin{array}{c}
   \phi_{(3,1)} \\ \phi_{(1,2)}
	 \end{array}\right) 
\label{eq:2x2-mass}
\end{equation}
should be diagonalized simultaneously, in order to identify 
the mass-eigen-basis with diagonal kinetic terms.
It is done by 
\begin{equation}
 \left( \begin{array}{c}
   \phi_{(3,1)} \\ \phi_{(1,2)}
	 \end{array} \right) = 
 \left( \begin{array}{cc}
  1 & 0 \\ -\epsilon & 1
	\end{array}\right)
 \left( \begin{array}{c}
   \hat{\phi}_{(3,1)} \\ \hat{\phi}_{(1,2)}
	 \end{array} \right),
\label{eq:base-transf}
\end{equation}
where $\hat{\phi}_{(3,1)}$ and $\hat{\phi}_{(1,2)}$ forms 
the basis we want. 
The effective superpotential in the 
$(\hat{\phi}_{(3,1)}, \hat{\phi}_{(1,2)})$ basis is obtained by 
substituting (\ref{eq:base-transf}) to the effective superpotential 
written in terms of $(\phi_{(3,1)}, \phi_{(1,2)})$.
Because $\hat{\phi}_{(1,2)}$ fields only pick up interactions that 
$\phi_{(1,2)}$ have, and because (\ref{eq:singl-trilinear}) of 
$\phi_{(3,1)}$ fields is the only cubic term of the $\SU(5)_{\rm GUT}$ 
singlets, the $\hat{\phi}_{(1,2)}$ fields are not involved in any cubic 
interactions among singlets in the superpotential.
This observation may well be taken 
as a prediction that singlet chiral multiplets appearing 
in the NMSSM do not have trilinear terms, apart from those 
possibly generated by M5-brane instantons with highly 
suppressed coefficients. 

The $H^{1,2}(X; \C)$ moduli fields not only have trilinear couplings 
to vector-like pairs in the $\SU(5)_{\rm GUT}$ visible sector on 
$S \subset B_3$, but also to any vector-like pairs $\Psi+\bar{\Psi}$ 
on other irreducible pieces of the discriminant.  
Thus, the singlet field $S$ of the NMSSM may have an effective 
superpotential of the form 
\begin{equation}
 \Delta W = \lambda \; S H_u H_d + \xi \; S \Psi \bar{\Psi}, 
\end{equation}
which is exactly the theory considered in \cite{NT}.

\section{Dimension-4 Proton Decay Problem Revisited}
\label{sec:Dim4}

Up to know, we have only imposed the conditions that the 
chiral multiplets of right-handed neutrinos are 
characterized as $\SU(5)_{\rm GUT}$ singlets that have 
Yukawa couplings of the form (\ref{eq:Yukawa-515}).  
This is, however, not all that is known from phenomenology. 

So far, we have not introduced an $R$-parity or anything that 
replaces it. Thus, in generic F-theory compactifications that result 
in supersymmetric extensions of the Standard Model, we should 
expect the dimension-4 proton decay operators 
\begin{equation}
 \Delta W = \bar{\bf 5} \; {\bf 10} \; \bar{\bf 5} 
  = \lambda^{''} \; \bar{D} \; \bar{U} \; \bar{D} + 
    \lambda' \; \bar{D} \; Q \; L + 
    \lambda \; L \; \bar{E} \; L
\label{eq:dim-4}
\end{equation}
to be generated with unsuppressed couplings. That is a totally 
unacceptable phenomenologically. The absence of dimension-4 proton decay operators 
implies that the complex structure for F-theory compactification
cannot be just generic, but somewhat special. Moduli fields 
in our vacuum must be fluctuations from such a special 
choice of complex structure, and chiral multiplets of right-handed 
neutrinos are among those fluctuations. The special choice of complex structure 
at our vacuum may have some special implications in physics of 
right-handed neutrinos. 

Whether the right-handed neutrinos have localized wavefunctions 
or not, and whether their Majorana masses and Yukawa couplings 
are localized or not are certainly part of questions of 
phenomenological interest. As we will see in this section, 
the answer to these questions are indeed different, for various 
mechanisms that solve the dimension-4 proton decay problem. 

\subsection{$R$-parity}
\label{ssec:R-parity}

Imposing $R$-parity is arguably the most conventional 
way to get rid of dimension-4 proton decay operators from 
the low-energy effective theory. If the Calabi--Yau 4-fold 
$X$ and a 4-form flux $G^{(4)}$ on it has a $\Z_2$ symmetry, then 
the symmetry manifests itself in the effective theory. 
By assumption, the $A_4$ singularity locus $S$ is mapped 
to itself by the $\Z_2$ transformation, and matter curves 
of various representations to themselves. 
Note that we consider $(X, G^{(4)})$ that has a $\Z_2$ symmetry 
in this scenario; it is not 
that we take a quotient of $(X, G^{(4)})$ by the $\Z_2$ symmetry. 

Because of the $\Z_2$ invariance of the matter curves $\bar{c}_{(R)}$ and 
the sheaves ${\cal F}_{(R)}$ on them, the $\Z_2$ transformation also acts 
on the zero modes---vector spaces of holomorphic sections 
$H^{p=0,1}(\bar{c}_{(R)}; {\cal F}_{(R)})$. 
The zero modes in a given representation $R$ of $\SU(5)_{\rm GUT}$, 
therefore, split into a direct sum of $\Z_2$-odd part and 
$\Z_2$-even part. We need to assume that 
\begin{eqnarray}
& & 
h^0_-(\bar{c}_{({\bf 10})}; {\cal F}_{({\bf 10})}) = 3, \;
h^0_-(\bar{c}_{(\bar{\bf 5})}; 
      {\cal F}_{(\bar{\bf 3} \subset \bar{\bf 5})}) = 3, \;
h^0_-(\bar{c}_{(\bar{\bf 5})}; 
      {\cal F}_{(\bar{\bf 2} \subset \bar{\bf 5})}) = 3, \;
h^1_-(\bar{c}_{(\bar{\bf 5})}; 
      {\cal F}_{(\bar{\bf 2} \subset \bar{\bf 5})}) = 0,  \nonumber \\
& & 
h^0_+(\bar{c}_{({\bf 10})}; {\cal F}_{({\bf 10})}) = 0, \;
h^0_+(\bar{c}_{(\bar{\bf 5})}; 
      {\cal F}_{(\bar{\bf 3} \subset \bar{\bf 5})}) = 0, \;
h^0_+(\bar{c}_{(\bar{\bf 5})}; 
      {\cal F}_{(\bar{\bf 2} \subset \bar{\bf 5})}) = 1, \;
h^1_+(\bar{c}_{(\bar{\bf 5})}; 
      {\cal F}_{(\bar{\bf 2} \subset \bar{\bf 5})}) = 1,  \nonumber 
\end{eqnarray}
where $\pm$ denotes $\Z_2$-even/odd part of the cohomology groups, 
and that all other cohomology groups vanish, 
because we need three copies of $(Q, \bar{U}, \bar{E})$-like 
chiral multiplets and $(\bar{D},L)$-like chiral multiplets that are 
odd under the $R$-parity (matter parity), and we need one pair 
of Higgs doublets that are even under the $\Z_2$ symmetry. 
Note that the Higgs doublets and the $(\bar{D},L)$-type fields are all 
localized along the same matter curve, $\bar{c}_{(\bar{\bf 5})}$, and
yet the curve $\bar{c}_{(\bar{\bf 5})}$ does not have to split into 
two (or more) irreducible pieces.
Codimension-3 singularity points for $A_4 \rightarrow E_6$ 
enhancement and those for $A_4 \rightarrow D_6$ enhancement are 
also mapped to the codimension-3 singularity points of the same type 
by the $\Z_2$ transformation. Yukawa couplings of the 
form (\ref{eq:dim-4}) will be generated at each codimension-3 
singularity point of $A_4 \rightarrow D_6$ enhancement, but they 
vanish as a result of $\Z_2$-odd nature of the wavefunctions of 
the three relevant chiral multiplets after contributions from 
all the codimension-3 singularity points of this type are summed up. 
The Yukawa couplings for down-type quarks and charged leptons, 
on the other hand, will remain non-zero, because the $\Z_2$ symmetry 
does not ensure cancellation of all the contributions to 
the Yukawa couplings of this type.

The chiral multiplets of right-handed neutrinos correspond to\footnote{
$H^{1,2}_-(X; \C)$ moduli fields also qualify, if the neutrino 
masses are Dirac. 
} $H^{3,1}_-(X; \C)$, because 
only the $\Z_2$-odd ones have trilinear couplings suitable for 
the neutrino Yukawa couplings $\Delta W = H_u \bar{N} L$. 
The Majorana mass terms of the complex structure moduli fields 
$H^{3,1}(X; \C)$ in (\ref{eq:M-mass-via-GVW}) consist of 
$H^{3,1}_+(X; \C)$--$H^{3,1}_+(X; \C)$ mass terms and 
$H^{3,1}_-(X; \C)$--$H^{3,1}_-(X; \C)$ mass terms, but 
the mass terms (\ref{eq:M-mass-via-GVW}) do not have 
a mixed $H^{3,1}_+(X; \C)$--$H^{3,1}_-(X; \C)$ mass term 
because of the $\Z_2$ symmetry of $(X, G^{(4)})$.
The $H^{3,1}_-(X; \C)$--$H^{3,1}_-(X; \C)$ mass terms 
in (\ref{eq:M-mass-via-GVW}) are identified with the Majorana 
masses of the right-handed neutrinos. 

In this scenario, the neutrino Yukawa couplings are localized along 
the matter curve 
$\bar{c}_{(\bar{\bf 5})}$ of $\SU(5)_{\rm GUT}$-$\bar{\bf 5}+{\bf 5}$ 
representations, but not further localization is expected along the
curve. The Majorana mass terms of the right-handed neutrinos come from 
the Gukov--Vafa--Witten superpotential, and are not localized anywhere
in the Calabi--Yau 4-fold, but come from the entire bulk of the geometry. 

The story developed in sections \ref{sec:Majorana} and
\ref{sec:nuYukawa-general} holds without any modification in the 
$\Z_2$ symmetry scenario. Apart from imposing (assuming) a $\Z_2$ 
symmetry in $(X, G^{(4)})$, one does not have to do anything
non-trivial. The $\Z_2$-odd part of the moduli fields automatically 
have neutrino Yukawa couplings, and the $H^{3,1}_-(X; \C)$ moduli, in 
particular, have Majorana masses from flux compactification. The
Majorana mass scale is just that of the right-handed 
neutrinos to account for the low-energy neutrino masses through 
the see-saw mechanism. There was nothing non-trivial in the 
scenario above, yet this is still quite remarkable. 

Traditionally in field theory model building in 3+1 dimensions, 
one tends to consider that it is not an easy task to generate 
the Majorana masses of right-handed neutrinos, if the gauge group 
is larger than $\SU(5)_{\rm GUT}$ at high-energy (or at microscopic
level). In SO(10) GUT models, for example, there is an option to 
introduce an exotic $\SO(10)$-${\bf 126}$ field that is $\Z_2$-even.
Its vev does not break the $\Z_2$ symmetry. Majorana mass terms are 
obtained if an effective theory has the following renormalizable 
coupling, 
\begin{equation}
 \Delta W = \vev{{\bf 126}_+} \; {\bf 16}_- \; {\bf 16}_-.
\label{eq:126}
\end{equation}
The situation does not change very much when the gauge group at 
the microscopic level is the $\SU(5)_{\rm GUT} \times \U(1)_\chi$ 
subgroup of $\SO(10)$. The ${\bf 126}$ 
field is replaced by an $\SU(5)_{\rm GUT}$ singlet that carries 
$+10$ units of the $\U(1)_\chi$ charge.\footnote{\label{fn:chi-charge}
It is conventional 
to set the normalization of $\U(1)_\chi$ charge so that ${\bf 10}$
carries $Q_\chi = -1$, $(\bar{D},L)$ has $Q_\chi = +3$ units of the 
$\U(1)_\chi$ charge, $H({\bf 5})$ [resp. $\bar{H}(\bar{\bf 5})$] 
$Q_\chi = +2$ [resp. $Q_\chi = -2$] units, and
right-handed neutrinos $\overline{N}$ carry $Q_\chi = -5$ units.}
An equally popular alternative is to introduce a less exotic 
field in the $\SO(10)$-$\overline{\bf 16}$ representation that is 
$\Z_2$-even. Majorana mass terms of right-handed neutrinos are
available, if an effective theory has the following non-renormalizable 
interactions:
\begin{equation}
 \Delta W = \vev{\overline{\bf 16}_+} \vev{\overline{\bf 16}_+} 
  {\bf 16}_- {\bf 16}_-.
\label{eq:++--}
\end{equation}
The origin of those fields and their interactions, as well as 
the $\Z_2$-parity assignment are usually not questioned in field theory 
model building and such assumptions are just introduced 
by hand for phenomenological convenience. In string phenomenology, 
however, these issues are ``the questions'' to be addressed. 
All necessary fields and interactions should be derived from known 
sources. Since charged matter fields of $\SU(5)$ GUT models  
originate from breaking of a symmetry $G$ larger than $\SU(5)_{\rm GUT}$, 
for example, in Heterotic $E_8 \times E_8$ string compactifications 
(with supergravity approximation), intersecting D-brane systems, and 
in deformation of singularity in M/F-theory compactifications \cite{KV}, 
a larger symmetry $G$ exists at this microscopic level in these
compactifications, and the origin of the necessary fields and 
interactions have been an issue with no clear answer

To see why/how this issue was overcome in generic F-theory 
compactifications without any difficulties, let us remind ourselves
of the following. As reviewed in section 2 
of \cite{Hayashi-2}, physics of $\SU(5)_{\rm GUT}$ GUT models in 
{\it generic} F-theory compactifications is described by a patchwork 
of gauge theories. The locus of $A_4$ singularity $S$ is covered by 
various patches, and individual patches have their own gauge groups 
containing $\SU(5)_{\rm GUT}$. In patches that contain the matter curve 
$\bar{c}_{({\bf 10})}$, the gauge group should be $G = \SO(10) \supset
\SU(5)_{\rm GUT}$ or larger. If a patch in $S$ contains the matter curve 
$\bar{c}_{(\bar{\bf 5})}$, then the gauge group in this patch has to be
$G = \SU(6)$ or larger. The gauge group in a patch containing a codimension-3
singularity point should be larger than $\SU(5)_{\rm GUT}$ at least by 
rank 2, that is, $G$ is at least $E_6$, $\SO(12)$ or $\SU(7)$, 
because 2 independent topological 2-cycles collapse at the
codimension-3 singularity point. The gauge groups in these field theory 
local models can be chosen even larger for higher level of approximation.
The field thoery local models defined on various patches on $S$ may
have different gauge groups $G$ with different ranks, but they can be 
glued together within an overlap of two neighboring patches, because 
a gauge theory with lower rank gauge group can provide reasonably well 
approximation when one of 2-cycles becomes large. 

An important point is that there is no common ``microscopic gauge group'' 
$G$ containing $G' \times \SU(5)_{\rm GUT}$ over the entire 
$A_4$ singularity locus $S$ in generic F-theory
compactifications.\footnote{F-theory compactifications with Heterotic
dual are an exception, because a unique gauge group $G = E_8$ and 
the structure group $G' = \SU(5)_{\rm str}$ are found.} 
The absence of the common structure group $G'$ means that the right-handed 
neutrinos and anything that effectively becomes the Majorana mass 
parameters cannot be characterized as something in definite
representation of the ``common structure group $G'$'' (or of the ``microscopic 
gauge group $G$''), and all the theoretical constraints associated with the 
representation theory of $G'$ or $G$ have gone. In generic F-theory 
compactifications, no clear line is drawn between the fields in the 
gauge theory sector and bulk gravity sector, as opposed to the (dual of) 
Heterotic string compactifications with supergravity approximation, or 
to the Type IIB Calabi--Yau orientifold compactifications. Complex 
structure moduli $H^{3,1}(X; \C)$ of bulk gravity can also be regarded 
locally as a part of Cartan subspace in local models on $S$ with 
$G = \SU(6)$ gauge groups. This class of $\SU(5)_{\rm GUT}$-neutral 
fields can be identified with right-handed neutrinos, because they 
have a natural $\SU(6)$ gauge interactions (\ref{eq:Yukawa-515}) that 
can be identified with the neutrino Yukawa couplings, and yet, 
the gauge invariance of $G = \SU(6)$ constrains the interactions 
of these right-handed neutrinos only in a local geometry, and 
does not prevent them from having the Majorana mass terms 
in the Gukov--Vafa--Witten superpotential. 
This is the heart of the trick that solve the long-standing
problem in generic F-theory compactifications. 
 
\subsection{Reducible Limit of Spectral Surface}
\label{ssec:split-surf}

Some other solutions to the dimension-4 proton decay problem in F-theory 
have already been discussed in the literature; the $\Z_2$ symmetry 
scenario in section \ref{ssec:R-parity} is not the first one. 
In sections~\ref{ssec:split-surf}--\ref{ssec:Rparity-violation}, 
we will elaborate on these alternative scenarios, and discuss 
phenomenological consequences in these scenarios. 
In section \ref{ssec:split-surf}, we will work on a scenario
\cite{TW-1}, where spectral surfaces are reducible, and an extra 
U(1) symmetry is used to bring the dimension-4 proton decay operators  
under control, instead of the $\Z_2$ symmetry.  

Defining equation of local geometry of Calabi--Yau 4-fold $X$ is given by 
\begin{eqnarray}
 y^2 & = & x^3 + (a_5 x y + a_4 z x^2 + a_3 z^2 y + a_2 z^3 x + a_0 z^5)
  \nonumber \\
  & & \quad + (A_5 z y + A_4 z^2 x^2 + A_3 z^3 y + A_2 z^4 x + A_0 z^6)
   + \cdots
\label{eq:local-def}
\end{eqnarray}
for generic F-theory compactifications that leave $\SU(5)$ 
GUTs \cite{6authors}. 
$(x,y)$ are the coordinates of the elliptic fiber of $X$. 
The discriminant of this elliptic fibration is given by 
\begin{equation}
 \Delta = z^5 \left[ \frac{1}{16} a_5^4 P^{(5)} 
  + \frac{z}{16} a_5^2 \left\{ 
    12 \left(a_4 + \frac{a_5 A_5}{2}\right) P^{(5)} 
  - a_5^2 \tilde{R}^{(5)} \right\} + {\cal O}(z^2) \right], 
\label{eq:discriminant}
\end{equation}
where 
\begin{equation}
 P^{(5)} = a_0 a_5^2 - a_2 a_5 a_3 + a_4 a_3^2,
\end{equation}
and the definition of $\tilde{R}^{(5)}$ is given later in this article.
An irreducible component $S$ of the discriminant locus is found along 
$z = 0$; and hence $z$ is the normal coordinate of $S$ in $B_3$. 
We will take a set of local coordinates $(u,v)$ on a local patch of $S$.
$a_r(u,v)$ and $A_r(u,v)$ ($r=0,2,\cdots, 5$) are locally 
regarded as functions on $S$, and ellipses stand for terms 
higher order in the series expansion of $z$.
$a_r$'s, $A_r$'s and all others are regarded globally as holomorphic 
sections of appropriate line bundles; defining a divisor $t$ 
(or equivalently $\eta$) on $S$ through 
$c_1(N_{S|B_3}) \equiv t \equiv (6K_S + \eta)$ \cite{Rajesh}, 
relevant line bundles are 
\begin{equation}
 a_r \in \Gamma(S; {\cal O}(r K_S + \eta)) = 
         \Gamma(S; {\cal O}(-(6-r)K_S + t)), \qquad 
 A_r \in \Gamma(S; {\cal O}(-(6-r)K_S) ).
\label{eq:line-bdls}
\end{equation}

In the first line of (\ref{eq:local-def}), $(a_5 xy + \cdots + a_0 z^6)$
part closely resembles the defining equation of a spectral surface of 
an $\SU(5)$ vector bundle in Heterotic $E_8 \times E_8$ string
compactification on an elliptic fibered Calabi--Yau 3-fold $Z$,  
\begin{equation}
 5\sigma + {\rm div} (a_5 xy + a_4 x^2 + a_3 y + a_2 x + a_0). 
\label{eq:spec-surf-Het}
\end{equation}
Here, $\sigma$ is the zero section of the elliptic fibration 
\begin{equation}
\pi_Z: Z \rightarrow S, 
\end{equation}
and $a_r$ ($r = 0,2,\cdots,5$) are regarded as holomorphic sections 
of line bundles on $S$ given as in (\ref{eq:line-bdls}) for some 
divisor $\eta$ of $S$.
This resemblance is at the heart of the duality between Heterotic string and 
F-theory compactifications \cite{KMV-BM, CD, DW-1, Hayashi-1, Hayashi-2}. 
Although not all of F-theory compactifications have a Heterotic dual, 
physics along the $A_4$ singularity locus $S$ at $z=0$ for generic 
F-theory compactifications is still qualitatively similar to that of 
Heterotic string compactification.  

\subsubsection{SO(10) Scenario}

In Heterotic $E_8 \times E_8$ string compactification, the structure 
group of a rank-5 vector bundle $V_5$ may be reduced from $\SU(5)$ 
to its subgroup $\SU(4) \times \U(1)_\chi$, and 
$V_5 = U_4 \oplus L_\chi$, where $U_4$ and $L_\chi$ are rank-4 and
rank-1 bundles, respectively. This can be achieved, for example, by setting 
$a_5 = 0$ in (\ref{eq:spec-surf-Het}). 
The spectral surface $C_{V_5} = C_{({\bf 10})}$ becomes reducible then:
\begin{equation}
 \left[4\sigma + {\rm div}(a_4 x^2 + \cdots a_0) \right] + \sigma.
\end{equation}
$\U(1)_\chi$ symmetry transformation commutes with the structure 
group $\SU(4) \times \U(1)_\chi$, and hence remains as a global symmetry 
of the effective action. 
One can do the same thing in generic F-theory compactifications 
by setting $a_5 = 0$ in (\ref{eq:local-def}) \cite{TW-1}. 
In this limit, the spectral surface of the Higgs bundle
for fields in the $\SU(5)_{\rm GUT}$-{\bf 10} representation, 
$C_{({\bf 10})}$, becomes reducible: 
\begin{equation}
C_{({\bf 10})} \rightarrow C_{({\bf 16})} + \sigma, \quad 
\left[a_5 + a_4 \xi + a_3 \xi^2 + \cdots = 0\right] \rightarrow 
\left[a_4 + a_3 \xi + \cdots = 0 \right] + [\xi = 0],
\end{equation}
where $\sigma$ is the zero section of the canonical bundle 
$\pi_{\mathbb{K}_S}: \mathbb{K}_S \rightarrow S$, and $\xi$ is 
the coordinate of the rank-1 fiber vector space of the canonical 
bundle $K_S$.

Now, with $a_5 = 0$, the discriminant $\Delta$ begins with an 
${\cal O}(z^7)$ term, and we have a split $D_5$ singularity along $S$.
There is an $\SO(10)$ gauge theory defined globally on $S$. The $\SO(10)$
symmetry can be broken down to $\SU(5)_{\rm GUT} \times \U(1)_\chi$ or 
$G_{SM} \times \U(1)_\chi$ 
[$G_{SM} \equiv \SU(3)_C \times \SU(2)_L \times \U(1)_Y$] 
subgroup by turning on fluxes. The massless vector fields may or may 
not become massive through the St\"{u}ckelberg interactions, but 
the associated global $\U(1)_\chi$ symmetry remains unbroken (unless 
spontaneous symmetry breaking is triggered; we will come back to this 
issue in section \ref{ssec:Rparity-violation}). The dimension-4 
proton decay operators are not allowed in the presence of unbroken 
$\U(1)_\chi$ symmetry, because the operators (\ref{eq:dim-4}) have 
$Q_\chi = +5 \neq 0$ units of the $\U(1)_\chi$ charge (see footnote 
\ref{fn:chi-charge}). 
Note that we do not need a Heterotic dual for this scenario to 
make sense; in other words, we do not need a rank-5 Higgs bundle 
defined globally on $S$. The essence here is to have a 
$\U(1)_\chi \subset \SO(10)$ gauge symmetry on $S$, which is 
characterized also as a reducible limit of the spectral surface 
$C_{({\bf 10})}$. This characterisation, $a_5 = 0$, 
involves information only of local geometry along $S$, and 
fully generic F-theory compactifications can be considered (regardless 
of whether Heterotic dual exists or not), as long as this condition 
is satisfied.

Let us now give some thoughts on the candidates of the right-handed 
neutrinos in this scenario. We now consider a limit where 
the spectral surface $C_{({\bf 10})}$ consists (at least) of 
two irreducible components. Along the intersection of the two 
components, the spectral surface $C_{({\bf 10})}$ is singular. 
Thus, we have to reconsider the argument in section
\ref{sec:nuYukawa-general}, where the spectral surface was implicitly 
assumed to be smooth. 

Suppose that the Higgs bundle $(\varphi, V)$ [i.e. Higgs sheaf
${\cal V}$]
is decomposed into a direct sum $(\varphi_1 \oplus \varphi_2, V_1 \oplus V_2)$ 
[i.e., ${\cal V} = i_{C_1 *} {\cal N}_1 \oplus i_{C_2 *} {\cal N}_2$].
Instead of $H^0(C_{({\bf 10})}; K_{C_{({\bf 10})}})$ and 
$H^1(C_{({\bf 10})}; \C)$ in (\ref{eq:moduli-cohomology-specsurf}), 
we find that ${\rm Ext}^1 ({\cal V}, {\cal V})$ in (\ref{eq:moduli-ext})
is the suitable characterization of the moduli fields, when the spectral 
surface consists of two irreducible pieces $C_1 + C_2$. 
The extension group can be decomposed as follows in such a 
reducible limit:
\begin{eqnarray}
 {\rm Ext}^1 ({\cal V}, {\cal V}) & = &
 {\rm Ext}^1 (i_{1*} {\cal N}_1, i_{1*} {\cal N}_1 )
+ {\rm Ext}^1 (i_{2*} {\cal N}_2, i_{2*} {\cal N}_2 ) \nonumber \\
& & + {\rm Ext}^1 (i_{1*} {\cal N}_1, i_{2*} {\cal N}_2 )
+ {\rm Ext}^1 (i_{2*} {\cal N}_2, i_{1*} {\cal N}_1 ).
\label{eq:moduli-intersection}
\end{eqnarray}
The last two components are essentially localized along the intersection 
of the two surfaces, $C_1$ and $C_2$.
Irreducible pieces of the Higgs sheaf $i_{C_i *} {\cal N}_i$ ($i=1,2$) 
on $\mathbb{K}_S$ can be treated as if they were 7-branes in
Calabi--Yau orientifolds; the spectral surfaces $C_i$ 
play the role of the supports of the ``7-branes'' and ${\cal N}_i$ 
the ``gauge bundles'' on them. The localized pieces of the moduli 
are like ``open strings'' connecting the ``7-branes''.

For the case of practical interest, we consider a limit where 
the Higgs sheaf for the $\SU(5)_{\rm GUT}$-{\bf 10} representation
fields ${\cal V}_{({\bf 10})}$ becomes reducible:
\begin{equation}
 {\cal V}_{({\bf 10})} \rightarrow 
 i_{C_{({\bf 16})}*} {\cal N}_{({\bf 16})} + 
 i_{\sigma * } {\cal L}_\chi.
\end{equation}
${\cal L}_\chi$ is the line bundle on $S$ whose structure group is 
$\U(1)_\chi$.
The two localized components of the moduli become 
\begin{eqnarray}
 {\rm Ext}^1(i_{\sigma *} {\cal L}^{-Q_\chi}_\chi, i_{C_{({\bf 16})}*} 
  {\cal N}_{({\bf 16})}) & = & H^0 (\bar{c}_{({\bf 16})}; 
  {\cal N}_{(\bf 16)}\otimes {\cal L}_\chi^{Q_\chi}\otimes K_S) 
 \nonumber \\
 & & = H^0(\bar{c}_{({\bf 16})}; 
     ({\cal L}_{({\bf 16})} \otimes {\cal L}_\chi^{Q_\chi})
    \otimes {\cal O}(K_S + r_{({\bf 16})}/2)), \label{eq:RHN-41} \\
 {\rm Ext}^1(i_{C_{({\bf 16})}*} {\cal N}_{({\bf 16})}, i_{\sigma *}
  {\cal L}^{- Q_\chi}_\chi) & = & H^0 (\bar{c}_{({\bf 16})}; 
  {\cal N}_{(\bf 16)}^{-1} \otimes {\cal L}_\chi^{- Q_\chi} 
     \otimes {\cal O}(C_{({\bf 16})}))
 \nonumber \\
 & & = H^0(\bar{c}_{({\bf 16})}; 
     ({\cal L}_{({\bf 16})} \otimes {\cal L}_\chi^{Q_\chi})^{-1}
    \otimes {\cal O}(K_S + r_{({\bf 16})}/2)),  \label{eq:RHNc-41}
\end{eqnarray}
where ${\cal L}_{({\bf 16})}$ is defined by 
${\cal N}_{({\bf 16})} = {\cal O}(r_{({\bf 16})}/2) \otimes 
{\cal L}_{({\bf 16})}$, and 
$r_{({\bf 16})} \equiv K_{C_{({\bf 16})}} - \pi_C^* K_S = 
 C_{({\bf 16})} - \pi_C^* K_S$ 
is the ramification divisor of the projection 
$\pi_C : C_{({\bf 16})} \rightarrow S$.
$\bar{c}_{({\bf 16})}$ is regarded as the intersection of the 
two irreducible pieces of the spectral surface, 
$\sigma \cdot C_{({\bf 16})}$, although it is also seen as the matter 
curve of $\SO(10)$ models. 
The $\SU(5)_{\rm GUT}$-neutral component in the $\SO(10)$-{\bf 16} 
representation has $Q_\chi = -5$ units of the $\U(1)_\chi$ charge. 
If there are such chiral multiplets (\ref{eq:RHN-41}) 
[denoted by $\overline{N}$] and their vector-like partners 
(\ref{eq:RHNc-41}) [denoted by $\overline{N}^c$],
then their products $\overline{N}\overline{N}^c$ are sections of 
a line bundle 
\begin{equation}
 {\cal O}(2K_S + r_{({\bf 16})})|_{\bar{c}_{({\bf 16})}} = 
{\cal O}(K_S + C_{({\bf 16})})|_{\bar{c}_{({\bf 16})}} = 
{\cal O}(5K_S + \eta)|_{\bar{c}_{({\bf 16})}}; 
\end{equation}
in the last step of the equation above, we used the fact that all the 
terms $a_4$, $a_3 \xi$ etc. appearing in the defining equation 
of $C_{({\bf 16})}$ are sections of ${\cal O}(4K_S + \eta)$.  
Using the exact sequence 
\begin{equation}
 0 \rightarrow {\cal O}_S (K_S) \rightarrow 
   {\cal O}_S (K_S + C_{({\bf 16})}) \rightarrow 
   i_{\bar{c}_{({\bf 16})}*} {\cal O}(K_S + C_{({\bf 16})}) \rightarrow 0
\end{equation}
and its long exact sequence 
\begin{equation}
 0 \rightarrow H^0(S; K_S) \rightarrow H^0(S; {\cal O}(5K_S + \eta)) 
   \rightarrow H^0(\bar{c}_{({\bf 16})}; {\cal O}(K_S + C_{({\bf 16})})) 
   \rightarrow H^1(S; K_S)
\label{eq:long}
\end{equation}
one finds that the space of $\overline{N}\overline{N}^c$, the third term 
in (\ref{eq:long}), is identified with the second term, if 
$h^{2,0}(S) = h^{0,1}(S) = 0$.
Thus, non-vanishing expectation values of $(\overline{N}\overline{N}^c)$ 
correspond to non-vanishing $a_5 \in \Gamma(S; {\cal O}(5K_S + \eta))$, 
and to recombination of the two irreducible pieces $\sigma$ and
$C_{({\bf 16})}$ of the spectral surface.\footnote{The observation so
far improves a similar argument made earlier in \cite{TW-1}.} 

At the reducible limit of the spectral surface, 
the $(\bar{D},L)$-type matter is on the matter curve 
$\bar{c}_{(\bf 16)}$ of SO(10) models ($a_4 = 0$), and the 
$H_u \subset H({\bf 5})$ chiral multiplet is on the 
matter curve $\bar{c}_{\bf vect.}$ ($a_3 = 0$). Thus, the neutrino 
Yukawa couplings are likely to be localized at the 
intersection points of the two matter curves, $(a_3, a_4) = (0,0)$. 
A trilinear interaction 
\begin{equation}
 \Delta W = {\bf 16} \; {\bf 16} \; {\bf vect.}
\end{equation}
is generated at this type of codimension-3 singularity points 
\cite{BHV-1, Hayashi-2}.
The moduli chiral multiplets $\overline{N}$ in (\ref{eq:RHN-41}) are 
also regarded as the ordinary ``right-handed neutrinos'' in the 
spinor representation of $\SO(10)$, and these 
$\overline{N}$ fields do have the neutrino Yukawa couplings 
with $L$ and $H_u$ as a part of the interactions above.  
Thus, the fields (\ref{eq:RHN-41}) are regarded indeed as the 
right-handed neutrinos. 

$\SU(5)_{\rm GUT}$-neutral fields coming from 
${\rm Ext}^1(i_{C_{({\bf 16})}*} {\cal N}_{({\bf 16})}, 
             i_{C_{({\bf 16})}*} {\cal N}_{({\bf 16})})$ 
(or equivalently, the bulk $H^{3,1}$ and $H^{1,2}$ moduli), on 
the other hand, are neutral under the $\U(1)_\chi$ symmetry. Since 
$H_u \subset H({\bf 5})$ and $L \subset (\bar{D},L)$ have 
$Q_\chi = +2$ and $Q_\chi = +3$ units of the $\U(1)_\chi$ charges, 
$\U(1)_\chi$-neutral fields cannot have Yukawa couplings with these 
two fields. Therefore, only (\ref{eq:RHN-41}) moduli can be 
identified with right-handed neutrinos in this scenario.

\subsubsection{SU(6) Scenario}

An alternative to the $a_5 = 0$ scenario (or $\SO(10)$ scenario) 
is to consider another factorization limit of the spectral surface 
\begin{eqnarray}
 & &  
 5\sigma + {\rm div}(a_5 xy + a_4 z x^2 + a_3 y + a_2 x + a_0) 
 \nonumber \\ & \rightarrow & 
 \left[2\sigma + {\rm div}(p_2 x + p_0) \right] + 
 \left[3\sigma + {\rm div}(q_3 y + q_2 x + q_0) \right]
\label{eq:cond-3}
\end{eqnarray}
in Heterotic language. This is to consider a limit where the structure 
group of rank-5 vector bundle $V_5$ becomes 
$\SU(3) \times \SU(2) \times \U(1)_{\tilde{q}_7}$, and 
$V_5 = U_3 \oplus U_2$, where $U_3$ and $U_2$ are rank-3 and rank-2
bundles, respectively. In F-theory language, 
this is to require that for a pair of divisors $\eta'$ and $\eta''$ 
on $S$ that satisfy $\eta' + \eta''= \eta$, there exist 
\begin{equation}
 p_r \in \Gamma(S; {\cal O}(r K_S + \eta')) \quad 
 (r=0,2), \qquad 
 q_r \in \Gamma(S; {\cal O}(r K_S + \eta'')) \quad 
 (r=0,2,3),  
\end{equation}
and $a_r$ ($r=0,2,3,4,5$) are given by 
\begin{eqnarray}
& &  a_0 = p_0 q_0, \qquad a_2 = p_0 q_2 + p_2 q_0, \qquad a_4 = p_2
 q_2, \label{eq:cond-1}\\
& & a_3 = p_0 q_3, \qquad a_5 = p_2 q_3. \label{eq:cond-2}
\end{eqnarray}
No conditions are imposed on other sections such as 
$A_r$'s in (\ref{eq:local-def}) \cite{TW-1, DW-2}.
Under this choice of complex structure of $X$, the defining equation 
of the $\bar{c}_{(\bar{\bf 5})}$ matter curve of $\SU(5)_{\rm GUT}$ 
models vanish identically on $S$:
\begin{equation}
 P^{(5)} = 
 p_0 p_2 q_3^2 (p_2 q_0 - (p_2 q_0 + p_0 q_2) + p_0 q_2) = 0, 
\end{equation}
and the discriminant $\Delta$ begins at the order of $z^6$.
Split $A_5$ singularity is along the surface $S$, and an $\SU(6)$ 
gauge theory is globally defined on $S$. The $\SU(6)$ symmetry 
may be broken down to $\SU(5)_{\rm GUT} \times \U(1)_{\tilde{q}_7}$
or $G_{SM} \times \U(1)_{\tilde{q}_7}$ by turning on fluxes. 
Since the $\U(1)_{\tilde{q}_7}$ symmetry is defined globally on $S$, 
and commutes with the structure group of the Higgs bundle in F-theory 
compactifications, it remains as a global symmetry of low-energy 
effective theory unless it is spontaneously broken by a vev of a field 
with non-vanishing $\U(1)_{\tilde{q}_7}$ charge. The dimension-4 
proton decay operators are absent when this global symmetry 
remains unbroken, just like in the $\SO(10)$ ($a_5 = 0$) scenario.
The essence here is the $\U(1)_{\tilde{q}_7}$ gauge symmetry on $S$, 
nothing else in the bulk of Calabi--Yau 4-fold $X$. Thus, the SU(6)
scenario can be considered in fully generic F-theory compactifications, 
regardless of whether Heterotic dual exists or not.

Moduli ($\SU(5)_{\rm GUT}$-neutral) fields are decomposed 
as in (\ref{eq:moduli-intersection}) in this scenario as well. 
Here, the spectral surface $C_{(\bar{\bf 5})}$ for the fields in 
$\SU(5)_{\rm GUT}$-$\bar{\bf 5}$ representation reduces to 
\begin{equation}
 C_{(\bar{\bf 5})} \rightarrow C_{(\bar{\bf 6})} + C_{(\wedge^2 {\bf
  6})} + \sigma,
\label{eq:C5bar-split}
\end{equation}
and localized moduli are found on the curve $\bar{c}_{(\bar{\bf 6})}$ 
given by the intersection\footnote{
Although the intersection of $\sigma$ and $C_{(\bar{\bf 6})}$ can 
be regarded as that of two irreducible pieces in (\ref{eq:C5bar-split}), 
this intersection can also be regarded as 
the matter curve of $\SU(6)$-$\bar{\bf 6}$ representation 
in $\SU(6)$ unified theories. The discriminant is given by 
\begin{equation}
 \Delta = \frac{z^6}{16} \; p_2^3 q_3^4 S_{(6K_S + 3\eta' + 2\eta'')} +
  {\cal O}(z^7)
\end{equation}
under the choice of complex structure in
(\ref{eq:cond-1}--\ref{eq:cond-2}). Here, $S_{(6K_S + 3\eta' +
2\eta'')}$ is a section of ${\cal O}(6K_S + 3\eta' + 2\eta'')$, and 
is given by $p_{0,2}$, $q_{0,2,3}$ and $A_{0,2,3,4,5}$; its explicit 
form is so messy that we do not write it here, but the zero locus 
of $S_{(6K_S + 3\eta' + 2 \eta'')}$ is the matter curve
$\bar{c}_{(\bar{\bf 6})}$.
The matter curve $\bar{c}_{(\wedge^3 {\bf 6})}$ is given by 
$(p_2 = 0) \in |2K_S + \eta'|$, while 
$(q_3 = 0) \in |3K_S + \eta''|$ is the matter curve 
$\bar{c}_{(\wedge^2 {\bf 6})}$. 
All the terms in $S_{(6K_S + 3\eta' + 2\eta'')}$ contain either $p_2$ 
or $q_3$, and hence the curve $\bar{c}_{(\bar{\bf 6})}$ (where 
$(\bar{D},L)$-type zero modes are localized) pass through 
all the intersection points of the matter curves 
$\bar{c}_{(\wedge^3 {\bf 6})}$ (where $\SU(5)_{\rm GUT}$-{\bf 10} 
fields are localized) and $\bar{c}_{(\wedge^2 {\bf 6})}$ (where  
$\SU(5)_{\rm GUT}$-$\bar{H}(\bar{\bf 5})$ is supported). 
This means that the down-type / charged lepton Yukawa couplings are 
not necessarily suppressed; see \cite{TW-1} for more details. } 
of $C_{(\bar{\bf 6})}$ and $\sigma$.
Let us denote the $\SU(5)_{\rm GUT}$-singlet chiral multiplets 
in the $\SU(6)$-{\bf 6} representation as $\overline{N}$, 
and those in the $\SU(6)$-$\bar{\bf 6}$ representation as
$\overline{N}^c$. In this scenario, the up-type Higgs $H_u$ 
comes from the $H({\bf 5}) \subset {\bf adj.}$ of $\SU(6)$ in the bulk 
of $S$, and lepton doublet $L \subset (\bar{D},L)$ chiral multiplets
are localized along the matter curve $\bar{c}_{(\bar{\bf 6})}$ \cite{TW-1}.
The curve-curve-bulk Yukawa couplings 
\begin{equation}
 \Delta W = \bar{\bf 6} \; {\bf adj.} \; {\bf 6} \supset 
  L \; H_u \; \overline{N}
\end{equation}
is generated all along the curve $\bar{c}_{(\bar{\bf 6})}$ \cite{BHV-1}, 
and this Yukawa couplings and the chiral multiplets $\overline{N}$ 
can be identified with the neutrino Yukawa couplings and the
right-handed neutrinos in this scenario. 
Just like in the $\SO(10)$ ($a_5 = 0$) scenario, singlets from 
$H^{3,1}$ or $H^{1,2}$ that are not associated with the singularity 
of the spectral surface do not have appropriate $\U(1)_{\tilde{q}_7}$ 
charges, and hence cannot have trilinear couplings for neutrino 
Yukawas. Thus, they are not identified with the right-handed 
neutrinos. 

\subsubsection{Majorana masses}

Both the $\SO(10)$ and $\SU(6)$ scenarios above leave an unbroken 
global U(1) symmetry. 
Such a U(1) symmetry is a powerful and reliable way to make sure that 
the dimension-4 proton decay operators are absent, but it is actually 
too powerful. Right-handed neutrinos in the two scenarios above are 
charged under the global $\U(1)_\chi$ or $\U(1)_{\tilde{q}_7}$ symmetry, 
and hence they cannot have Majorana mass terms. Without relying upon 
the see-saw mechanism of right-handed neutrinos, one has to resort 
to the see-saw mechanism of higgsino/wino in the (bilinear) R-parity 
violating scenario \cite{HS, Barbieri}, 
or to assume that the neutrino Yukawa couplings 
are somehow sufficiently small; $\lambda^{(\nu)} < 10^{-12}$.
Although the $H^{3,1}$ moduli irrelevant to the intersection of the
spectral surfaces still have Majorana mass terms from 
(\ref{eq:M-mass-via-GVW}), they cannot have neutrino Yukawa couplings 
because the $\U(1)$ charge do not match. 

One may not exclude a possibility  
that the Majorana masses are generated for fields like (\ref{eq:RHN-41})
through M5-brane 
instanton effects. M5-brane instantons are much like D3-brane 
instantons in Type IIB string theory, and a recent review 
article is available \cite{BCKW}. 
Usually such an amplitude involves a small exponential factor, 
coming from the volume of a divisor of $B_3$ that ``D3-branes'' are 
wrapped. Since the right-handed neutrinos are localized along 
a curve in $S$ in either one of these $\SO(10)$ or $\SU(6)$ scenarios, 
the Kaluza--Klein scale of $S$ and $B_3$, that is, $R_{\rm GUT}$ and
$R_6$, set the scale of the volume of the divisor, unless there are 
some collapsed divisors in $B_3$. As we will see 
in section \ref{sssec:FIdeform}, such an exponential factor is too
small, when the volume is estimated by using $R_{\rm GUT}$ and $R_6$. 
Thus, moduli of geometry have to be tuned so that $B_3$ has a divisor 
with small volume, in order to generate large enough Majorana masses of 
the right-handed neutrinos. 

\subsection{Reducible Limit of $\SU(5)_{\rm GUT}$-$\bar{\bf 5}$ Matter Curve}
\label{ssec:split-curve}

\subsubsection{Just the reducible limit of the matter curve}

It was suggested in \cite{BHV-2, DW-2} that the dimension-4 proton decay 
problem may be solved by considering a reducible limit of the matter
curve $\bar{c}_{(\bar{\bf 5})}$:
\begin{equation}
 \bar{c}_{(\bar{\bf 5})} = \bar{c}_{(\bar{D}L)} + \bar{c}_{(H)}.
\label{eq:curve-split}
\end{equation}
The idea is that the curve $\bar{c}_{(\bar{D}L)}$ supports only the 
three chiral multiplets of matter $(\bar{D},L)$, and 
the up-type and down-type Higgs doublets are supported in the other 
curve\footnote{
An option of taking a further reducible limit of 
$\bar{c}_{(H)} \rightarrow \bar{c}_{(Hu)} + \bar{c}_{(Hd)}$ has been
discussed as a solution to the dimension-5 proton decay problem 
in \cite{BHV-2}. Our discussion in this section \ref{ssec:split-curve} 
is applied, in such a limit, mostly to $\bar{c}_{(\bar{D}L)}$ and 
$\bar{c}_{(Hd)}$. See also discussion at the end of 
section \ref{ssec:split-curve}. Note also that the dimension-5 proton 
decay problem is not as serious in such string compactifications as in 
GUT models on 3+1 dimensions; this is because not much is known about 
the Yukawa couplings involving Kaluza--Klein colored Higgsinos
\cite{Kuriyama, BHV-2}. } $\bar{c}_{(H)}$. At all the codimension-3 singularity points 
for the enhancement $A_4 \rightarrow D_6$, the matter curve 
$\bar{c}_{(\bar{\bf 5})}$ forms a double point, and the idea is to
assume that the two branches of the matter curve passing through the 
double point correspond to $\bar{c}_{(\bar{D}L)}$ and
$\bar{c}_{(H)}$. Mathematically, this assumption is equivalent to 
factorization of the defining equation of the curve 
$\bar{c}_{(\bar{\bf 5})}$ as follows:
\begin{equation}
 P^{(5)}  = a_0 a_5^2 - a_2 a_5 a_3 + a_4 a_3^2 = 
(p_0 a_5 - p_2 a_3)(q_0 a_5 - q_2 a_3).
\label{eq:split-5}
\end{equation}
Here, 
$p_r \in \Gamma (S; {\cal O}(r K_S + \eta'))$ ($r=0,2$),  
$q_r \in \Gamma (S; {\cal O}(rK_S + \eta''))$ ($r=0,2$) and the divisors
$\eta'$ and $\eta''$ on $S$ satisfy $\eta' + \eta'' = \eta$.
As noted in \cite{DW-2}, this condition is actually equivalent to 
the condition (\ref{eq:cond-1}) alone without (\ref{eq:cond-2}), and 
hence can be regarded as a generalization of the $\SU(6)$ scenario 
in section \ref{ssec:split-surf}.

The assumptions in section \ref{ssec:split-surf} surely get rid of 
dimension-4 proton decay operators, but the unbroken U(1) symmetry 
were too powerful because they forbid the Majorana mass terms of 
right-handed neutrinos altogether at perturbative level. 
The condition (\ref{eq:cond-1}) alone is certainly more general, 
and the Majorana masses of right-handed neutrinos may be generated. 
A reducible limit of the matter curve is a weaker condition than the 
reducible limit of the spectral surface.  
But now we do not necessarily have such a U(1) symmetry in the 
effective theory, and it is not absolutely clear whether the 
dimension-4 proton decay operators are absent in the effective theory.
We therefore study in this section \ref{ssec:split-curve} whether 
the reducible limit of the matter curve (\ref{eq:curve-split}) is 
a sufficient condition for the absence of the dimension-4 proton 
decay operators.

It is important to note that reducible limit of the matter curve 
(\ref{eq:curve-split}) does not immediately imply that the sheaf 
cohomology on the curve also splits as in 
\begin{equation}
 H^0(\bar{c}_{(\bar{\bf 5})}; {\cal F}_{(\bar{\bf 5})}) \rightarrow 
  H^0(\bar{c}_{(\bar{D}L)}; {\cal F}_{(\bar{D}L)})
\oplus 
  H^0(\bar{c}_{(H)}; {\cal F}_{(H)}).
\label{eq:split-cohomology}
\end{equation}
For this splitting to take place, one needs to make sure in the 
reducible limit of the curve that the sheaf ${\cal F}_{(\bar{\bf 5})}$
also becomes 
\begin{equation}
 {\cal F}_{(\bar{\bf 5})} \rightarrow 
 i_{(\bar{D}L)*} {\cal F}_{(\bar{D}L)} \; \oplus \;  
 i_{(H)*} {\cal F}_{(H)}, 
\label{eq:split-sheaf}
\end{equation}
where 
\begin{equation}
i_{(\bar{D}L)}: \bar{c}_{(\bar{D}L)} \hookrightarrow 
  \bar{c}_{(\bar{\bf 5})} , \qquad 
i_{(H)}: \bar{c}_{(H)} \hookrightarrow 
  \bar{c}_{(\bar{\bf 5})}.
\end{equation}
Sections of a line bundle ${\cal F}_{(\bar{D}L)}$ 
[resp. ${\cal F}_{(H)}$] form a rank-1 fiber at a given point 
of the matter curve $\bar{c}_{(\bar{D}L)}$ [resp. $\bar{c}_{(H)}$], 
but rank of the the sheaf ${\cal F}_{(\bar{\bf 5})}$ suddenly jumps up 
to 2 at the intersection point of the two curves in the case of 
(\ref{eq:split-sheaf}). The question is whether this condition is 
realized automatically at the reducible limit of the matter 
curve (\ref{eq:curve-split}). 

Let us call $(p_0 a_5 - p_2 a_3) = 0$ piece as 
$\bar{c}_{(\bar{D}L)}$, and $(q_0 a_5 - q_2 a_3) = 0$ as 
$\bar{c}_{(H)}$. The two curves surely intersect at points in 
$S$ where $(a_5, a_3) = (0,0)$. That is where $A_4$ singularity 
is enhanced to $D_6$. We know that the condition (\ref{eq:split-sheaf})
is satisfied there, at least at the level of analyses 
in \cite{Hayashi-1, Hayashi-2}. The other type of intersection points 
are found where 
\begin{equation}
 p_0: p_2 = q_0: q_2 = a_3: a_5,
\label{eq:common-0}
\end{equation}
but $(a_5, a_3) \neq (0,0)$. Since (\ref{eq:common-0}) consists of 
two conditions, (\ref{eq:common-0}) is satisfied at finite number 
of isolated points on $S$ generically. 
Is the condition (\ref{eq:split-sheaf}) satisfied at this type 
of codimension-3 singularities?

Let us study this problem by constructing a field theory model of 
local geometry around a point of this type.
The singularity of local geometry is observed better in a new 
set of coordinates,  
\begin{align}
\displaystyle 
\tilde x &= x + \frac{a_3}{a_5} z^2 + \left( \frac{2}{a_5} \right)^2
 \left( \frac{a_2}{2} - \frac{a_3 a_4}{a_5} + \frac{1}{4} 
 (a_5 A_3 - a_3 A_5) \right) z^3,   \\
\tilde y &= y - \frac{1}{2} \biggl( 
 (a_5 + A_5 z)x + a_3 z^2 + A_3 z^3 \biggl).
\end{align}
In this new set of local coordinates, the defining 
equation (\ref{eq:local-def}) becomes 
\begin{equation}
\tilde y^2 = \frac{a_5^2}{4} \tilde x^2 + \frac{P^{(5)}}{a_5^2} z^5
 - \frac{\tilde R^{(5)}}{a_5^2} z^6 + \mathcal{O}(z^7) + \mathcal{O}(\tilde x^3)
 + \mathcal{O}(z) \tilde x^2 + \mathcal{O}(z^4) \tilde x.
\label{eq:A5-canonical} 
\end{equation}
$P^{(5)}$ is defined as before, and $\tilde{R}^{(5)}$ is given by 
\begin{equation}
 \tilde{R}^{(5)} \equiv  \left(a_2 - \frac{2 a_3 a_4}{a_5} \right)^2
  + \left(a_2 - \frac{2 a_3 a_4}{a_5} \right)(a_5 A_3 - a_3 A_5)
   - a_3^2 A_4 - a_5^2 y_*^2,
\end{equation}
where $y_*^2 \equiv x_*^3 + A_2 x_* + A_0$ and $x_* \equiv - (a_3/a_5)$.
The last four terms of (\ref{eq:A5-canonical}) drop under the scaling 
\begin{equation}
(\tilde x, \tilde y,z) = (\lambda^3 x_0, \lambda^3 y_0, \lambda z_0) 
\quad \lambda \rightarrow 0,
\end{equation}
because their weights are higher than 6. 
The equation (\ref{eq:A5-canonical}) describes a deformation of 
$A_5$ singularity surface in the space with $(\tilde{x}, \tilde{y},z)$ 
local coordinates. The undeformed singularity is $A_4$ at a generic 
point on $S$, but it is enhanced to $A_5$ on either one of the 
matter curves $\bar{c}_{(\bar{D}L)}$ and $\bar{c}_{(H)}$, 
because $P^{(5)} = (p_0 a_5 - p_2 a_3)(q_0 a_5 - q_2 a_3)$ vanishes.
At the intersection points of the two curves, the first two terms 
of $\tilde{R}^{(5)}$ vanish, because
\begin{equation}
 (a_2 a_5 - 2 a_3 a_4) = p_2 (q_0 a_5 - q_2 a_3) + q_2 (p_0 a_5 - p_2 a_3).
\end{equation}
The last two terms in $\tilde{R}^{(5)}$, however, do not have 
a reason to vanish at such intersection points, and hence 
$\tilde{R}^{(5)} \neq 0$ generically. The singularity in the 
$(\tilde{x},\tilde{y},z)$ space remains $A_5$ at this type of 
intersection points, without being enhanced to $A_6$.
The absence of further enhancement can also be seen easily 
from the discriminant (\ref{eq:discriminant}).
The coefficient of the $z^6$ term becomes 
$-a_5^4 \tilde{R}^{(5)}/16 \rightarrow a_5^4 (a_3^2 A_4 + a_5^2
y_*^2)/16$, 
which is the same as the last two terms of $\tilde{R}^{(5)}$.
It does not vanish generically, and hence the discriminant is not 
enhanced to $z^7$ at this type of codimension-3 singularity point. 
It is not that we failed to find an appropriate set of local 
coordinates and/or scaling limit to detect enhanced singularity, 
because the discriminant remains $\Delta \sim z^6$; the singularity 
remains $A_5$.

Local geometry around this type of intersection points, therefore, can be 
modelled by an $\SU(6)$ gauge theory. Let us take a set of 
local coordinates $(u,v)$ on $S$, so that 
\begin{equation}
  u \equiv \frac{p_0}{p_2} - \frac{a_3}{a_5}, \qquad 
  v \equiv \frac{q_0}{q_2} - \frac{a_3}{a_5}. 
\end{equation}
The background field value of $\varphi$ is chosen as\footnote{
In the language of elliptic fibered compactification of Heterotic 
string, the behavior of $\varphi$ is understood as follows. 
The stable degeneration limit of F-theory compactifications corresponds  
to a limit of Heterotic compactifications where supergravity
approximation is good, and we restrict our attention to this region 
of the moduli space. In this limit, $A_4$ is small, and $f_0 = A_2$ 
and $g_0 = A_0$ determine the complex structure of elliptic fiber 
$y^2 = x^3 + f_0 x + g_0$ of Heterotic compactification.
The spectral surface (\ref{eq:spec-surf-Het}) determines 
five points in a given elliptic fiber. For small $(u,v)$, 
the $(x,y)$ coordinates of the two among the five points 
behave as 
\begin{equation}
(x,y)_\pm \sim (x_*, \pm y_*) + 
\mp \frac{p_2 q_2}{a_5} \frac{uv}{y_*^2} 
  \left(1, \pm \frac{3x_*^2 + f_0}{2y_*}\right), 
\nonumber 
\end{equation}
and three other points are determined by 
$(a_5/p_2 q_2)y + x - x_* = 0$ on the elliptic curve.
Thus, the group-law sum of the two points $p_{\pm} = (x,y)_\pm$ 
become $\xi_{p_+ \boxplus p_-} \simeq (p_2 q_2/a_5) uv/y_*^2$; 
$\xi \simeq x/y$ is the local coordinate of the elliptic 
fiber near the infinity point. The addition theorem of elliptic
functions is used here. Values of $\xi$ of nine other 
points of the form $p_j \boxplus p_k$ are not close to zero. 
Thus, the spectral surface $C_{\wedge^2 V_5}$ has only one smooth 
layer $\xi \simeq (p_2 q_2/a_5) uv/y_*^2$ near the zero locus $\xi = 0$, 
and nine other layers are far away from $\xi = 0$.
} 
\begin{equation}
 \vev{\varphi} \propto \diag 
 (\overbrace{0,\cdots,0}^5, uv \; du \wedge dv).
\end{equation}
The zero modes in the $\SU(5)_{\rm GUT}$-$\bar{\bf 5}$ representation 
originate from the $1 \times 5$ lower-left block in the $6 \times 6$ 
matrix representation of $\SU(6)$. An important point is that 
the zero mode wavefunction $(\psi, \chi)$ is in a single component 
representation of the background;  
$\rho(\vev{\varphi}) \propto uv \; du \wedge dv$ acts on the zero mode 
wavefunction as a $1 \times 1$ matrix.
Zero mode wavefunctions $(\psi, \chi)$ will approximately be Gaussian 
$e^{-|u|^2}$ near the matter curve $\bar{c}_{(\bar{D}L)}$ ($u=0$), 
and will be Gaussian $e^{-|v|^2}$ near the matter curve $\bar{c}_{(H)}$ 
($v=0$), but the wavefunctions on the both branches 
cannot take different values at the intersection point. 
Both $(\psi, \chi)|_{\bar{c}_{(\bar{D}L)}}$ along 
$(0, {}^\forall v)$ on $\bar{c}_{(\bar{D}L)}$ and 
$(\psi, \chi)|_{\bar{c}_{(H)}}$ along $({}^\forall u, 0)$ 
on $\bar{c}_{(H)}$ should approach the same value 
$(\psi,\chi)|_{(u,v) = (0,0)}$ at the intersection point. 
A line bundle ${\cal N}_{(\bar{\bf 5})}$ is on the smooth spectral 
surface given by $\xi \propto uv$, and the sheaf ${\cal F}_{(\bar{\bf 5})}$ 
on the curve 
$\bar{c}_{(\bar{\bf 5})} = \bar{c}_{(\bar{D}L)} + \bar{c}_{(H)}$
is a restriction of a line bundle ${\cal N}_{(\bar{\bf 5})} \otimes K_S$
on $\bar{c}_{(\bar{\bf 5})}$ given by $\xi = 0$ on the spectral
surface. The sheaf ${\cal F}_{(\bar{\bf 5})}$ remains strictly rank-1 
at any points of the curve $\bar{c}_{(\bar{\bf 5})}$, and the rank 
does not jump up to 2 at the intersection point. The zero modes 
in this case are not regarded as locally free fluctuations on 
$\bar{c}_{(\bar{D}L)}$ and locally free fluctuations on $\bar{c}_{(H)}$
that are mutually independent locally, but the wavefunctions on the 
two curves are constrained to have the same value at the intersection 
points. 

Thus, the reducible limit of the matter curve alone does not 
guarantee that the zero modes of the $(\bar{D},L)$ type and 
Higgs type split into the two distinct curves.
Therefore, we conclude that taking the reducible limit of the matter 
curve alone does not help remove the dimension-4 proton decay 
operators.

\subsubsection{tuning more parameters}
\label{sssec:more-tuning}

With a little more tuning of coefficients of the defining 
equation (\ref{eq:local-def}), the coefficient of the 
$z^6$ term ($\propto a_3^2 A_4 + a_5^2 y_*^2$) can be made vanish 
at each of such intersection points. The number of such points 
is given topologically by 
\begin{equation}
 (5K_S + \eta + \eta') \cdot (5K_S + \eta + \eta'') - 
 (5 K_S + \eta) \cdot (3 K_S + \eta).
\end{equation}
There may be many of them, and the same number of complex structure 
moduli may have to be tuned by hand, yet one might accept this tuning 
in order to avoid the dimension-4 proton decay. 

Now the field theory local model is an $\SU(7)$ gauge theory. 
By rescaling the local coordinates $(u, v)$ if necessary, 
the defining equation of the spectral surface $C_{(\bar{\bf 5})}$ 
can be made 
\begin{equation}
 A \xi^2 + (u + v) \xi + uv = 0, 
\label{eq:spec-surf-conic}
\end{equation} 
and the $\xi = 0$ locus on $C_{(\bar{\bf 5})}$, that is $uv = 0$, 
is the matter curve $\bar{c}_{(\bar{\bf 5})}$. 
$A$ is a coefficient that we do not specify; rescaling of local
coordinates $(u,v)$ and $\xi$ cannot absorb this $A$.
Local geometry of $X$ is given by $\tilde{y}^2 - \tilde{x}^2 \simeq 
z^5 (A z^2 + (u + v)z + uv)$, and two topologically independent 
2-cycles in the $(\tilde{x}, \tilde{y},z)$ space vanish on 
$u=0$ ($\bar{c}_{(\bar{D}L)}$) and $v=0$ ($\bar{c}_{(H)}$) respectively.
Thus, there are two extra topologically independent vanishing 2-cycles at the 
intersection point $(u,v) = (0,0)$, and it may be possible that the 
independent degrees of zero modes are locally associated with these 
independent 2-cycles. If this is proved to be true,\footnote{As the 
spectral surface (\ref{eq:spec-surf-conic}) is singular at 
$(\xi,u,v)=(0,0,0)$, structure of the sheaves ${\cal N}_{(\bar{\bf 5})}$ 
and ${\cal F}_{(\bar{\bf 5})}$ is not obvious there.} then 
the splitting of the matter fields in $\SU(5)_{\rm GUT}$-$\bar{\bf 5}$ 
representation as desired in (\ref{eq:split-cohomology}, 
\ref{eq:split-sheaf}) may follow, as {\it classification of zero-mode 
degrees of freedom}. 

It should be noted, however, that the spectral surface 
$C_{(\bar{\bf 5})}$ given by (\ref{eq:spec-surf-conic}) is 
still a single irreducible surface unless $A = 1$. The matter curve is 
reducible, but the spectral surface is not. As we have learnt 
in \cite{Hayashi-2}, zero-mode wavefunctions of charged matter fields 
are regarded as single component $(\psi, \chi)$ wavefunctions on the 
spectral surface. Even for a zero-mode whose wavefunction can be
characterized by a holomorphic section on the curve $u=0$ [resp. $v=0$]
in $C_{(\bar{\bf 5})}$, we cannot imagine that the wavefunction 
$(\psi, \chi)$ is absolutely zero in any open subset of 
$C_{(\bar{\bf 5})}$; if it were, then it would be zero everywhere 
on $C_{(\bar{\bf 5})}$. 
Thus, zero modes that may be classified as 
``$H^0(\bar{c}_{(\bar{D}L)}; {\cal F}_{(\bar{D}L)})$'' should also have 
non-vanishing 
wavefunctions along the $\bar{c}_{(H)}$ curve (and vice versa).

If we are to phrase this statement in language of field theory local 
models, that will be as follows. 
One can choose $\SU(7)$ as the gauge group of field theory local 
models around points satisfying (\ref{eq:common-0}). Along the 
matter curve $\bar{c}_{(H)}$ [resp. matter curve
$\bar{c}_{(\bar{D}L)}$] away from the $(u,v) = (0,0)$ points, however, 
field theory models with an $\SU(6)_H \subset \SU(7)$ [resp. 
$\SU(6)_M \subset \SU(7)$] gauge group can provide a reasonable
approximation. An argument in the previous paragraph means that 
zero modes that may be classified as 
``$H^0(\bar{c}_{(\bar{D}L)}; {\cal F}_{(\bar{D}L)})$'' also have 
non-vanishing single component $(\psi,\chi)$ wavefunction in 
the $\SU(6)_H$ gauge theory along the curve $\bar{c}_{(H)}$.

Now, it is at the codimension-3 singularity points $(a_3, a_5) = (0,0)$ 
that the trilinear couplings for the dimension-4 proton decay
(\ref{eq:dim-4}) are potentially generated. $\SO(12)$ gauge theories 
can be chosen as field theory local models of geometry around this 
type of singularity. Two branches of the matter curve
$\bar{c}_{(\bar{\bf 5})}$ pass through this type of points, and under 
the assumption (\ref{eq:split-5}), the two branches correspond to 
$\bar{c}_{(\bar{D}L)}$ and $\bar{c}_{(H)}$ respectively. 
Field theory local models with $\SU(6)_{H} \subset \SO(12)$ 
[resp. $\SU(6)_M \subset \SO(12)$] gauge group can provide a good
approximation along the matter curve $\bar{c}_{(H)}$ 
[resp. $\bar{c}_{(\bar{D}L)}$] away from the $(a_3, a_5) = (0,0)$
points. Any zero modes in the $\SU(5)_{\rm GUT}$-$\bar{\bf 5}$
representation are assigned two-component $(\psi,\chi)$ wavefunctions 
in the $\SO(12)$ local models, and one of the two components correspond 
to the single component $(\psi, \chi)$ wavefunction in the 
$\SU(6)_H \supset \SU(5)_{\rm GUT}$ gauge theory, and the other
component to the single component wavefunction in the 
$\SU(6)_M \supset \SU(5)_{\rm GUT}$ gauge theory. 
The trilinear couplings (\ref{eq:dim-4}) are calculated by overlap 
integration in the $\SO(12)$ gauge theory, and the overlap integration 
picks up both of the two components. Thus, the couplings would vanish 
if zero modes that may be classified as 
``$H^0(\bar{c}_{(\bar{D}L)}; {\cal F}_{(\bar{D}L)})$'' had vanishing 
wavefunction in the $\SU(6)_H$ gauge theory. We have learnt in the
previous paragraph, however, that this is not the case. 
Therefore, the trilinear couplings are indeed generated, even after 
tuning the complex structure of a 4-fold $X$ so that 
$(a_3^2 A_4 + a_5^2 y_*^2)$ vanish at all the points satisfying 
(\ref{eq:common-0}).

Of course the value of the $(\bar{D},L)$-like wavefunction may well 
be small in the $\SU(6)_H$ gauge theory along $\bar{c}_{(H)}$ far away 
from the points satisfying (\ref{eq:common-0}). 
The question is whether such a mixing of the wavefunction is small 
enough to satisfy a phenomenological constraint, 
\begin{equation}
 \sqrt{\lambda'' \lambda'} \lesssim 10^{-13},   
\label{eq:dim4-limit}
\end{equation}
which comes from experimental lower bounds on the proton lifetime.
To make an estimate of the size of the trilinear couplings generated 
in this scenario, we make a conservative assumption that 
the zero-mode wavefunctions decrease as in Gaussian profile far away from 
the points satisfying (\ref{eq:common-0}).
Then, the wavefunction of the $(\bar{D},L)$-like matter fields 
along the $\bar{c}_{(H)}$ branch at a $(a_5,a_3) = (0,0)$ point
is expected to be of order 
\begin{eqnarray}
 e^{- \left(\frac{L}{d}\right)^2} & > &
 e^{- \left(\frac{R_{\rm GUT}}{d}\right)^2}
 = e^{- \sqrt{\frac{R_{\rm GUT}^4}{g_s l_s^4}} \frac{1}{(M_* d)^2} } 
  \nonumber \\
 & \simeq & 10^{- 2.1 \frac{1}{(M_* d)^2}} 
 \simeq 10^{- 13. \frac{1/(2\pi)}{(M_*d)^2} }
 \simeq 10^{- 84. \left(\frac{1/(2\pi)}{M_* d}\right)^2 }.
\label{eq:non0-dim4}
\end{eqnarray}
Here, $d$ is the width of the Gaussian profile, 
$L$ is a distance to a point where $(a_5,a_3) =(0,0)$, which is 
always smaller than $R_{\rm GUT}$. 
Since the size of $S$ is finite, not infinitely large, wavefunctions 
cannot be smaller than the value given above; the ratio between 
the Kaluza--Klein radius and the width of localized wavefunctions sets 
the smallest possible value of trilinear couplings (c.f. \cite{HSW}).  
For a rough estimate at the first try, we have used 
$(R_{\rm GUT} M_*)^2$ for $(R_{\rm GUT}/d)^2$, and we know 
its value from (\ref{eq:val-alpha_GUT}).
Given the three crude estimates above, which correspond to 
$d \sim 1/M_*$, $1/(\sqrt{2\pi} M_*)$ and $1/(2\pi M_*)$, 
it is therefore crucial to know the relation between $d$ and $1/M_*$
more precisely, to see if this scenario can be a viable solution 
to the dimension-4 proton decay problem. 

Let us suppose that the local defining equation is 
\begin{equation}
 y^2 \simeq x^2 + z^5 (f u + z + \cdots),
\label{eq:D7-intersect}
\end{equation}
where local coordinates $u, z$ are made dimensionless by some unit 
length $l_*$, and $f$ is a dimensionless numerical coefficient. 
This equation describes a geometry near a matter curve at $u=0$. 
To be a little more general, we can think of a spectral surface given by 
\begin{equation}
 f u + \xi + \cdots = 0,
\end{equation}
where the fiber coordinate $\xi$ of $\mathbb{K}_S$ is also 
dimensionless. It is natural to imagine that the dimensionless 
coefficient $f$ is of order unity, although it should be an issue 
to be confirmed ultimately by flux compactification. 
Since the geometry in (\ref{eq:D7-intersect}) allows for an 
interpretation as an intersecting D7--D7 system, we can determine 
the $2\alpha\varphi_{12}$ field vev for (\ref{eq:D7-intersect}), 
without an ambiguity in the normalization. Using the fact that 
the mass of an open string state stretching a distance $D$ is 
\begin{equation}
 m = \frac{1}{2\pi \alpha'} D,
\end{equation}
we find\footnote{See the appendix of \cite{Hayashi-2} for details 
on the normalization convention.} that 
$2\alpha \varphi_{12} = (4\pi \alpha')^{-1} f u'$, 
where $u' \equiv u l_*$ is the local coordinate with a physical 
mass dimension $-1$ restored. Corresponding Gaussian wavefunction 
is $e^{- (4\pi\alpha')^{-1} f |u'|^2}$. The typical width parameter, 
therefore, turns out to be $d \sim \sqrt{4 \pi \alpha'}$. 
If we further make a crude approximation\footnote{See footnote
\ref{fn:gs1}.} $g_s \sim {\cal O}(1)$, 
then $d \sim 1/(\sqrt{\pi} M_*)$, and\footnote{This parameter 
$(d/R_{\rm GUT})^2 \sim \sqrt{\alpha_{\rm GUT}}/\pi$ 
may also sets the hierarchical scale of flavor physics 
(c.f. \cite{HV-Nov08}).} 
\begin{equation}
 \left(\frac{R_{\rm GUT}}{d}\right)^2 \sim \pi (R_{\rm GUT} M_*)^2
  = \frac{\pi}{\sqrt{\alpha_{\rm GUT}}} \simeq \frac{1}{0.06} \simeq 15..
\end{equation}
Thus, we have a lower bound on the $R$-parity violating trilinear couplings:
\begin{equation}
 \lambda, \lambda', \lambda'' \gtrsim 10^{-7}.
\label{eq:dim4-LBd}
\end{equation} 
Given the so many crude approximations we have made (especially 
$g_s \sim {\cal O}(1)$ and $f \sim {\cal O}(1)$), the inconsistency 
between the phenomenological limit (\ref{eq:dim4-limit}) and 
the lower bound (\ref{eq:dim4-LBd}) should not be taken as an argument 
excluding the scenario in this section \ref{sssec:more-tuning}. 
One should also keep in mind, however, that the lower bound 
(\ref{eq:dim4-LBd}) is based on an inequality $L < R_{\rm GUT}$ 
that is virtually never saturated, and furthermore, wavefunctions 
do not always damp as fast as in the Gaussian profile; 
see \cite{Hayashi-2} and the appendix of this article.

If this scenario is phenomenologically acceptable, then the neutrino 
Yukawa couplings are generated around the codimension-3 singularity
points at the points satisfying (\ref{eq:common-0}), 
because that is where the wavefunctions of lepton doublets 
$L\subset (\bar{D}, L)$ and $H_u \subset H({\bf 5})$ 
are not small. Right-handed neutrinos are identified either with 
$H^{3,1}$ or $H^{1,2}$ in this scenario. The Yukawa couplings can be 
calculated by the $\SU(7)$ gauge theory local model, and the
wavefunctions of right-handed neutrinos can be dealt with as in 
the prescription given in section \ref{sec:nuYukawa-general}.
Majorana masses are generated for the $H^{3,1}$ moduli from flux 
compactification, as we have explained in section \ref{sec:Majorana}.

\subsubsection{yet another limit of reducible spectral surface}

In order to safely remove the mixing of the wavefunction altogether 
between the $H_d \subset \bar{H}(\bar{\bf 5})$-like matter and 
$L \subset (\bar{D},L)$-like matter fields, one should consider
a limit where the spectral surface $C_{(\bar{\bf 5})}$ is
reducible:
\begin{equation}
 C_{(\bar{\bf 5})} = C_{(\bar{D}L)} + C_{(H)}, 
\label{eq:surface-split}
\end{equation}
where the $\xi = 0$ loci of $C_{(\bar{D}L)}$ and $C_{(H)}$ become 
the matter curves $\bar{c}_{(\bar{D}L)}$ and $\bar{c}_{(H)}$ in 
(\ref{eq:curve-split}), respectively. 
The zero-mode wavefunctions of the $L \subset (\bar{D},L)$-like matter 
becomes absolutely zero on the $C_{(H)}$ piece of $C_{(\bar{\bf 5})}$ 
in this case, and hence the couplings of the dimension-4 proton 
decay operators vanish.\footnote{
The factorization limit of the spectral surface (\ref{eq:surface-split})
here does not have an easy interpretation within the $E_8$ gauge
theory in the dual Heterotic 
language. This is not a reducible limit of a vector bundle in 
an $E_8$; it is easy to see this because the tuning required in 
section \ref{sssec:more-tuning} is to set 
$y_*^2 = x_*^3 + f_0 x_* + g_0 = 0$ at the points satisfying 
(\ref{eq:common-0}), and this condition involves not just moduli 
of the spectral surface, but also the complex structure parameter 
$f_0$ and $g_0$ of the elliptic fibration. }
  
In the $\SU(7)$ gauge theory which models the local geometry around 
points satisfying (\ref{eq:common-0}), 
this factorisation of the spectral surface is realized by further 
tuning complex structure so that $A = 1$ in (\ref{eq:spec-surf-conic}).
The two pieces $C_{(\bar{D}L)}$ ($\xi + u = 0$) and $C_{(H)}$ 
($\xi + v = 0$) intersect along $u= v = - \xi$, and form a double curve 
singularity. Since the gauge group of this local model is $A_N$-type, 
a natural Type IIB interpretation exists; this double curve singularity 
is nothing but D7--D7 intersection. 

Now remember that the $H^{3,1}$ and $H^{1,2}$ moduli fields are 
captured as 
${\rm Ext}^1(i_* {\cal N}_{({\bf 10})}, i_* {\cal N}_{({\bf 10})})$ 
in $\mathbb{K}_S$ in the field theory local models, where 
$i_* {\cal N}_{(\bf 10)}$ [resp. ${\cal N}_{(\bf 10)}$] is the 
Higgs sheaf [resp. line bundle supported on the spectral surface 
$C_{(\bf 10)}$] for the fields in the $\SU(5)_{\rm GUT}$-{\bf 10}
representation. The double curve singularity in the spectral surface 
$C_{(\bar{\bf 5})}$ (other than those in the local models around
codimension-3 singularity points of $A_4 \rightarrow D_6$ enhancement) 
for the fields in the $\SU(5)$-$\bar{\bf 5}$ 
representation indicates that the spectral surface $C_{(\bf 10)}$ 
for the fields in the $\SU(5)_{\rm GUT}$-${\bf 10}$ representation 
also has a double curve singularity. $C_{(\bf 10)}$ is not irreducible, 
but it also splits into $C_1 + C_2$. Then, the Higgs sheaf 
${\cal V} = i_* {\cal N}_{(\bf 10)}$ also splits as 
$i_{1*} {\cal N}_1 + i_{2*} {\cal N}_2$, and the moduli fields also
split as in (\ref{eq:moduli-intersection}).
The last two components in (\ref{eq:moduli-intersection}) are localized 
at the intersection of the two pieces, or intuitively, the D7--D7 
intersection curve.
In the absence of ramification of the spectral surface, we purely 
have an $\SU(7)$ field theory local model. Open string interaction 
generates neutrino Yukawa couplings for the one of the last two 
components above, which correspond to the off-diagonal pieces in 
the 2 by 2 block of the 7 by 7 matrix of $\SU(7)$. 
We have thus arrived at the picture assumed in \cite{BHV-2}.

It is not obvious whether there is a global unbroken $\U(1)$ symmetry 
as an explanation for the absence of dimension-4 proton decay 
in this scenario with a factorized spectral surface. This is 
a crucial question because the Majorana masses of the right-handed 
neutrinos are forbidden as long as such an unbroken $\U(1)$ symmetry 
exists. At least, in the local field theory models with $\SO(12)$ and 
$\SU(7)$ gauge groups, the spectral surface for 
the fields in the $\SU(5)$-$\bar{\bf 5}$ representation does not ramify, 
and one can find $\U(1)$ symmetry transformations in these gauge
theories, where the lepton doublets and $H_d$ have distinct charges 
under the $\U(1)$'s.\footnote{Whether an associated U(1) gauge symmetry 
remains massless and anomaly free in low-energy effective theory is 
yet another (and often global) issue, and we will not discuss here.}  
It is not obvious, however, whether one should 
maintain such a $\U(1)$ symmetry in all the field theory local models 
of the patches covering the $A_4$ singularity surface $S$; the
factorization limit of the spectral surface (\ref{eq:surface-split}) 
is sufficient in removing all the dimension-4 proton decay operators, 
and it is not clear if the factorization limit immediately implies 
the existence of a $\U(1)$ symmetry in the effective theory. If it does
not, then we do not strictly need a symmetry.\footnote{It certainly goes 
against a common sense of field theory model building to claim that 
certain operators are absent without an explanation in terms of
symmetry, but such things may or may not happen in string theory. 
We do not have any arguments in favor of or against such a mechanism 
of vanishing couplings without a symmetry reason.} 

Even more controversial is whether the factorization 
limit (\ref{eq:surface-split}) is well-defined. 
The spectral surface of Higgs bundle is defined in F-theory 
compactifications only in field theory {\it local models}. 
One can choose field theory local models with $\SU(6)$ or $\SO(10)$ 
gauge groups along the matter curves, and local models can be chosen 
with gauge groups $\SU(7)$, $\SO(12)$ and $E_6$ at codimension-3 
singularity points. These choices, however, are just a 
the minimal choice preserving essential features of the local 
geometries. For higher level of approximation, local models can be
replaced by gauge theories with higher-rank gauge groups.
For example, in F-theory compactifications with Heterotic dual, 
one can choose $E_8$ gauge theory as local models at any patches 
of $S$ (or even globally on $S$), not just the minimal rank-1 or 
rank-2 extension of the common $\SU(5)_{\rm GUT}$ over $S$.
Even in F-theory compactifications with Heterotic dual, however, 
it is a good approximation to cut out the rank-5 Higgs bundle 
from the rest only in the stable degeneration limit. If the 
complex structure moduli are not necessarily in this limit, then 
taking just the rank-5 part into the field theory formulation on $S$ 
and discarding all the rest is not a systematically justified 
approximation. How can one extend the gauge group of the local 
models beyond $E_8$ to achieve a higher level of approximation 
within the field theory formulation? The situation is essentially 
the same in generic F-theory compactifications; field theory local 
models can capture local geometry of $X$ near the $A_4$ singularity 
surface $S$, but the field theory formulation does not offer a 
systematic way (order by order) to capture the entire geometry 
of $X$ for higher level of approximation. 
The field theory formulation can capture only a local geometry 
that is approximately a deformed ADE singularity fibered over a local 
patch of $S$. The whole geometry of $X$ is compact and is not an ALE
fibration on $S$.
Since the spectral surfaces of the Higgs bundles can be defined 
only within the field theory local models, the reducibility 
(factorization) of the spectral surfaces can also be defined order 
by order in this approximation, which will never be able to cover 
the entire geometry of $X$. At this moment, we do not have 
a clear idea\footnote{The factorization (reducibility) of the 
spectral surface in 
the $\SU(7)$ local models is the same as the factorization of the 
discriminant locus. Thus, it might seem at first sight that the 
factorization condition of the spectral surface can be replaced 
by the factorization condition of the discriminant locus. 
But these conditions are actually totally different. 
As explained in section 4.3 of \cite{Hayashi-2}, the spectral 
surface for the $\SU(5)_{\rm GUT}$-$\bar{\bf 5}$ representation fields 
consists of two irreducible pieces in the $\SO(12)$ local models around 
the codimension-3 singularities with enhanced $D_6$ singularity, whereas 
the corresponding discriminant locus consists of a single irreducible
piece.}  how to define the factorization limit (\ref{eq:surface-split}) 
rigorously. For the real-world physics, however, the constraints from 
phenomenology (\ref{eq:dim4-limit}) always leave a room for very small 
couplings for the dimension-4 proton decay operators.
Thus, it may be an option to enforce 
factorization limit in gauge theory local models with higher-rank 
gauge groups so that sufficiently high level of approximation is 
achieved, and trilinear couplings as small as $10^{-13}$ can be discussed.

Before closing this section \ref{ssec:split-curve}, we comment on a variation of the scenario that has been discussed 
so far. We have discussed this scenario along the line of 
(\ref{eq:curve-split}), where neither $H_u \subset H({\bf 5})$ 
nor $H_d \subset \bar{H}(\bar{\bf 5})$ are localized in the irreducible 
curve $\bar{c}_{(\bar{D}L)}$. As a solution to the dimension-4 proton 
decay problem, however, only the distinction between $H_d$ and the 
three lepton doublets is essential. Thus, $H_u$ may originate from 
the same curve as the lepton doublets. That is, we can consider another 
reducible limit,\footnote{The factorization 
condition of the spectral surface may be relaxed to the level of 
tuning in section \ref{sssec:more-tuning}.} 
\begin{equation}
 \bar{c}_{(\bar{\bf 5})} \rightarrow \bar{c}_{(\bar{D}LHu)} + 
 \bar{c}_{(Hd)}, \qquad 
 C_{(\bar{\bf 5})} \rightarrow C_{(\bar{D}LHu)} + C_{(Hd)}.
\end{equation}
This is theoretically possible; what was discussed in \cite{Ibanez} 
is essentially the same thing from theoretical perspectives. 
In this new factorization scenario, neutrino Yukawa couplings 
are generated just as in section \ref{sec:nuYukawa-general}, and 
the $H^{3,1}(X; \C)$ moduli have Majorana masses, just as in 
section \ref{sec:Majorana}. There is nothing to worry about the Majorana
masses in the absence of possible protection by a $\U(1)$ symmetry. 
Neutrino Yukawa couplings have contributions all along the curve 
$\bar{c}_{(\bar{D}LHu)}$, and the Majorana masses come from the 
entire bulk of $B_3$.
The $\Delta W = S H_u H_d$ interaction of the NMSSM, on the other hand, 
is localized at the points satisfying (\ref{eq:common-0}) in this
factorization limit, if such a massless singlet chiral multiplet $S$ 
exists in the spectrum.

Phenomenology of supersymmetry breaking terms in the scenarios 
in this section \ref{ssec:split-curve} is beyond the scope of 
this article. 

\subsection{$R$-parity Violating Scenarios}
\label{ssec:Rparity-violation}

An unbroken $\U(1)$ symmetry is powerful in removing the 
dimension-4 proton decay operators, but it also forbids 
the Majorana mass terms of right-handed neutrinos. That would 
be the executive summary of section \ref{ssec:split-surf}, and 
may be also of the $C_{(\bar{D}L)} + C_{(HuHd)}$ splitting scenario 
in section \ref{ssec:split-curve}. 
If we could find a field $\phi$ with even units of charge of 
an unbroken $\U(1)$ symmetry, then an unbroken $\Z_2$ symmetry would be 
found after spontaneous breaking of the $\U(1)$ by a vev of the 
field $\vev{\phi}$, but one still has to find how trilinear couplings 
$\Delta W = \phi \overline{N} \overline{N}$ like (\ref{eq:126})
would be generated. 
We have not found a way to discover such fields and such couplings.\footnote{
A caveat in this argument is mentioned at the end of 
section \ref{ssec:split-surf}.}
If one throws away the $\U(1)$ symmetry altogether and just 
impose a $\Z_2$ symmetry from the beginning, then that is 
the $\Z_2$ parity scenario in section \ref{ssec:R-parity}. 

Actually there is a caveat in this argument, however. 
Suppose that we begin with a compactification that leaves  
an unbroken U(1) symmetry. This U(1) symmetry can be broken 
spontaneously by vev's only of chiral multiplets with, 
for example, positive U(1) charges. Let us denote such chiral 
multiplets as $\phi_+$. We assume that all the fields with 
negative charges under the U(1) symmetry do not have non-vanishing 
vev's.
Suppose that right-handed neutrinos $\bar{N}$ have a negative 
U(1) charge, while the U(1) charge of the dimension-4 proton 
decay operators $\bar{\bf 5} \; {\bf 10} \; \bar{\bf 5}$ is positive. 
Then the Majorana masses of right-handed neutrinos are allowed 
by the spontaneously broken symmetry, because the effective Majorana 
mass parameter $M_R$ can involve the vev's with positive U(1) charges, 
$M_R \sim \vev{\phi_+}^n$. 
On the other hand, the dimension-4 proton decay operators are 
still forbidden by the broken symmetry, because of the absence of 
chiral multiplet vev's with negative U(1) charges.\footnote{This
selection rule is applied to, and only to, renormalizable operators 
in low-energy effective theories below the Kaluza--Klein scale. 
See \cite{TW-LSP, Kuriyama} for the discussion.} 
Supersymmetric D-term condition can be satisfied for this U(1) 
symmetry, because the vev's of the positively charge field can 
balance against a negative Fayet--Iliopoulos parameter $\xi$ 
proportional to $\omega \wedge F$. 

This scenario was proposed in \cite{TW-1}, and studied in detail 
in \cite{Kuriyama}. Study in \cite{TW-1, Kuriyama} was done mostly in 
language of Heterotic string compactification. 
Although two of the authors (RT and TW) made an effort to provide 
a description of this scenario in F-theory language as well 
in \cite{TW-1}, various theoretical aspects of F-theory
compactifications and the duality between Heterotic string and F-theory 
were not as clearly understood back then as they are now, and the 
translation from Heterotic description to F-theory description was 
not completed there. With a better theoretical understanding of 
F-theory compactification, we now provide a little better version. 

In the language of Heterotic $E_8 \times E_8$ string compactification 
on a Calabi--Yau 3-fold $Z$, the key idea of \cite{TW-1} was to use 
a rank-5 vector bundle $V_5$ with a structure 
\begin{eqnarray}
& & 0 \rightarrow L_\chi \rightarrow V_5 \rightarrow U_4 \rightarrow 0 
 \qquad \qquad {\rm or} \label{eq:ext-41}\\
& & 0 \rightarrow U_2 \rightarrow V_5 \rightarrow U_3 \rightarrow 0.
 \label{eq:ext-32}
\end{eqnarray}
Here, $L_\chi$ [resp. $U_2$] is a rank-1 [resp. rank-2] sub-bulde of $V_5$, 
whose structure group is $\SU(5)_{\rm str} \subset E_8$.
Zero mode chiral multiplets in the $\SU(5)_{\rm GUT}$-$\bar{\bf 5}$
representation are $H^1(Z; \wedge^2 V_5)$ in general Heterotic string 
compactifications, but for $V_5$ with such a sub-bundle as above, 
a subspace 
\begin{equation}
 H^1(Z; L \otimes V_5) \subset H^1(Z; \wedge^2 V_5) 
 \qquad {\rm or} \qquad 
 H^1(Z; U_2 \otimes V_5) \subset H^1(Z; \wedge^2 V_5)
\label{eq:matter-subsp}
\end{equation}
is well-defined, and this subspace is identified with the 
$(\bar{D},L)$-type matter fields; 
the $H_d \subset \bar{H}(\bar{\bf 5})$ field, on the other hand, is 
regarded as a generic element of $H^1(Z; \wedge^2 V_5)$.
Chiral multiplets ${\bf 10} = (Q, \bar{U},\bar{E})$ 
are identified with $H^1(Z; U_2) \subset H^1(Z; V_5)$ in the 
case (\ref{eq:ext-32}).
All of the down-type/charged lepton Yukawa couplings and 
the dimension-4 proton decay operators originate from the 
product 
\begin{equation}
 H^1(Z; \wedge^2 V_5) \times H^1(Z; V_5) \times H^1(Z; \wedge^2 V_5)
 \rightarrow H^3 (Z; \wedge^5 V_5) = H^{0,3}(Z; \C)
\end{equation}
in the Heterotic string superpotential
\begin{equation}
 \Delta W_{\rm Het} = \int_Z \Omega \wedge \tr \left(
   A dA + i \frac{2}{3} A A A \right).
\label{eq:Het-super}
\end{equation}
The product vanishes when both of the $H^1(Z; \wedge^2 V_5)$ elements 
are in the subspace (\ref{eq:matter-subsp}), and hence the dimension-4 
proton decay operators are absent. See \cite{TW-1, Kuriyama} for more 
about this scenario. 

The extension structures (\ref{eq:ext-41}, \ref{eq:ext-32}) can be 
regarded as spontaneous breaking of $\U(1)$ symmetries.\footnote{
Not all of vector bundles with sub-bundles may admit such 
an interpretation. The following discussion, therefore, should be taken
for granted only for bundles that can be constructed that way.}
A rank-5 vector bundle $L_\chi \oplus U_4$ [resp. $U_2 \oplus U_3$] 
has a structure group 
$\SU(4) \times \U(1)_\chi \subset \SU(5)_{\rm str}$
[resp. $\SU(2) \times \SU(3) \times \U(1)_{\tilde{q}_7}$]. 
The structure group in both cases has a U(1) factor. 
The rank-5 bundle $V_5$ is unstable, if the Fayet--Iliopoulos parameter 
$\xi \propto \int_Z \omega \wedge \omega \wedge F$ of the $\U(1)$ 
symmetry does not vanish. In the case (\ref{eq:ext-41}), for example, 
we assume that $\xi_\chi$ is negative, and chiral multiplets 
$\overline{N}^c \in H^1(Z; L_\chi \otimes \overline{U}_4) \subset 
 H^1(Z; {\rm adj}(V_5))$ with positive $\U(1)_\chi$ charge $Q_\chi$ absorb 
the Fayet--Iliopoulos parameter. This is why $L_\chi$ remains a 
well-defined sub-bundle, but $U_4$ does not. 

Heterotic string compactification has an F-theory dual, 
when the Calabi--Yau 3-fold $Z$ is an elliptic fibration 
\begin{equation}
\pi_Z: Z \rightarrow S 
 \label{eq:ZtoS}
\end{equation}
over a complex surface $S$.
In the following, we start from a Heterotic compactification 
on such a Calabi--Yau 3-fold $Z$ with a vector bundle $V_5$ 
constructed from a pair of vector bundle ($L_\chi$, $U_4$) or 
($U_2$, $U_3$) as above, and find out its F-theory dual description. 

Bilinear $R$-parity violation \cite{HS} is generated in the scenario 
explained above \cite{Kuriyama}. An order of magnitude estimate of the 
bilinear $R$-parity violating parameters was given in \cite{Kuriyama}, 
using weakly coupled Heterotic string compactification. 
$R$-parity violating decay of gravitino dark matter has been discussed 
as one of the possible explanations of the recent cosmic ray 
anomalies \cite{Pamela}. 

\subsubsection{sub-bundle with vanishing first Chern class in the fiber}
\label{sssec:FIdeform}

An F-theory dual description of this scenario becomes quite different, 
depending on whether $c_1(L_\chi) = - c_1 (U_4)$ 
[resp. $c_1(U_2) = - c_1(U_3)$] vanishes in the elliptic fiber direction 
of (\ref{eq:ZtoS}), or it is strictly negative \cite{TW-1}.
Let us begin with the case with vanishing first Chern classes 
in the fiber direction.

In this case, bundles $U_{1,2}$ ($U_1 \equiv L_\chi$ hereafter) 
and $U_{4,3}$ may be given by Fourier--Mukai transform 
separately. The spectral data, $(C_{k}, {\cal N}_k)$ ($k = 1,2,4,3$), 
for Heterotic compactification can readily be used for 
the spectral data of Higgs bundles in F-theory 
compactification \cite{Hayashi-2}.
The case $V_5 = U_4 \oplus U_1$ (with a vanishing Fayet--Iliopoulos 
parameter $\xi_\chi$ even on the base space $S$) then corresponds 
to the $\SO(10)$ scenario in section \ref{ssec:split-surf}, and 
$V_5 = U_3 \oplus U_2$ to the $\SU(6)$ scenario in 
section \ref{ssec:split-surf}.

Discussion in section \ref{ssec:split-surf} corresponds to 
Higgs bundles in F-theory compactification where 
both $\omega \wedge F$ and $[ \varphi, \overline{\varphi}]$ 
vanish in the the first one of the BPS conditions (\ref{eq:BPS}).
More general, however, is Higgs bundles where the first condition 
is satisfied as a combination 
$\omega \wedge F - |\alpha|^2 [\varphi, \overline{\varphi}]/2$, but 
not separately. 
This corresponds to non-vanishing Fayet--Iliopoulos parameter 
for the $\U(1)_\chi$ [$\U(1)_{\tilde{q}_7}$] symmetry;  
non-vanishing Fayet--Iliopoulos parameter is equivalent to 
$\omega \wedge F \neq 0$. The Fayet--Iliopoulos parameter 
triggers spontaneous breaking of the $\U(1)$ symmetry;  
if the parameter is negative [resp. positive], then chiral 
multiplets with positive [resp. negative] U(1) charge have 
tachyonic masses, and develop non-vanishing expectation values, 
which makes the $[\varphi, \overline{\varphi}]$ term non-zero. 
In the end, stable minimum with vanishing D-term potential is 
equivalent to a $(A,\varphi)$ field configuration on $S$ satisfying 
the BPS conditions. 

In the $\SO(10)$ scenario, for example, only the positively 
charged chiral multiplets $\overline{N}^c$ need to develop 
non-zero vev's, in order to cancel the negative Fayet--Iliopoulos 
parameter, and $\vev{\overline{N}}$ remains zero. 
The $\vev{\varphi}$ configuration now have non-zero off-diagonal 
entries in the $5 \times 5$ matrix representation as 
in (\ref{eq:varphi-vac}), 
but not in a way the spectral surface is affected.  
As discussed in section \ref{ssec:split-surf}, 
only the vev of $\vev{\overline{N} \overline{N}^c}$ modify the 
spectral surface; $\vev{\overline{N}^c}$ alone do not. The spectral 
surface only extracts information associated with eigenvalues of 
$\vev{\varphi}$, but symmetry breaking pattern in Higgs bundle 
is not always encoded only by the eigenvalues\footnote{See 
also \cite{DKS}, where nilpotent Higgs vev is discussed in 
Type IIB compactifications.} of $\vev{\varphi}$.

In the $\SO(10)$ scenario, a rank-1 bundle $L_\chi$ remains 
a $\varphi$-invariant sub--Higgs-bundle of the rank-5 Higgs bundle on $S$. 
In the $\SU(6)$ scenario, a rank-2 sub-Higgs-bundle remains
$\varphi$-invariant. 
The rank-5 Higgs bundle in these scenarios are, so to speak, 
constructed by the parabolic construction on $S$; see section 5 
of \cite{FMW} for the parabolic construction of vector bundles on 
elliptic fibered Calabi--Yau manifolds $Z$. 

Majorana mass terms of right-handed neutrinos can be generated 
in such a vacuum as long as an effective theory has a
non-renormalizable term 
\begin{equation}
 \Delta W = \overline{N} \overline{N} \overline{N}^c \overline{N}^c.
\label{eq:4N}
\end{equation}
Either $\overline{N}^c$ or $\overline{N}$ develop non-vanishing 
expectation values to balance the non-vanishing Fayet--Iliopoulos
parameter $\xi \propto \omega \wedge F$, and then the counter part 
acquire Majorana masses \cite{TW-1, Kuriyama}. This term also lifts 
the D-flat direction of the U(1) symmetry \cite{DSWW1}.

Two independent origins of the effective interaction (\ref{eq:4N}) 
have been discussed \cite{TW-1, Farragi, Kuriyama} in language 
of Heterotic string compactification. 
One of them \cite{Farragi, Kuriyama} is that the interaction 
(\ref{eq:4N}) is generated in effective theory, when Kaluza--Klein 
modes are integrated out of a field theory description. 
The superpotential (\ref{eq:Het-super}) contains 
interactions not only of Kaluza--Klein zero modes, but also Kaluza--Klein 
mass terms (the mass parameter coming from the derivative $d$) and 
trilinear interactions involving the Kaluza--Klein
modes. 
\begin{equation}
 W = \sum_I M_I \Phi_I \Phi_I + 
    \sum_I \lambda_I \overline{N}^c \Phi_I \overline{N} + \cdots.
\label{eq:2&3}
\end{equation}
Non-renormalizable terms like (\ref{eq:4N}) are generated 
in the effective theory, when the Kaluza--Klein modes $\Phi_I$ are 
integrated out (c.f. \cite{WittenSU(3)});  
\begin{equation}
 \Delta W = \left(\sum_I \frac{\lambda_I^2}{M_I} \right)
 \; \; \overline{N} \overline{N}  \overline{N}^c \overline{N}^c.
\end{equation}
There is no reason to doubt that the same thing happens in F-theory 
compactifications, because the superpotential for F-theory 
(\ref{eq:F-super}) is essentially the same as that of the Heterotic 
string compactification (\ref{eq:Het-super}); only difference is that 
the F-theory superpotential (\ref{eq:F-super}) has Kaluza--Klein modes 
on $S$ alone, not on the Kaluza--Klein modes on Calabi--Yau 3-folds, 
but the Kaluza--Klein modes on $S$ should be enough in generating 
interactions like (\ref{eq:4N}). Thus, the Majorana 
mass terms are generated for the right-handed neutrinos in F-theory 
compactifications also in the $\SO(10)$ and $\SU(6)$ scenarios with 
a non-vanishing Fayet--Iliopoulos parameter. 

The mass scale of the Majorana mass is given approximately by 
\begin{equation}
 M_R \sim \frac{(\lambda \vev{N})^2}{M_{\rm KK}} \sim \lambda^2
  \frac{\xi}{M_{\rm KK}},
\label{eq:MR-estimate-4NKK}
\end{equation}
where $\xi$ is the Fayet--Iliopoulos parameter. 
The Fayet--Iliopoulos parameter has been calculated in Type IIB 
Calabi--Yau orientifold compactifications. Applying the result of \cite{Louis} 
naively for a case a Calabi--Yau 3-fold $\widetilde{B}$ is like 
$C \times S$, and 
${\rm vol}(\widetilde{B}) = {\rm vol}(C) \times {\rm vol}(S) = 
R_{\perp}^2 R_{\rm GUT}^4$ (like in the tubular model of 
Fig.~\ref{fig:(an)-isotropic}~(b)), and restoring proper 
dimensionality and $g_s$ dependence, we obtain 
\begin{equation}
 \xi \sim M_{\rm Pl}^2 \frac{1}{\pi} 
     \frac{g_s l_s^4}{R_{\rm GUT}^2 R_{\perp}^2}.
\end{equation}
Since this expression comes in a combination $g_s l_s^4 = 1/M_*^4$, 
which remains constant (relatively to $l_{11}$) everywhere in $B_3$ in 
generic F-theory compactifications, we dare to use this expression 
for F-theory compactifications that are not necessarily Type IIB
Calabi--Yau orientifolds. This estimate of the Fayet--Iliopoulos 
parameter is simplified by using
(\ref{eq:val-alpha_GUT}--\ref{eq:val-M-Pl}) as
\begin{equation}
 \xi \sim 4 M_*^4 R_{\rm GUT}^2 \sim 
  \frac{4}{\alpha_{\rm GUT}} \frac{M_{\rm GUT}^2}{c^2}.
\end{equation}
This result perfectly agrees with the Heterotic result in \cite{TW-1}, 
up to a factor of ${\cal O}(1)$ that we did not care about here.
This result does not depend on $R_{\perp}$ or on the geometry in 
the direction transverse to $S$. Canonically normalized zero modes 
$|\overline{N}^c|$ [resp. $|\overline{N}|$] develop vev's of order 
$\sqrt{|\xi|}$, meaning that the original Higgs bundle that corresponds to 
$U_4 \oplus U_1$ [resp. $U_3 \oplus U_2$] receives an order-one 
correction to become a Higgs bundle with the extension structure 
as in (\ref{eq:ext-41}, \ref{eq:ext-32}). 
An estimate of the Majorana masses of right-handed neutrinos is 
obtained by plugging the estimate of $\xi$ in
(\ref{eq:MR-estimate-4NKK}). Typical value of the trilinear couplings 
$\lambda_I$ in (\ref{eq:2&3}) are of order $g_{\rm GUT}$ with
suppression factors coming from overlap integration of normalized 
wavefunctions (c.f \cite{HSW} for more details). The overlap integration 
tends to be smaller in the trilinear couplings like those 
in (\ref{eq:2&3}) (c.f. \cite{Kuriyama}), 
because the overlap integral involves two almost ``flat 
wavefunctions'' for two zero modes and one ``higher Fourier mode'' 
for the Kaluza--Klein states. Two $g_{\rm GUT}$ coming from $\lambda^2$ 
cancel those in $1/\alpha_{\rm GUT}$ in $\xi$, and the mass scale of 
right-handed neutrinos in this scenario is somewhere around the GUT
scale with a suppression factor coming from the overlap integrals.
This result fits very well with phenomenological 
expectation (\ref{eq:MR-bound}).

The other known mechanism for generating (\ref{eq:4N}) is the 
world-sheet instanton effect in the language of Heterotic string 
compactification \cite{DSWW1, TW-1}. Although such world-sheet instanton 
effects are known to cancel for certain class of (0,2) Heterotic
compactifications, there may be other choices of geometries other than 
in such a class, and one does not have to rule this possibility out. 
A world-sheet instanton contribution from a curve $\Sigma$ in the base 
surface $S$ in Heterotic theory corresponds in F-theory \cite{WittenM5} 
to M5-brane (Euclidean D3) instanton contribution from a divisor 
of $B_3$ that is projected on to the same curve of $S$ in the $\P^1$ fibration 
\begin{equation}
 \pi_{B_3}: B_3 \rightarrow S.
\label{eq:P1-fib}
\end{equation}
An exponential suppression factor associated with this non-perturbative 
effect is of order 
\begin{equation}
 \exp \left.\left(
   - \frac{{\rm vol}(\Sigma)}{2\pi \alpha'}\right)\right|_{\rm Het}
 = e^{- \left[(2\pi)M_*^4 R_{\rm GUT}^2 R_{\perp}^2 \right]|_{\rm F}}
 = e^{- \frac{2\pi}{\alpha_{\rm GUT}}\frac{1}{\epsilon^2}}
 \simeq 10^{-660},
\label{eq:M5-estimate}
\end{equation}
and is too small to be relevant for phenomenology. 
However, this estimate is very crude, and does not take account 
of a possibility that there may be a collapsed divisor in $B_3$. 
Thus, this estimate does not completely exclude a possibility 
that the effective interactions (\ref{eq:4N}) are generated by 
M5-brane instanton effects. 

We have so far discussed this $R$-parity violating scenario 
in F-theory compactification, by starting from a class of 
Heterotic string compactifications, and translating into 
F-theory language using the duality. Rank-5 vector bundles 
in Heterotic compactification on a Calabi--Yau 3-fold $Z$ are 
translated into rank-5 Higgs bundles on a base 2-fold $S$ 
for F-theory compactification. The extension structure of the 
vector bundles (and hence the subspace structure 
of the zero modes) in Heterotic theory is carried over to F-theory 
compactification as the extension structure of the Higgs bundles
(and as the subspace structure of the zero modes).
This way of understanding, however, raises a question 
whether this scenario is possible only in F-theory compactification 
with Heterotic dual. 
In F-theory compactification with a Heterotic dual, a Higgs bundle with 
a fixed rank can be defined globally on $S$. 
Generic F-theory compactifications, however, have field theory local models 
only locally on $S$, and physics associated with $S$ (that is, GUT physics)
has to be recovered by gluing those local models together. 
Since the nature of gluing process is at most approximate, there is 
a concern that the notion of sub-Higgs-bundle may not be well-defined
globally on $S$. If so, that would be a problem, given the severe
constraint on the couplings of the dimension-4 proton decay operators. 

In each local field theory model of a generic F-theory compactification, 
however, there is a well-defined $\U(1)$ symmetry: $\U(1)_\chi$ in 
the $\SO(10)$ scenario, and $\U(1)_{\tilde{q}_7}$ in the $\SU(6)$
scenario. The absence of the dimension-4 proton decay operators is 
guaranteed by the $\U(1)$ symmetry broken only by positively 
[resp. negatively] charged fields, and hence the $R$-parity violating 
scenario in this section \ref{sssec:FIdeform} is available not just 
for F-theory compactifications with Heterotic dual.\footnote{The
estimate of the suppression factor of M5-brane instanton effect 
can be a little more moderate in generic F-theory compactifications, 
as $(1/\epsilon^{\gamma = 1})^2$ in (\ref{eq:M5-estimate}) may be 
replaced by $(1/\epsilon^{\gamma=1/3})^2$ in the homogeneous model. 
This does not make a practical difference, though, as the exponential 
suppression factor remains extremely small.} 

\subsubsection{sub-bundle with non-vanishing first Chern class in the fiber}

Let us now study the F-theory dual description of the scenario 
explained at the beginning of this section \ref{ssec:Rparity-violation}, 
in the case the first Chern class in the elliptic fiber direction does
not vanish. For stability of $V_5$, 
$c_1(U_{1,2})|_{T^2} = - c_1(U_{4,3})|_{T^2}$ is negative.\footnote{
Section 5 of \cite{FMW} explains in detail how to construct a vector 
bundle $V_5$ on an elliptic fibered Calabi--Yau manifold, using 
a vector bundle ${\cal W}_k$ for $U_{4,3}$ and ${\cal W}_{5-k}^*$ 
for $U_{1,2}$. The bundles ${\cal W}_k$ and ${\cal W}_{n-k}^*$ satisfy 
$c_1({\cal W}_k)|_{T^2} = - c_1({\cal W}^*_{n-k})|_{T^2} = 1$.
} We will consider a region of the moduli space where the K\"{a}hler
class of the $T^2$ fiber is smaller than those of the base $S$.
This is where the Heterotic--F theory duality with 16 SUSY charges 
can be promoted to the duality with smaller number of SUSY charges 
adiabatically. 
 
Since the bundles $U_{1,2}$ and $U_{4,3}$ have non-vanishing 
first Chern classes in the fiber direction, they are not 
given by spectral cover construction. It is thus non-trivial to see 
even such a thing as whether the charged matter fields are localized 
in the F-theory dual description. Let us begin with addressing this question.

Because $c_1(U_{1,2})$ in the fiber direction is negative, 
and $c_1(U_{4,3})|_{T^2}$ positive, $R^0\pi_{Z*} U_{1,2}$ and 
$R^1 \pi_{Z*} U_{4,3}$ vanish. 
Since $V_5$ restricted on a fiber becomes a flat bundle, and is 
non-trivial generically, $R^0 \pi_{Z*} V_5 = 0$. Thus, this 
exact sequence follows: 
\begin{equation}
  0 \longrightarrow R^0 \pi_{Z*} U_{4,3} \longrightarrow R^1 \pi_{Z*} U_{1,2}
    \longrightarrow R^1 \pi_{Z*} V_5 \longrightarrow 0.
\label{eq:short}
\end{equation}
The support of the sheaves $R^0\pi_{Z*} U_{4,3}$ and 
$R^1 \pi_{Z*} U_{1,2}$ is the entire $S$, but the support 
of $R^1 \pi_{Z*} V_5$ can be a curve in $S$.
The map from $R^0 \pi_{Z*} U_{4,3}$ to $R^1 \pi_{Z*} U_{1,2}$ 
is to multiply a global section of 
$R^1 \pi_{Z*} {\rm adj.}V_5 = 
 R^1 \pi_{Z*} (\overline{U}_{4,3} \otimes U_{1,2})$, which describes 
how $U_{4,3}$ are extended by $U_{1,2}$ to become $V_5$ in each fiber. 
Remember \cite{FMW} that the moduli of this global holomorphic section is 
identified with that of the spectral surface, when 
$U_{4,3}$ and $U_{1,2}$ are bundles ${\cal W}_{4,3}$ and 
${\cal W}_{1,2}^*$ constructed in \cite{FMW} and 
${\cal W}_k \oplus {\cal W}_{5-k}^*$ ($k=4,3$) is minimally unstable. 
The support of $R^1 \pi_{Z*} V_5$ is where the extension of 
$U_{4,3}$ in the $T^2$ fiber by $U_{1,2}$ becomes less non-trivial.

Massless matter fields in the $\SU(5)_{\rm GUT}$-{\bf 10} representation
are identified with $H^1(Z; V_5) \simeq H^0(S; R^1\pi_{Z*} V_5)$. Using 
the exact sequence (\ref{eq:short}), one finds a following long exact
sequence where $H^0(S; R^1\pi_{Z*} V_5)$ is in:
\begin{equation}
\begin{split}
 0 & \rightarrow H^0(S; R^0 \pi_{Z*} U_{4,3}) \rightarrow 
   H^0(S; R^1 \pi_{Z*} U_{1,2}) \rightarrow H^0(S; R^1 \pi_{Z*} V_5) \\
 & \quad \quad \quad \rightarrow H^1(S; R^0 U_{4,3}) \rightarrow 
   H^1(S; R^1 \pi_{Z*} U_{1,2}).
\end{split}
\end{equation}
Thus, $H^1(Z; V_5) \simeq H^0(S; R^1 \pi_{Z*} V_5)$ has a subspace 
\begin{equation}
{\rm Coker}( H^0(S; R^0 \pi_{Z*} U_{4,3}) \rightarrow 
             H^0(S; R^1 \pi_{Z*} U_{1,2}) ), \label{eq:coker} 
\end{equation}
and the quotient by this subspace is 
\begin{equation}
{\rm Ker}( H^1(S; R^0 \pi_{Z*} U_{4,3}) \rightarrow 
             H^1(S; R^1 \pi_{Z*} U_{1,2}) ). \label{eq:ker} 
\end{equation}
The vector space of the zero modes, $H^0(S; R^1 \pi_{Z*} V_5)$, has 
a subspace structure like (\ref{eq:matter-subsp}).
A chain complex (\ref{eq:short}) is regarded as the essence of this 
subspace structure in F-theory language. 

Similar argument can be repeated for the zero modes in the 
$\SU(5)_{\rm GUT}$-$\bar{\bf 5}$ and -{\bf 5} representations.\footnote{
A sequence of sheaves on $S$ similar to (\ref{eq:short}) is derived 
for $R^1 \pi_{Z*} \wedge^2 V_5$, following the same line of argument as
for $R^1\pi_{Z*} V_5$. However, we have no idea how to construct 
something like a principal-bundle version of such sequences for
different representations.} 
The subspace structure of the zero modes (\ref{eq:matter-subsp}) 
does follow for these representations. The $(\bar{D},L)$-type fields 
are identified with this subspace. Since the trilinear couplings 
(\ref{eq:dim-4}) vanish at least in Heterotic string compactifications 
with supergravity approximation, the same should be true in the stable 
degeneration limit of F-theory compactifications that have Heterotic dual.

Massless moduli fields coming from 
$H^1(S; R^0\pi_{Z*} U_4 \otimes U_1^{-1})$ have trilinear couplings 
with $L \subset (\bar{D},L) \in H^1(Z; U_1 \otimes V_5)$ and 
$H_u \subset H({\bf 5}) \in H^1(Z; \wedge^2 \overline{U}_4)$ in the 
case with the structure (\ref{eq:ext-41}). Thus, they are identified 
with the right-handed neutrinos $\overline{N}$. Moduli fields 
$H^0(S; R^1 \pi_{Z*} \overline{U}_4 \otimes U_1)$, denoted as 
$\overline{N}^c$, do not have the couplings to be identified with 
the neutrino Yukawa couplings. 
In the case with the structure (\ref{eq:ext-32}), on the other hand, 
$H^0(S; R^1 \pi_{Z*} U_2 \otimes \overline{U}_3)$ is identified with 
the right-handed neutrinos $\overline{N}$ \cite{TW-1}. 

Many issues are still beyond the scope of this article. We have not 
discussed whether the dimension-4 proton decay operators are still 
protected in moduli space not necessarily at the stable degeneration 
limit. Such issues as how important the world-sheet instanton effects 
would become in small $T^2$-fiber limit (in Heterotic language) or 
how to define the subspace structure more rigorously despite the 
``approximate'' nature of the field theory formulation of F-theory 
remain totally unaddressed. 
More detailed study of the structure of (\ref{eq:short}--\ref{eq:ker}) 
and that for $R^1\pi_{Z*} \wedge^2 V_5$ is also desirable. 

 \section*{Acknowledgements}  

We thank Teruhiko Kawano, with whom we maintained close 
communications during the entire period of this work.
TW thanks Kentaro Hori and Kyoji Saito for useful comments.
This work is supported by PPARC (RT), and by WPI Initiative, 
MEXT, Japan (TW).


\appendix

\section{Branch Cut, Orbifold and 2-Form VEV}

Non-Abelian gauge theories arising from F-theory compactification 
are described conveniently by using the field theory on 8-dimensions 
introduced by \cite{KV, DW-1, BHV-1}. Local geometry of Calabi--Yau 4-fold $X$ 
and base 3-fold $B_3$ containing a divisor $S$, the discriminant locus 
for the non-Abelian gauge fields in 3+1-dimensions, is encoded as 
the choice of gauge group in a local region of $S$ and a field vev 
in the field theory that models the local geometry \cite{KV, DW-1,BHV-1, 
Hayashi-2}. Such filed theories constructed in local patches of $S$ 
are glued together to reproduce all the information encoded in 
the geometry of $X$ and $B_3$. It is of crucial importance, 
therefore, to properly translate the geometry into the field vev 
of field theory local models. 

The choice of the field background for F-theory compactification 
on a Calabi--Yau 3-fold (that is, the base manifold is a 2-fold, 
$S$ is a curve and low-energy effective theory is on 5+1-dimensions)
is discussed in \cite{KV}. The codimension-1 locus where 1-form 
field $\varphi$ on $S$ vanishes is called matter locus. Nothing 
else to discuss. Upon compactification to 3+1 dimensions, however, 
one more extra complication appears \cite{Hayashi-2}. Local geometry 
of $X$ around $S$ is regarded as fibered space with the fiber being 
a surface with (partially) deformed ADE singularity, but the topological 
2-cycles in the fiber have monodromies, and these monodromies introduce 
branch cuts and monodromies under the Weyl group in the field thoery. 
Branch locus and matter locus are both codimension-1 loci in a surface
$S$,  and hence their intersections become points. 
These points are codimension-3 in $B_3$.
It is inevitable to have these codimension-3 singularity points in 
compactification down to 3+1 dimensions. 
How to choose the vev of $\varphi$ field around such codimension-3 
singularity points actually still remains a bit of an issue. 

The choice of $\varphi$ vev around codimension-3 singularities 
was one of the main issues in \cite{Hayashi-2}. 
Reference \cite{HV-Nov08-rev} further noted that the field theory 
local modes on $S$ with branch cuts and the Weyl-group monodromies 
has equivalent descriptions on the covering space $\widetilde{S}$ of $S$.
The latter description does not involve branch cuts or twists, 
and the former description is obtained by taking a quotient 
of the latter.
The $\varphi$ field vev configuration 
in \cite{Hayashi-2} and that of \cite{HV-Nov08-rev} are largely 
the same under this identification, but there still is a difference. 
This is why the choice of $\varphi$ for a given geometry around codimension-3 
singularity points still remains an issue. 
As we will note at the end of this appendix, this difference 
in $\vev{\varphi}$ leads directly to difference in phenomenology. 
Thus, this is not just an academic issue. 

Let us first briefly review the relation between the field theory 
local models with branch cuts \cite{Hayashi-2} and the description 
on the covering space \cite{HV-Nov08-rev}. We choose simplest case 
for illustrative purpose: a local geometry given by 
\begin{equation}
 y^2 = x^2 + (z^2 + 2 u z + v + u^2) z^N,
\end{equation}
where $(u,v)$ are local coordinates of $S$. 
This local geometry can be modelled by a field theory on a local 
region of $S$ with $\SU(N+2)$ gauge group \cite{Hayashi-2}, and 
it was claimed in \cite{Hayashi-2} that the field vev 
$\vev{\varphi} = \vev{\varphi_{uv}} du \wedge dv$ should be chosen as 
\begin{equation}
 \vev{\varphi_{uv}}|_{2 \times 2} = \diag (\tau_+, \tau_-); \qquad 
  \tau_{\pm} = - u \pm \sqrt{- v}.
\label{eq:our-choice}
\end{equation}
$\tau_{\pm}$ above corresponds to the two roots of 
$(z^2 + 2 u z + v + u^2) = 0$.  The branch locus is $v = 0$.
The spectral surface 
is 
\begin{equation}
 \xi^2 + 2 u \xi + v + u^2 = 0.
\end{equation}

Because the monodromy around the branch locus $v=0$ is the 
Weyl reflection $\Z_2 = \mathfrak{S}_2$ of the $\SU(2) \subset \SU(N+2)$
structure group, one can describe the the same field theory in 
the covering space $\widetilde{S}$; the field theory with branch locus 
above can be regarded as a $\Z_2$ orbifold of a theory on the covering
space \cite{HV-Nov08-rev}.\footnote{It is certainly a concern whether 
``twisted sectors'' have to be introduced or not. Nothing is known about 
this issue up until now, however. } Let the local coordinates of 
$\widetilde{S}$ be $(s,t)$, and the vev was chosen in
\cite{HV-Nov08-rev} as 
\begin{equation}
 \vev{\varphi}|_{2 \times 2} = \diag (s, t) \; ds \wedge dt.
\label{eq:lin-vev}
\end{equation}
This background has a $\Z_2$ symmetry transformation 
acting on $\widetilde{S}$ as exchange of the coordinates $s$ and $t$, 
accompanied by the Weyl-group transformation
\begin{equation}
 \varphi_{st} =
 \left(\begin{array}{cc}
       s & \\ & t
 \end{array}\right) \rightarrow 
   \left( \begin{array}{cc}
	  & 1 \\ 1 & 
   \end{array}\right) 
    \left(\begin{array}{cc}
       s & \\ & t
 \end{array}\right) 
   \left( \begin{array}{cc}
	  & 1 \\ 1 & 
   \end{array}\right). 
\end{equation}
Thus, one can think of a quotient of the free $\SU(N+2)$ 
gauge theory on $\widetilde{S}$ under the $\Z_2$ symmetry 
transformation. To see the relation between the description 
on the covering space $\widetilde{S}$ and the quotient space $S$, 
it is convenient to take the following coordinates on $\widetilde{S}$:
\begin{equation}
 u = - \frac{1}{2}(s+t), \qquad \tilde{v} = \frac{1}{2}(s - t). 
\end{equation}
The $\Z_2$ transformation flips the sign of $\tilde{v}$, while 
$u$ remains invariant, and hence a point $(u, \tilde{v})$ in 
$\widetilde{S}$ is sent to $(u,v) = (u, - \tilde{v}^2)$ in $S$
under the $\Z_2$ quotient map.
The filed vev (\ref{eq:lin-vev}) becomes 
\begin{equation}
 \vev{\varphi}|_{2 \times 2} = \diag(\tau_+, \tau_-) \; 2 du \wedge d
  \tilde{v} = \diag(\tau_+, \tau_-) \; du \wedge dv 
    \left(\frac{-1}{\sqrt{-v}}\right).
\end{equation}
This $\varphi$ vev configuration is almost the same as the one in 
(\ref{eq:our-choice}), but differs by a factor $v^{-1/2}$.
For a given point $(u,v)$ in $S$, the difference is the overall 
scaling between $\mathfrak{h}/W \otimes \C$-valued $\diag(\tau_+, \tau_-)$ 
and $\diag(\tau_+, \tau_-)/\sqrt{-v}$, which cannot be determined by 
the dictionary in \cite{KM}.

We have discussed in the main text how to obtain the 2-form field 
vev $\vev{\varphi}$ from the 4-form $\Omega$ on the original 
Calabi--Yau 4-fold. Discussion leading 
to (\ref{eq:z-pm}, \ref{eq:varphi-vac-2})
revealed 
that (\ref{eq:our-choice}) is the right choice. The deformation 
parameters of ADE singularities are regarded as sections of
$\mathfrak{h}/W \otimes K_S$, and hence the deformation parameters 
should be identified with the coefficients of the holomorphic top 
form made out of the local coordinates of $S$, not with those 
of the top form made of the local coordinates of the covering space 
$\widetilde{S}$.

The choice of the background configuration of $\varphi$ is relevant 
to phenomenology, because it determines the asymptotic behavior 
of the zero mode wavefunctions away from the matter curves. 
If the background were (\ref{eq:lin-vev}), we knew that the 
zero mode wavefunctions would be $e^{-|s|^2}$ and $e^{-|t|^2}$ 
in the covering space $\widetilde{S}$, and they become a doublet 
wavefunction $(e^{-|-u + \sqrt{-v}|^2}, e^{-|-u-\sqrt{-v}|^2})$ 
on the $\Z_2$ quotient $S$. In an asymptotic region where 
$|u|$ remains small, but $|v|$ becomes large, this zero mode 
wavefunction would decrease as $e^{-|v|}$, not as fast as in the 
Gaussian profile $e^{-|v|^2}$. It turns out, however, that 
(\ref{eq:our-choice}) is the right choice for the $\varphi$ field 
background, and the wavefunction falls as $e^{-|v|^{3/2}}$ in the 
asymptotic region, which is a little faster than the $e^{-|v|}$  
fall off, but still slower than in the Gaussian profile. 
As one can see in the discussion around (\ref{eq:non0-dim4}), 
how fast zero mode wavefunctions fall off in regions away from the 
matter curves is an important issue in phenomenology. Such a 
difference in the wavefunction profile in the asymptotic region 
also affects the phenomenological analysis of flavor pattern 
in \cite{HSW, HSW-2} as well.

\end{document}